\documentclass[12pt]{article}

\usepackage[
    geometry=yurie,
    tikz=off,
    biblatex=yurie,
    hyperref=yurie
]{yurie}
\usepackage{yurie-math}
\usepackage{yurie-phy}

\usepackage{float}

\numberwithin{equation}{section}

\allowdisplaybreaks

\definecolor{Incoming}{rgb}{0.8,0,0}
\definecolor{Outgoing}{rgb}{0,0,0.8}
\definecolor{Tachyonic}{rgb}{0.5,0.15,0.8}

\renewcommand{\inn}{{\color{Incoming}\mathsf{i}}}
\renewcommand{\out}{{\color{Outgoing}\mathsf{o}}}
\renewcommand{\io}{{\color{Incoming}\mathsf{i}}/{\color{Outgoing}\mathsf{o}}}
\renewcommand{\oi}{{\color{Outgoing}\mathsf{o}}/{\color{Incoming}\mathsf{i}}}

\newcommand{\mumu}[2]{\mu_{#1} \dotsm \mu_{#2}}
\newcommand{\nunu}[2]{\nu_{#1} \dotsm \nu_{#2}}
\newcommand{\rhorho}[2]{\rho_{#1} \dotsm \rho_{#2}}

\newcommand{\symL}{{\color{brown}\boldsymbol{(}}}
\newcommand{\symR}{{\color{brown}\boldsymbol{)}}}

\newcommand{\chit}{\chi_{\mathsf{t}}}
\newcommand{\chibt}{\bar{\chi}_{\mathsf{t}}}

\newcommand{\deltaK}[1]{\delta_{\mathrm{K}}(#1)}
\newcommand{\deltaComplex}[1]{\delta_{\CC}(#1)}
\newcommand{\deltaMC}[1]{\delta^{(4)}(#1)}

\newcommand{\tsign}[1]{{\color{Tachyonic}#1}}
\newcommand{\ts}[1]{{\color{Tachyonic}s_{#1}}}

\newcommand{\cpwflimit}{N}

\newcommand{\shadow}{\mathcal{S}}

\newcommand{\coTwo}[1]{\mathcal{C}(#1)}
\newcommand{\coThree}[1]{\mathcal{C}(#1)}
\newcommand{\coOPE}[1]{\mathcal{C}(#1)}
\newcommand{\coBlock}[1]{\mathscr{C}(#1)}

\newcommand{\gmShort}[1]{\Gamma\left[#1\right]}
\newcommand{\betaShort}[1]{B\left(#1\right)}
\newcommand{\mgShort}[2]{\Gamma\left[\genfrac..{0pt}{}{#1}{#2}\right]}

\newcommand{\massMellin}{\gamma}

\newcommand{\pseries}{1+i\, \RR}

\newcommand{\region}{R}

\newcommand{\become}{$\to$\xspace}

\newcommand{\Tall}{\texttt{\textcolor{Tachyonic}{T}}\xspace}
\newcommand{\Tplus}{\texttt{\textcolor{Tachyonic}{T}}\textsuperscript{\tsign{+}}\xspace}
\newcommand{\Tminus}{\texttt{\textcolor{Tachyonic}{T}}\textsuperscript{\tsign{-}}\xspace}
\newcommand{\Minn}{\texttt{\textcolor{Incoming}{M}}\xspace}
\newcommand{\Mout}{\texttt{\textcolor{Outgoing}{M}}\xspace}
\newcommand{\Oinn}{\texttt{\textcolor{Incoming}{O}}\xspace}
\newcommand{\Oout}{\texttt{\textcolor{Outgoing}{O}}\xspace}
\newcommand{\Sinn}{\texttt{\textcolor{Incoming}{\~{O}}}\xspace}
\newcommand{\Sout}{\texttt{\textcolor{Outgoing}{\~{O}}}\xspace}

\newcommand{\oom}{\Oout{}\Oout{}\Minn}
\newcommand{\mom}{\Mout{}\Oout{}\Minn}
\newcommand{\too}{\Tall{}\Oinn{}\Oout}
\newcommand{\tom}{\Tall{}\Oinn{}\Mout}
\newcommand{\tot}{\Tall{}\Oinn{}\Tall}
\newcommand{\soooS}{\Sinn{}\Oinn{}\Oout{}\Oout}
\newcommand{\moooS}{\Minn{}\Oinn{}\Oout{}\Oout}
\newcommand{\toooS}{\Tall{}\Oinn{}\Oout{}\Oout}
\newcommand{\moooD}{\Minn{}\Oout{}\Oout{}\Oout}

\DeclareSymbolFont{lettersA}{U}{txmia}{m}{it}
\SetSymbolFont{lettersA}{bold}{U}{txmia}{bx}{it}
\DeclareFontSubstitution{U}{txmia}{m}{it}
\DeclareMathSymbol{\phiup}{\mathord}{lettersA}{30}

\newcommand{\scalar}{\phiup}
\newcommand{\gluon}{\mathsf{g}}
\newcommand{\gluonS}{\mathsf{s}}

\newcommand{\graviton}{\mathsf{h}}

\ExplSyntaxOn

\cs_new_protected:Npn \yurie_tachyon_sign:n #1
{
    \str_case:nnF {#1}
    {
        {+}{\scalebox{0.6}{{\color{Tachyonic}\ensuremath{\pmb{+}}}}}
        {-}{\scalebox{0.6}{{\color{Tachyonic}\ensuremath{\pmb{-}}}}}
        {\pm}{\scalebox{0.6}{{\color{Tachyonic}\ensuremath{\pmb{\pm}}}}}
        {\mp}{\scalebox{0.6}{{\color{Tachyonic}\ensuremath{\pmb{\mp}}}}}
    }
    {{\color{Tachyonic}#1}}
}

\cs_new_protected:Npn \yurie_narrow_sign_process:n #1
{
    \tl_set:Nn \l_tmpa_tl {#1}
    \tl_replace_all:Nnn \l_tmpa_tl {+} {\mathord{+}}
    \tl_replace_all:Nnn \l_tmpa_tl {-} {\mathord{-}}
    \tl_use:N \l_tmpa_tl
}

\RenewDocumentCommand{\deltaMC}{m}{
    \delta^{(4)}( \yurie_narrow_sign_process:n {#1} )
}

\RenewDocumentCommand{\tsign}{m}{
    \yurie_tachyon_sign:n {#1}
}

\RenewDocumentCommand{\gmShort}{m}{
    \Gamma\mleft[ \yurie_narrow_sign_process:n {#1} \mright]
}

\RenewDocumentCommand{\betaShort}{m}{
    B\mleft( \yurie_narrow_sign_process:n {#1} \mright)
}

\RenewDocumentCommand{\mgShort}{m m}{
    \Gamma\mleft[
        \genfrac..{0pt}{}{
            \yurie_narrow_sign_process:n {#1}
        }{
            \yurie_narrow_sign_process:n {#2}
        }
    \mright]
}

\RenewDocumentCommand{\coTwo}{g}{
    \IfNoValueTF{#1}
    {\mathcal{C}}
    {\mathcal{C}({#1})}
}

\RenewDocumentCommand{\coThree}{g}{
    \IfNoValueTF{#1}
    {\mathcal{C}}
    {\mathcal{C}({#1})}
}

\RenewDocumentCommand{\coBlock}{g}{
    \IfNoValueTF{#1}
    {\mathscr{C}}
    {\mathscr{C}({#1})}
}

\ExplSyntaxOff

\newcommand{\RCARepository}{https://github.com/ReikoAntoneva/RCA/}

\begin{document}

\begin{titlepage}

    \title{Regular celestial amplitudes}

    \author{Reiko Liu, Wen-Jie Ma$^{a,b}$}

    \date{}

    \maketitle\thispagestyle{empty}

    \address{a}{Fudan Center for Mathematics and Interdisciplinary Study, Fudan University, Shanghai, 200433, China}

    \address{b}{Shanghai Institute for Mathematics and Interdisciplinary Sciences (SIMIS), Shanghai, 200433, China}

    \email{
        reiko.antoneva@foxmail.com,
        wenjie.ma@simis.cn
    }

    \vfill

    \begin{abstract}
        Conventional massless celestial amplitudes are distributional and fail to realize the celestial OPE --- most sharply in the non-MHV paradox, where OPEs predict nonzero celestial amplitudes with helicities $-{+}{+}+$ that are known to vanish at tree level.
        To resolve this, we introduce regular celestial amplitudes.
        We demonstrate that at tree-level, these amplitudes are non-distributional and, crucially, consistent with the celestial OPE.
        This suggests a revised dictionary: CCFT correlators are the regular, not conventional, celestial amplitudes.
    \end{abstract}

    \vfill

\end{titlepage}


\begingroup
\hypersetup{linkcolor=black}
\tableofcontents
\endgroup


\newpage

\section{Introduction}

Celestial holography proposes a duality between scattering amplitudes in four-dimensional flat spacetime and two-dimensional conformal field theory (CFT) on the celestial sphere \cite{Cheung:2016iub,Pasterski:2016qvg,Pasterski:2017kqt,Strominger:2017zoo,Raclariu:2021zjz,Pasterski:2021rjz,Pasterski:2021raf,McLoughlin:2022ljp}, providing a novel framework for understanding scattering process and even quantum gravity. The fundamental observables in this framework are celestial amplitudes, which are obtained by performing a change of basis on scattering amplitudes and transform covariantly under the conformal group. While this change of basis can be formally implemented in any quantum field theory, the profound claim of celestial holography lies in the identification of celestial amplitudes with correlation functions of a putative celestial CFT (CCFT). It is this second step that elevates the construction from a mere change of basis to a genuine holographic duality, indicating the existence of a rich two-dimensional structure governing flat-space scattering.

Operator product expansion (OPE) is the foundational algebraic structure of CFTs and underlies the modern conformal bootstrap.
In this framework, the dynamical data of a CFT is encoded in its OPE coefficients and the spectrum of primary operators; consequently, all local correlators can be constructed from OPEs.
Therefore, for CCFT to be a genuine CFT, it must admit a well-defined celestial OPE that captures the dynamics of bulk scattering.
The celestial OPE has been extensively studied in \eg \cite{Fan:2019emx,Pate:2019lpp,Lam:2017ofc,Garcia-Sepulveda:2022lga,Atanasov:2021cje,Chang:2021wvv,Fan:2022kpp,Chang:2023ttm,Fan:2023lky,Liu:2024lbs,Liu:2024vmx,Fan:2021isc,Fan:2021pbp,Chang:2022jut,Chang:2022seh,Surubaru:2025qhs,Himwich:2025bza,Pranzetti:2025flv,Liu:2025dhh,Sleight:2023ojm,Iacobacci:2024nhw,Malherbe:2025qex}.
An important discovery is that, for massless particles, the leading terms in the OPEs are universally determined by the soft and collinear behaviors of scattering amplitudes \cite{Fan:2019emx,Pate:2019lpp}.
These leading terms generate infinite-dimensional symmetry algebras which are expected to constrain the bulk scattering, see \eg \cite{He:2014laa,Kapec:2016jld,Pate:2019mfs,Banerjee:2020zlg,Banerjee:2020vnt,Guevara:2021abz,Himwich:2021dau,Strominger:2021mtt,Banerjee:2021cly,Banerjee:2021dlm,Ball:2021tmb,Adamo:2021lrv,Fan:2022vbz,Banerjee:2023jne,Himwich:2023njb,Banerjee:2023bni,Banerjee:2023zip,Agrawal:2024sju,Ball:2024oqa,Guevara:2025tsm,Banerjee:2025grp}.

Despite the elegant algebraic structure of celestial OPEs, a persistent inconsistency remains between its predictions and the actual behavior of celestial amplitudes—most notably in Yang-Mills and Einstein gravity.
In particular, the gluon and graviton OPEs do not produce the expected contributions in three- and four-point celestial amplitudes. More critically, these OPEs imply nonzero values for certain non-MHV correlators that are known to vanish at tree level. For example, consider the simplest non-MHV case $\vev{\gluon_{-}\gluon_{+}\gluon_{+}\gluon_{+}}$. Applying the OPE to the last two operators yields a nonvanishing three-point correlator $\vev{\gluon_{-}\gluon_{+}\gluon_{+}}$, which in turn predicts a nonzero four-point correlator.
This tension indicates that the known celestial OPE does not form a closed algebra at the level of celestial amplitudes.

Another major challenge in celestial holography arises from the distributional nature of celestial amplitudes: due to momentum conservation, lower-point celestial amplitudes localize only on specific kinematic configurations.
For example, in the massless three-point case, momentum conservation forces the relevant momenta to be collinear, and consequently the celestial amplitudes contain Dirac delta functions of celestial coordinates \cite{Pasterski:2017ylz,Chang:2022seh}. Similarly, for massless two-to-two scattering, the celestial amplitude is supported only on the equator of the celestial sphere.
It is precisely this distributional feature that becomes the primary obstruction to applying standard CFT techniques.\footnote{
    This distributional behavior can be smeared out by shadow transform and/or lightray transform, see \eg \cite{Crawley:2021ivb,Fan:2021isc,Fan:2021pbp,Chang:2022jut,Chang:2022seh,Furugori:2023hgv,Surubaru:2025qhs,Himwich:2025bza,Pranzetti:2025flv,Liu:2025dhh,Sharma:2021gcz,Jorge-Diaz:2022dmy,Banerjee:2022hgc,Banerjee:2024hvb,Banerjee:2025oyu}.
}

In this paper, we resolve the above inconsistencies by introducing regular celestial amplitudes. We present explicit constructions of the regular celestial amplitudes and demonstrate through several concrete examples that: (1) they are consistent with the celestial OPE; (2) they exhibit the standard form of correlators in CFT. This resolution provides a modified dictionary for massless particles in celestial holography: the local correlators in celestial CFT are not the conventional celestial amplitudes, but the regular counterparts.

This paper is organized as follows.
Section \ref{sec: conventions} fixes conventions.
Section \ref{sec: conformal basis} reviews the massless and massive conformal bases, constructs the tachyonic conformal basis, and discusses their massless limits.
Section \ref{sec: celestial OPE and regular celestial amplitudes} diagnoses the inconsistency between celestial OPEs and conventional celestial amplitudes and introduces regular celestial amplitudes.
Section \ref{sec: scalar phi3 theory} computes tree-level regular celestial amplitudes in massless $\phi^3$ theory and verifies their consistency with the celestial OPE.
Section \ref{sec: Yang-Mills and Gravity} extends the analysis to Yang--Mills and Einstein gravity theories.


\section{Conventions}
\label{sec: conventions}

In this section, we summarize the conventions and notations.

\textbf{Celestial kinematics.}
\begin{itemize}
    \item
        The bulk spacetime is $4d$ Lorentzian with metric signature $(-,+,+,+)$ and the boundary is $2d$ Euclidean celestial sphere.
    \item
        Bulk coordinates and spin are denoted by $X$ and $\ell$, respectively. Boundary coordinates are denoted by $(z,\zb)$, and conformal weights by $(h,\hb)$, which are related to the conformal dimension and spin by $h=(\Delta+J)/2$ and $\hb=(\Delta-J)/2$.
        Furthermore, we introduce a multi-label notation for sums and differences of these quantities as follows:
        \begin{equation}
            \Delta_{a_{1}\cdots a_{n}}
            \equiv
            \sum_{i=1}^{n}\Delta_{a_{i}}
            \, ,
            \quad
            \Delta_{a_{1}\cdots a_{n},b_{1}\cdots b_{m}}
            \equiv
            \sum_{i=1}^{n}\Delta_{a_{i}}-
            \sum_{j=1}^{m}\Delta_{b_{j}}
            \, .
        \end{equation}
    \item
        Generic momenta are denoted by $P$, while massless, massive and tachyonic momenta are denoted by $q$, $p$ (with $p^2=-m^2$) and $k$ (with $k^2=m^2$), respectively.
        Their explicit parametrizations are
        \begin{align}
            \label{eq: momentum parametrization}
            &
            q=\omega\hat{q}
            =
            \omega(1+z \zb,z+\zb,-i (z-\zb),1-z \zb)
            \, ,
            \TextInMath{for} \omega\geq 0
            \, ,
            \\
            &
            p=m\phat
            =
            \frac{m}{2y}(1+z \zb+y^2,z+\zb,-i (z-\zb),1-z \zb-y^2)
            \, ,
            \TextInMath{for} y>0
            \, ,
            \nn
            \\
            &
            k=m\khat
            =
            \frac{m}{2y}(1+z \zb-y^2,z+\zb,-i (z-\zb),1-z \zb+y^2)
            \, ,
            \TextInMath{for} y\in\RR
            \, .
            \nn
        \end{align}
        The polarization vectors are chosen as $\epsilon_{+}=\pp_{z}\qhat$ and $\epsilon_{-}=\pp_{\zb}\qhat$. The corresponding vielbeins are $\set{\epsilon_{+},\epsilon_{-},\hat{n},\qhat}$ for massless momentum and $\set{\epsilon_{+},\epsilon_{-},\khat,\phat}$ for massive or tachyonic momentum, where $\hat{n}=(1,0,0,-1)$.
        Furthermore, for tensor products of identical vectors we use the abbreviation
        \begin{equation}
            \epsilon_{a,\mumu{1}{n}}
            \eqq
            \epsilon_{a,\mu_{1}}\cdots\epsilon_{a,\mu_{n}}
            \, .
        \end{equation}
    \item
        The integration measure on the boundary is $\int d^{2}z\eqq \int d\Re z\,  d\Im z$.
        The measures on mass shells are defined as follows:
        \begin{align}
            \label{eq: mass-shell measure}
            &
            \intt{[dq]}
            \eqq
            \intt{\frac{d^{3}q}{q^{0}}}\theta(q^{0})
            =
            \intt{d^{4}q}2\delta(q \co q)\theta(q^{0})
            =
            \intt{d^{2}z}\intrange{4\omega d\omega}{0}{\oo}
            \, ,
            \\
            &
            \intt{[d\phat]}
            \eqq
            \intt{\frac{d^{3}\phat}{\phat^{0}}}\theta(\phat^{0})
            =
            \intt{d^{4}\phat}2\delta(\phat \co \phat+1)\theta(\phat^{0})
            =
            \intt{d^{2}z}\intrange{y^{-3}dy}{0}{\oo}
            \, ,
            \nn
            \\
            &
            \intt{[d\khat]}
            \eqq
            \intt{\frac{d^{3}\khat}{|\khat^{0}|}}
            =
            \intt{d^{4}\khat}2\delta(\khat \co \khat-1)
            =
            \intt{d^{2}z}\intrange{|y|^{-3}dy}{-\oo}{+\oo}
            \, ,
            \nn
        \end{align}
        and $\int [dp]=m^{2}\int [d\phat]$, $\int [dk]=m^{2}\int [d\khat]$.
\end{itemize}

\textbf{CFT kinematics.}
\begin{itemize}
    \item
        An operator is abbreviated as $\op_{i}\eqq \op_{\Delta_{i},J_{i}}(z_{i},\zb_{i})$ if there is no confusion.
        The conformal structures are denoted by double brackets as
        \begin{align}
            \label{eq: conformal structure}
            &
            \vevv{\op_{1}\op_{2}}
            =
            z_{1,2}^{-2 h_1}
            \zb_{1,2}^{-2 \bar{h}_1}
            \, ,
            \\
            &
            \vevv{\op_{1}\op_{2}\op_{3}}
            =
            z_{1,2}^{h_{3,12}} z_{2,3}^{h_{1,23}} z_{1,3}^{h_{2,13}}
            \zb_{1,2}^{\bar{h}_{3,12}} \zb_{2,3}^{\bar{h}_{1,23}} \zb_{1,3}^{\bar{h}_{2,13}}
            \, ,
            \nn
            \\
            &
            \vevv{\op_{1}\op_{2}\op_{3}\op_{4}}
            =
            z_{1,2}^{-h_{12}}
            z_{1,3}^{h_{4,3}}
            z_{1,4}^{h_{23,14}}
            z_{2,4}^{h_{1,2}}
            z_{3,4}^{-h_{34}}
            \bar{z}_{1,2}^{-\bar{h}_{12}}
            \bar{z}_{1,3}^{\bar{h}_{4,3}}
            \bar{z}_{1,4}^{\bar{h}_{23,14}}
            \bar{z}_{2,4}^{\bar{h}_{1,2}}
            \bar{z}_{3,4}^{-\bar{h}_{34}}
            \, .
            \nn
        \end{align}
        Then the two- and three-point coefficients $\cC$ are defined via
        \begin{equation}
            \vev{\op_{1}\op_{2}}
            =
            \vevv{\op_{1}\op_{2}}
            \,
            \coTwo{\op_{1}\op_{2}}
            \, ,
            \quad
            \vev{\op_{1}\op_{2}\op_{3}}
            =
            \vevv{\op_{1}\op_{2}\op_{3}}
            \,
            \coThree{\op_{1}\op_{2}\op_{3}}
            \, ,
        \end{equation}
        and the four-point correlators can be decomposed as
        \begin{equation}
            \vev{\op_{1}\op_{2}\op_{3}\op_{4}}
            =
            \vevv{\op_{1}\op_{2}\op_{3}\op_{4}}
            \,
            \cG(\chi,\chib)
            \, ,
        \end{equation}
        where the stripped correlator $\cG$ depends only on the cross-ratios $\chi=\frac{z_{1,2}z_{3,4}}{z_{1,3}z_{2,4}}$ and $\chib=\frac{\zb_{1,2}\zb_{3,4}}{\zb_{1,3}\zb_{2,4}}$.
    \item
        The shadow transform of a primary operator is defined as
        \begin{equation}
            \label{eq: shadow transform}
            \shadow[\op_{\Delta,J}] (z,\zb)
            \eqq
            \intt{d^{2}z'}
            (z-z')^{2h-2}
            (\zb-\zb')^{2\hb-2}
            \op_{\Delta,J}(z',\zb')
            \, ,
        \end{equation}
        which has the conformal weights $\wave{\Delta}\eqq 2-\Delta$, and $\wave{J}\eqq -J$.
        This transform is invertible for generic $\Delta\in\CC$, given by
        \begin{equation}
            \label{eq: inverse shadow transform}
            \op_{\Delta,J} (z,\zb)
            =
            N_{\Delta,J}
            \intt{d^{2}z'}
            (z-z')^{-2h}
            (\zb-\zb')^{-2\hb}
            \shadow[\op_{\Delta,J}](z',\zb')
            \, ,
        \end{equation}
        where $N_{\Delta,J} = -\pi^{-2}(\Delta -J-1) (\Delta +J-1)$.
        With the above normalization, the star-triangle relation is
        \begin{equation}
            \label{eq: star-triangle relation}
            \vevv{\op_{1}\op_{2}\shadow[\op_{\Delta_{3},J_{3}}]}
            =
            \pi
            \mg{
                2 h_3-1,
                h_{2,13}+1,
                \bar{h}_{1,23}+1
            }{
                2-2 \bar{h}_3,
                h_{23,1},
                \bar{h}_{13,2}
            }
            \vevv{\op_{1}\op_{2}\op_{\wave\Delta_{3},\wave{J}_{3}}}
            \, ,
        \end{equation}
        and the prefactor is called the shadow coefficient.
\end{itemize}

\section{Conformal basis and massless limit}
\label{sec: conformal basis}

The celestial amplitudes are defined by expanding the scattering amplitudes in the conformal basis rather than the plane-wave basis \cite{Pasterski:2016qvg,Pasterski:2017kqt}. The conformal basis consists of a set of wavefunctions that: (1) transform as conformal primary operators on the boundary; (2) satisfy the equations of motion in the bulk.
For lower spins $\ell=0,1,2$, the massless and massive bases were constructed in \cite{Pasterski:2017kqt}. For arbitrary spins, the massless shadow and massive bases were constructed in \cite{Law:2020tsg,Chang:2022seh}.
The scalar tachyonic basis was introduced in \cite{Chang:2023ttm} to establish the split representation of bulk propagators. For half-integer-spin particles, the discussion can be found in \cite{Iacobacci:2020por,Narayanan:2020amh}.

In this section, we introduce the conformal bases for massless, massive, and tachyonic particles of arbitrary integer spin, and then study the massless limits of the massive and tachyonic bases.

\textbf{Notations.}
To emphasize the distributional nature of the conformal basis $\Phi$, we prefer to present it via its pairing with a test scattering amplitude $\cT$ in momentum space, rather than by an explicit coordinate-space expression:
\begin{equation}
    (\Phi,\cT)=\int\cdots
    \, .
\end{equation}
For massless and massive particles we distinguish incoming and outgoing bases by $\inn$ and $\out$, respectively, and their momenta enter the amplitude with sign $(-)$ for incoming and $(+)$ for outgoing. For example, the pairing for an incoming massive particle is $(\Phi^{\inn,m},\cT(-p))$, see \eqref{eq: CPWF massive}.
Tachyonic particles lack a distinction between incoming and outgoing. We therefore treat all tachyons as outgoing, and their momenta always carry a positive sign $(+)$. As a compensation, the tachyonic basis $(\Phi^{\tsign{s},im},\cT(k))$ carries an additional $\ZZ_{2}$ quantum number $\tsign{s}$ that distinguishes different branch cut prescriptions, see \eqref{eq: CPWF tachyonic}.
In summary, we have the following color scheme:
\\[0.5em]
\centerline{
    \textcolor{Incoming}{incoming},
    \quad
    \textcolor{Tachyonic}{tachyonic},
    \quad
    \textcolor{Outgoing}{outgoing}.
}

\subsection{Massless/massive conformal basis}
\label{sec: massless/massive basis}

For a massless particle with bulk spin $\ell$, there are two equivalent conformal bases: the Mellin basis $\Phi_{\Delta,J}^{\io,\ell}$ and the shadow basis $\wave\Phi_{\Delta,J}^{\io,\ell}$. They both satisfy the Fronsdal equations and are related by the shadow transform \eqref{eq: shadow transform} as
\begin{equation}
    \label{eq: massless basis shadow relation}
    \wave\Phi_{\Delta,J}^{\io,\ell}
    =
    \shadow[\Phi_{2-\Delta,-J}^{\io,\ell}]
    \, .
\end{equation}
Due to the matching of the boundary rotation group and the massless little group $\sogroup(2)$, the conformal spin of the bases coincides with the bulk polarization $J=\pm\ell$.
The conformal dimension is initially taken on the principal series $\Delta\in \pseries$ and in practice can be analytically continued to the complex plane.

\textbf{Massless Mellin basis.}
In the momentum space, the scattering amplitude $\cT$ expanded in the Mellin basis is
\begin{equation}
    \label{eq: CPWF Mellin}
    (\Phi_{\Delta,J}^{\io,\ell},\cT)
    =
    \int_0^{\infty}d\omega\, \omega^{\Delta-1}
    \sum_{n=0}^{\ell}
    \frac{(\ell-n+1)_n}{n!(\ell-n+\Delta -1)_n}
    \pp_{J}^{n}
    \LRb{
        \qhat_{\symL\mumu{1}{n}}\epsilon_{J,\mumu{n+1}{\ell}\symR}
        \cT^{\mumu{1}{\ell}}(\mp q)
    }
    \, .
\end{equation}
Here $\partial_{J}=\partial_z$ for $J=+\ell$ and $\partial_{J}=\partial_{\zb}$ for $J=-\ell$.
$\symL\mumu{1}{\ell}\symR$ denotes tensor symmetrization with the conventional $1/\ell!$ normalization.

The readers should not confuse this Mellin basis \eqref{eq: CPWF Mellin} with the Mellin transform. For spinning particles, the $n=0$ term is the Mellin transform of the helicity amplitude $\epsilon \co \cT$, while the rest are derivatives of gauge Ward identities, and hence do not contribute to the conventional celestial amplitude, see \eg \cite{Pasterski:2017ylz}.
However, for the regular celestial amplitudes introduced later, these gauge terms are necessary for conformal covariance.

\textbf{Massless shadow basis}.
The shadow basis convoluted with scattering amplitude is
\begin{equation}
    \label{eq: CPWF shadow}
    (\wave\Phi_{\Delta,J}^{\io,\ell},\cT)
    =
    (-1)^{\ell}
    2^{\Delta-2}
    \intt{[dq']}
    (-q' \co \qhat)^{-\Delta}
    \cP_{\mumu{1}{\ell}}{\!}^{\nunu{1}{\ell}}(\qhat,\qhat')
    \epsilon_{J,\nunu{1}{\ell}}
    \cT^{\mumu{1}{\ell}}(\mp q')
    \, .
\end{equation}
Here the projector $\cP(\qhat,\qhat')$ selects the spin-$\ell$ representation of the massless little group $\sogroup(2)$ from bulk tensors.
For scalar the projector is trivially $1$. For spin-1 it projects onto the $2d$ subspace transverse to two lightlike momenta,
\begin{equation}
    \cP^{\mu\nu}(\qhat,\qhat')=
    g^{\mu \nu}
    -\frac{\qhat^{\mu} \qhat'^{\nu}}{\qhat \co \qhat'}
    -\frac{\qhat^{\nu} \qhat'^{\mu}}{\qhat \co \qhat'}
    \, .
\end{equation}
The higher spin projectors are built upon the spin-1 case as
\begin{equation}
    \label{eq: projector higher spin}
    \cP_{\mumu{1}{\ell}}{\!}^{\nunu{1}{\ell}}
    =
    \sum_{n=0}^{\lfloor \ell/2 \rfloor}
    \frac{
        (-1)^n
        (\ell-2 n+1)_{2 n}
    }{
        2^{2 n} n!
        (\ell-n+\frac{r}{2}-1)_n
    }
    \cP_{\symL\mu_{1}\mu_{2}}\cP^{\symL\nu_{1}\nu_{2}} {\cdots} \cP_{\mu_{2n-1}\mu_{2n}}\cP^{\nu_{2n-1}\nu_{2n}}
    \cP_{\mu_{2n+1}}{\!}^{\nu_{2n+1}} {\cdots} \cP_{\mu_{\ell}\symR}{\!}^{\nu_{\ell}\symR}
    \, ,
\end{equation}
where $r=2$ for $\cP(\qhat,\qhat')$ is the dimension of the transverse subspace. $\cP$ is traceless among $\mumu{1}{\ell}$ and $\nunu{1}{\ell}$, and singles out the spin-$\ell$ symmetric traceless representation from the $\ell$-fold tensor product.

\textbf{Massive basis.}
For a massive particle with mass $m$ and spin $\ell$, the massive basis $\Phi_{\Delta,J}^{\io,m,\ell}$ satisfying the Fierz-Pauli equations is given by
\begin{equation}
    \begin{aligned}
        \label{eq: CPWF massive}
        (\Phi_{\Delta,J}^{\io,m,\ell},\cT)
        &=
        (-1)^{J}
        2^{\Delta-2}
        m^{2-\Delta}
        \intt{[d\phat']}
        (-\phat' \co \qhat)^{-\Delta-\ell+|J|}
        \cT^{\mumu{1}{\ell}}(\mp p')
        \\
        &\peq \xx
        \cP_{\mumu{1}{\ell}}{\!}^{\rhorho{1}{\ell}}(\phat')\,
        \cP_{\rhorho{1}{|J|}}{\!}^{\nunu{1}{|J|}}(\qhat,\phat')\,
        \qhat_{\rhorho{|J|+1}{\ell}}\,
        \epsilon_{J,\nunu{1}{|J|}}
        \, .
    \end{aligned}
\end{equation}
There are two new projectors $\cP(\phat)$ and $\cP(\qhat,\phat)$ comparing with the shadow basis \eqref{eq: CPWF shadow}. For spin-1 the building blocks are
\begin{align}
    \label{eq: projector qp}
    &
    \cP^{\mu\nu}(\phat)=g^{\mu \nu}+\hat{p}^{\mu} \hat{p}^{\nu}
    \, ,
    \\
    &
    \cP^{\mu\nu}(\qhat,\phat)=
    g^{\mu \nu}
    -\frac{\hat{p}^{\mu} \hat{q}^{\nu}}{\hat{p} \co \hat{q}}
    -\frac{\hat{p}^{\nu} \hat{q}^{\mu}}{\hat{p} \co \hat{q}}
    -\frac{\hat{q}^{\mu} \hat{q}^{\nu}}{(\hat{p} \co \hat{q})^2}
    \, .
\end{align}
The higher spin projectors are given by \eqref{eq: projector higher spin} with transverse dimensions $r=3$ and $r=2$, respectively.

The conformal spins of the massive basis \eqref{eq: CPWF massive} take values in $J\in\set{-\ell,\dots,\ell}$, and they correspond to the different components of the massive little group $\sogroup(3)$ representation decomposed onto the boundary rotation group $\sogroup(2)$. This decomposition is done in three steps. First, $\cP(\phat)$ projects onto the spin-$\ell$ symmetric traceless representation of the massive little group. Second, contracting with $\ell-|J|$ factors of $\qhat$ selects the component with  magnetic number $|S_z|=|J|$. Finally, $\cP(\qhat,\phat)\,\epsilon_{J}$ projects that component onto the spin-$J$ representation of the boundary rotation group.

The shadow transform of the massive basis is proportional to itself with shadow conformal weights, and there is no further ``massive shadow basis''. In particular, for the extremal spins $J = \pm \ell$ we show that in Appendix \ref{app: shadow transform of massive and tachyonic bases},
\begin{align}
    \label{eq: CPWF massive shadow}
    \shadow[\Phi_{\Delta,J}^{\io,m,\ell}]
    =
    \LRa{\cpwflimit_{\Delta}^{m,\ell}}^{-1}
    \Phi_{2-\Delta,-J}^{\io,m,\ell}
    \, ,
\end{align}
where $\cpwflimit$ is the massless limit prefactor in \eqref{eq: massive to Mellin factor}.

\textbf{Remarks on normalization.}
In Appendix \ref{app: CPWF lower spin comparison}, we compare the massless Mellin basis \eqref{eq: CPWF Mellin} for spins $\ell=1,2$ with the coordinate-space representations in \cite{Pasterski:2017kqt} and provide the relative normalization factors.
Compared with \cite{Chang:2022seh}, we change the normalization as: $\Phi_{\text{there}} = (-1)^{\ell} 2^{1-\Delta} \pi^{-1} (1-\Delta+\ell)\, \Phi_{\text{here}}$ for the massless shadow basis \eqref{eq: CPWF shadow} and $\Phi_{\text{there}} = (-1)^{\ell} 2^{1-\Delta} m^{\abs{J}-\ell} \pi^{-1} (1-\Delta+\ell)\, \Phi_{\text{here}}$ for the massive basis \eqref{eq: CPWF massive}. Our normalizations simplify the massless shadow relation \eqref{eq: massless basis shadow relation} and the massless limit \eqref{eq: massless limit massive shadow}.

All the constructions on conformal bases can be directly generalized to symmetric traceless tensors in higher dimensions. One should be careful that, in the literature the higher-dimensional shadow transform can differ from the $2d$ one \eqref{eq: shadow transform} by a sign $(-1)^{\ell}$, then to preserve the shadow relation \eqref{eq: massless basis shadow relation} between massless bases, this sign factor should be dropped.

\subsection{Tachyonic conformal basis}

In this section we introduce the tachyonic conformal basis. Our aim is not to discuss tachyonic particles that are unitary representations of the Poincare group $\isogroup(1,3)$, but rather to prepare the necessary ingredients for the regular celestial amplitudes introduced later.

The tachyonic little group is $\sogroup(1,2)$ and its unitary representations are infinite-dimensional except the trivial one. Here we consider the nonunitary finite-dimensional representations of $\sogroup(1,2)$, which are analytic continuations of the massive ones. The corresponding tachyonic particles are characterized by the imaginary mass $im$ and integer spin $\ell$, and they satisfy the Fierz-Pauli equations with mass $im$.
Their conformal basis is obtained essentially by replacing the massive momenta in \eqref{eq: CPWF massive} by tachyonic ones,
\begin{equation}
    \begin{aligned}
        \label{eq: CPWF tachyonic}
        (\Phi_{\Delta,J}^{\tsign{s},im,\ell},\cT)
        &=
        (-1)^{J}
        2^{\Delta-2}
        m^{2-\Delta}
        \intt{[d\khat']}
        (-\khat' \co \qhat)_{\tsign{s}}^{-\Delta-\ell+|J|}
        \cT^{\mumu{1}{\ell}}(k')
        \\
        &\peq \xx
        \cP_{\mumu{1}{\ell}}{\!}^{\rhorho{1}{\ell}}(\khat')\,
        \cP_{\rhorho{1}{|J|}}{\!}^{\nunu{1}{|J|}}(\qhat,\khat')\,
        \qhat_{\rhorho{|J|+1}{\ell}}\,
        \epsilon_{J,\nunu{1}{|J|}}
        \, .
    \end{aligned}
\end{equation}
Here the projectors $\cP(\khat)$ and $\cP(\qhat,\khat)$ are built upon
\begin{align}
    \label{eq: projector qk}
    &
    \cP^{\mu\nu}(\khat)=g^{\mu \nu}-\hat{k}^{\mu} \hat{k}^{\nu}
    \, ,
    \\
    &
    \cP^{\mu\nu}(\qhat,\khat)=
    g^{\mu \nu}
    -\frac{\hat{k}^{\mu} \hat{q}^{\nu}}{\hat{k} \co \hat{q}}
    -\frac{\hat{k}^{\nu} \hat{q}^{\mu}}{\hat{k} \co \hat{q}}
    +\frac{\hat{q}^{\mu} \hat{q}^{\nu}}{(\hat{k} \co \hat{q})^2}
    \, ,
\end{align}
with transverse dimensions $r=3$ and $r=2$, respectively.

A new feature is that there is no distinction between incoming and outgoing, hence no $\mp$ sign enters in $\cT(k')$. Instead, due to the presence of branch cut, $(-\khat'\co \qhat)^{\cdots}$ is ill-defined as a distribution and should be regularized properly. Following Appendix \ref{app: regularization of distributions}, to label different regularization schemes, we introduce a $\ZZ_{2}$ quantum number $\tsign{s}$ and three sets of signed powers as
\begin{equation}
    \label{eq: signed power}
    x^{\lambda}_{\tsign{\pm i}}
    \equiv
    (x\pm i\varepsilon)^{\lambda}
    \, ,
    \quad
    x^{\lambda}_{\tsign{\pm}}
    \equiv
    \abs{x}^{\lambda}\theta(\pm x)
    \, ,
    \quad
    x^{\lambda}_{\tsign{0|1}}
    \equiv
    \abs{x}^{\lambda}\sign^{0|1}(x)
    \, .
\end{equation}
They are related by linear combinations and are convenient for different usages: $\tsign{\pm i}$ are suitable for computation due to holomorphicity, $\mathcolor{Tachyonic}{\pm}$ trivialize the massless limit in Section \ref{sec: massless limit}, and $\tsign{0|1}$ diagonalize the completeness relation in \cite{Chang:2023ttm}.
In particular, the relation from $\mathcolor{Tachyonic}{\pm}$ to $\tsign{\pm i}$ is
\begin{equation}
    \label{eq: tachyon basis change}
    \Phi^{\tsign{\pm i},im,\ell}_{\Delta,J}
    =
    \Phi^{\tsign{+},im,\ell}_{\Delta,J}
    +
    (-1)^{\ell-|J|}e^{\mp i\pi\Delta}
    \Phi^{\tsign{-},im,\ell}_{\Delta,J}
    \, .
\end{equation}

The shadow transform changes the conformal weights and the sign $\tsign{s}$ of the tachyonic basis. For the extremal spins $J = \pm \ell$ we show that in Appendix \ref{app: shadow transform of massive and tachyonic bases},
\begin{align}
    \label{eq: CPWF tachyonic shadow}
    \shadow[\Phi_{\Delta,J}^{\tsign{\pm},im,\ell}]
    =
    -\LRa{\cpwflimit_{\Delta}^{m,\ell}}^{-1}
    \Phi_{2-\Delta,-J}^{\tsign{\mp},im,\ell}
    \, ,
\end{align}
where $\cpwflimit$ is the massless limit prefactor in \eqref{eq: massive to Mellin factor}.

\subsection{Massless limit}
\label{sec: massless limit}

In \cite{Pasterski:2017kqt} it was noticed that the massive scalar conformal basis admits two distinct massless limits, leading to the Mellin basis and the shadow basis, respectively. The shadow-type limit was extended to arbitrary spins in \cite{Chang:2022seh}. In this section we discuss, for spinning particles, how these two limits interplay and depend on the asymptotic behavior of test amplitudes, and whether any other limiting procedure exists.

We first provide the main result.
For the massive basis with extremal spins $J=\pm \ell$, the shadow limit and the Mellin limit are
\begin{align}
    \label{eq: massless limit massive shadow}
    &
    \wave\Phi^{\io,\ell}_{\Delta,J}
    =
    \lim_{m \to 0}
    \Phi^{\io,m,\ell}_{\Delta,J}
    \, ,
    \TextInMath{for}
    \Re\Delta<1
    \, ,
    \\
    \label{eq: massless limit massive Mellin}
    &
    \Phi^{\io,\ell}_{\Delta,J}
    =
    \lim_{m \to 0}
    \cpwflimit_{\Delta}^{m,\ell}
    \Phi^{\io,m,\ell}_{\Delta,J}
    \, ,
    \TextInMath{for}
    \Re\Delta>1
    \, ,
\end{align}
and for the tachyonic basis with extremal spins $J=\pm \ell$, the two limits are
\begin{align}
    \label{eq: massless limit tachyonic shadow}
    &
    \wave\Phi^{\io,\ell}_{\Delta,J}
    =
    \lim_{m \to 0}
    \Phi^{\tsign{\mp},im,\ell}_{\Delta,J}
    \, ,
    \TextInMath{for}
    \Re\Delta<1
    \, ,
    \\
    \label{eq: massless limit tachyonic Mellin}
    &
    \Phi^{\io,\ell}_{\Delta,J}
    =
    \lim_{m \to 0}
    -\cpwflimit_{\Delta}^{m,\ell}
    \Phi^{\tsign{\pm},im,\ell}_{\Delta,J}
    \, ,
    \TextInMath{for}
    \Re\Delta>1
    \, ,
\end{align}
where the prefactor is
\begin{equation}
    \label{eq: massive to Mellin factor}
    \cpwflimit_{\Delta}^{m,\ell}
    =
    (-1)^{\ell}
    2^{2-2 \Delta}
    m^{2 \Delta -2}
    \pi^{-1}
    (\Delta+\ell-1)
    \, .
\end{equation}
As a crosscheck, applying the shadow transform \eqref{eq: shadow transform} to both sides of these limits agrees with the shadow relations \eqref{eq: massless basis shadow relation}, \eqref{eq: CPWF massive shadow} and \eqref{eq: CPWF tachyonic shadow}.

\textbf{Derivation.}
Now we explain the origin of these limits using the massive scalar as an example.
Kinematically, with the parametrization \eqref{eq: momentum parametrization} and the variable change $y=\frac{m}{2\omega}$, a massive momentum is related to a massless one by
\begin{equation}
    p(m,y,z)=q(\omega,z)+\frac{m^2}{4 \omega}\hat{n}
    \to
    q
    \TextInMath{as}
    m\to 0
    \, .
\end{equation}
Then the conformal basis of an outgoing massive scalar becomes
\begin{align}
    (\Phi^{\out,m}_{\Delta},\cT)(z)
    &=
    2^{2 \Delta}
    \intrange{d\omega}{0}{\oo}\!\!
    \intt{d^{2}z'}
    \omega^{\Delta+1}
    (
        4 \omega^2 \abs{z-z'}^{2}
        +m^2
    )^{-\Delta}
    \cT(m,p(m,\omega,z'))
    \\
    &=
    \intt{\frac{ds}{2\pi i}}\!\!
    \intrange{d\omega}{0}{\oo}\!\!
    \intt{d^{2}z'}
    2^{-2 s} m^{2 s}
    \omega^{-\Delta-2 s+1}
    \abs{z-z'}^{-2\Delta-2s}
    \mg{
        -s,
        \Delta+s
    }{
        \Delta
    }
    \cT(m,p)
    \, .
    \nn
\end{align}
In the second line we have used the Mellin-Barnes relation \eqref{eq: Mellin-Barnes relation} to separate the mass dependence.

We assume the test amplitude decays sufficiently fast in the UV and is regular in the massless limit, $\lim_{m\to 0}\cT(m,p)=\cT(q)$. In the small mass regime, the dominant contribution arises from the leading $s$-pole in the right half-plane, which naively is $s=0$ from $\gm{-s}$ and corresponds to the shadow limit. However, as explained in Appendix \ref{app: regularization of distributions}, the powers of $\abs{z-z'}$ and $\omega$ as distributions are also meromorphic in $s$. Using \eqref{eq: homogeneous distribution - localization} and \eqref{eq: homogeneous distribution - localization 2d}, there are three series of poles in the right half-plane: for $n\in\NN$,
\begin{equation}
    \label{eq: massless limit - massive scalar case - s-poles}
    s=n
    \, ,
    \quad
    s=n-\Delta+1
    \TextInMath{from}
    \abs{z-z'}^{-2\Delta-2s}
    \, ,
    \quad
    s=\frac{1}{2} (n-\Delta+2)
    \TextInMath{from}
    \omega^{-\Delta-2 s+1}
    \, .
\end{equation}
If $\Re\Delta<1$, the first series is dominant and we obtain the shadow limit \eqref{eq: massless limit massive shadow}. Conversely, if $\Re\Delta>1$, the second series is dominant, the $z$-dependence becomes distributional and we obtain the Mellin limit \eqref{eq: massless limit massive Mellin}.

For the tachyonic and/or spinning bases, the analysis is similar, and we only indicate here the differences compared to the massive scalar case.
For the tachyonic basis, the $y$-integral should be split into two regions, $y>0$ and $y<0$, with the respective changes of variables $y=\pm\frac{m}{2\omega}$. Then a tachyonic momentum is related to a massless one by
\begin{equation}
    k(m,y,z)=\pm q(\omega,z) \mp\frac{m^2}{4 \omega}\hat{n}
    \to
    \pm q
    \TextInMath{as}
    m\to 0
    \, ,
\end{equation}
and the test amplitude $\cT(k)$ becomes $\cT(\pm q)$ in the massless limit.
For the spinning basis, the shadow limit corresponds to the analytic structure $\gm{-s}$ again, but another analytic structure is
\begin{equation}
    \sum_{i=0}^{\ell}\rest\gm{1-s-\Delta-i}m^{2s+2i}
    \, .
\end{equation}
These terms contribute equally in the Mellin limit, and the resulting Mellin basis takes the form \eqref{eq: CPWF Mellin form2} instead of \eqref{eq: CPWF Mellin}.

\textbf{Remarks.}
In the discussion of the massless limit, we leave two issues to future work.
First, when $\Delta=1$, the first two series of $s$-poles in \eqref{eq: massless limit - massive scalar case - s-poles} coincide and produce a logarithmic behavior. We expect this is related to the conformally soft and Goldstone modes discussed in \eg \cite{Donnay:2018neh,Donnay:2020guq,Pasterski:2021fjn,Pasterski:2021dqe}.
Second, the third series in \eqref{eq: massless limit - massive scalar case - s-poles} seems to be harmless. However, when the test amplitude is singular in the IR, $\cT=\omega^{\lambda} \cT_{0}$ with $\lambda<-1$, this series gets corrected to $s=\frac{1}{2}(n-\Delta+2+\lambda)$. Then there exists a window $2+\lambda<\Re\Delta<-\lambda$ in which this series becomes dominant, leading to a distinct ``soft limit'',
\begin{equation}
    (\Phi^{\out,m}_{\Delta},\cT)(z)
    \sim
    2^{\Delta -\lambda -3}
    m^{-\Delta +\lambda +2}
    \mg{
        \frac{\Delta -\lambda -2}{2},
        \frac{\Delta +\lambda +2}{2}
    }{
        \Delta
    }
    \intt{d^{2}z'}
    \abs{z-z'}^{-\Delta -\lambda -2}
    \cT_{0}(\omega=0,z')
    \, .
\end{equation}

We also clarify the meaning of ``massless limit of amplitude''.
Consider a scattering process in which the first particle is massless with momentum $q$, and the remaining momenta are denoted by $P_{i}$.
The physical helicity amplitude $\epsilon \co \cT(q)$ is defined only on the singular subvariety $V = \{(q, P_{i}) \mid q^{2} = 0,\, P_{i}^{2} = -m_{i}^{2},\, q + \sum_{i} P_{i} = 0\}$ of momentum space.
In contrast, the amplitude $\cT(P)$ obtained from perturbative Feynman diagrams is defined on the entire momentum space, subject to gauge redundancy.
To define the massless limit of the conformal basis,
\begin{equation}
    (\Phi, \cT)
    \stackrel{?}{=}
    \lim_{m \to 0}
    (\Phi^{m}, \cT_{m})
    \, ,
\end{equation}
we do not modify the theory by giving this particle a mass, which can be problematic for massless bosons; instead, we choose a family of continuations $\cT_{m}$ of the physical amplitude in an infinitesimal neighborhood of $V$, satisfying $\lim_{m \to 0} \cT_{m} = \cT$. This continuation is not unique, and a natural choice is to take the perturbative amplitude $\cT(P)$ itself and then manually verify the results are gauge independent.

\section{Celestial OPE and regular celestial amplitudes}
\label{sec: celestial OPE and regular celestial amplitudes}
\subsection{Celestial OPE}
\label{sec: discussion on celestial OPE}

The OPE is the fundamental algebraic structure of a standard CFT, and its existence ensures that higher-point correlators can be completely determined by lower-point ones.
This has motivated extensive research on OPE in CCFTs, and key approaches include analyzing collinear limits of scattering amplitudes and examining constraints on OPE from translation and asymptotic symmetries \cite{Fan:2019emx,Pate:2019lpp,Guevara:2021abz,Himwich:2021dau}.
Taking massless $\phi^3$ theory as an example, the following scalar OPEs can be extracted from the collinear behavior:
\begin{align}
    \label{eq: scalar OPE from collinear}
    \scalar^{\out}_{\Delta_1}(z_1)
    \scalar^{\out}_{\Delta_2}(z_2)
    &\sim
    \frac{1}{4 \abs{z_{1,2}}^{2}}
    \betaShort{\Delta_1-1,\Delta_2-1}
    \scalar^{\out}_{\Delta_{12}-2}(z_2)
    \, ,
    \\
    \scalar^{\inn}_{\Delta_1}(z_1)
    \scalar^{\out}_{\Delta_2}(z_2)
    &\sim
    -\frac{1}{4 \abs{z_{1,2}}^{2}}
    \Bigl(
        \betaShort{\Delta_2-1,3-\Delta_{12}}
        \scalar^{\inn}_{\Delta_{12}-2}(z_2)
        +
        \betaShort{\Delta_1-1,3-\Delta_{12}}
        \scalar^{\out}_{\Delta_{12}-2}(z_2)
    \Bigr)
    \, .
    \nn
\end{align}
Despite their elegant form, a serious problem arises:
\emph{the above OPEs are incompatible with the known three-point and four-point celestial amplitudes, when analyzed with standard CFT techniques.}

Specifically, the OPE \eqref{eq: scalar OPE from collinear} predicts the following form of the three-point celestial amplitude,
\begin{align}
    \vev{\scalar^{\inn}_{\Delta_{1}}\scalar^{\out}_{\Delta_2}\scalar^{\out}_{\Delta_3}}
    =
    \frac{1}{4}
    \betaShort{\Delta_2-1,\Delta_3-1}
    \coTwo{\scalar_{\Delta_1}\scalar_{\Delta_{23}-2}}
    \vevv{\scalar_{1}\scalar_{2}\scalar_{3}}
    \, ,
\end{align}
where $\coTwo$ is the two-point coefficient and $\vevv{\scalar_{1}\scalar_{2}\scalar_{3}}$ is the three-point conformal structure \eqref{eq: conformal structure}.
However, three-point massless celestial amplitudes are usually considered to be zero due to momentum conservation. Even when carefully accounting for contributions from collinear and soft regions in momentum space \cite{Chang:2022seh}, or converting to Klein space \cite{Pasterski:2017ylz}, although the results are nonvanishing, they still do not match the predictions from collinear OPEs.
Furthermore, the OPE \eqref{eq: scalar OPE from collinear} also implies that, for the four-point celestial amplitude $\vev{\scalar^{\inn}_{\Delta_1}\scalar^{\inn}_{\Delta_2}\scalar^{\out}_{\Delta_3}\scalar^{\out}_{\Delta_4}}$, the exchange operator $\scalar^{\out}_{\Delta_{34}-2}$ should appear in the \schannel conformal block expansion. This prediction also fails, since the required \schannel expansion of this amplitude does not exist \cite{Chang:2022jut,Chang:2023ttm,Liu:2024lbs,Liu:2024vmx}.

To address these issues, we must first identify their underlying causes:
\begin{itemize}
    \item
        Two- and three-point celestial amplitudes deviate from the standard form \eqref{eq: conformal structure} of conformal correlators.
        Specifically, in the Mellin basis, the massless two-point amplitude is proportional to the delta function $\delta^{(2)}(z_{1,2})$, instead of the expected power function \cite{Pasterski:2017kqt}. The three-point amplitudes obtained in \cite{Pasterski:2017ylz,Chang:2022seh} also contain delta functions of boundary coordinates.
    \item
        The solution space of momentum conservation, as a real algebraic variety, is singular and has different components, and the delta function supported on it is ill-defined. Consequently, celestial amplitudes computed by directly solving momentum conservation constraints, fail to capture contributions from all components.
        For example, in \cite{Pasterski:2017ylz}, the gluon celestial amplitude $\vev{\gluon^{\inn}_{\Delta_1,-}\gluon^{\inn}_{\Delta_2,-}\gluon^{\out}_{\Delta_3,+}}$ in Klein space was computed using the anti-holomorphic solution:
        \begin{align}
            \label{eq: 3pt momentum conservation - antiholomorphic}
            \omega_1=-\frac{z_{23}}{z_{12}}\omega_3,
            \quad
            \omega_2=\frac{z_{13}}{z_{12}}\omega_3,
            \quad
            \zb_1=\zb_2=\zb_3
            \, .
        \end{align}
        But the three-point momentum conservation has other solutions, such as the single-soft one:
        \begin{align}
            \omega_1=0,
            \quad \omega_2=\omega_3,
            \quad z_2=z_3,
            \quad \zb_2=\zb_3
            \, ,
        \end{align}
        and the triple-soft one:
        \begin{align}
            \omega_1=\omega_2=\omega_3=0
            \, .
        \end{align}
        They have unequal dimensions and are mutually independent, hence the celestial amplitude obtained from \eqref{eq: 3pt momentum conservation - antiholomorphic} is incomplete.
\end{itemize}

We now resolve these issues.
\begin{itemize}
    \item
        To ensure that massless two-point celestial amplitudes take the standard form, we can adopt the shadow basis for one particle while keeping the Mellin basis for the other particle.
    \item
        To resolve the singularities and capture the complete contributions from momentum conservation, we can deform it into a smooth variety. This is computationally difficult, and we find a more practical strategy: recast the massless bases \eqref{eq: CPWF Mellin} and \eqref{eq: CPWF shadow} as phase space integrals of the form $\int d^{4}q \delta(q^{2})$, then smear out the lightcone delta function $\delta(q^{2})$ to obtain an integral over off-shell momentum $\int d^4q$. In this way, the momentum conservation gets smoothed by relaxing the on-shell conditions.
\end{itemize}

\subsection{Regular celestial amplitudes}
\label{sec: regular celestial amplitude}

In contrast to the conventional celestial amplitudes defined using the massless bases \eqref{eq: CPWF Mellin}, \eqref{eq: CPWF shadow} and the massive basis \eqref{eq: CPWF massive}, we define \emph{regular celestial amplitudes} by replacing the massless bases with any of the regularized conformal bases: power-type \eqref{eq: CPWF - power regularization}, mass-type \eqref{eq: CPWF - mass regularization}, and Gaussian-type \eqref{eq: CPWF - Guassian regularization}.
These regularized bases effectively smear the lightcone delta function, and tame the distributional behavior of the celestial amplitudes, as discussed in Section \ref{sec: discussion on celestial OPE}.

\textbf{Power regularization.}
We rewrite the shadow basis \eqref{eq: CPWF shadow} as
\begin{align}
    (\wave{\Phi}^{\io,\ell}_{\Delta,J},\cT)
    =
    \intt{
        d^4q'\,2\delta(q'^2)
        \theta(-\qhat \co q')
    }
    \wave{\Phi}^{\io,\mumu{1}{\ell}}_{\Delta,J}(\qhat,q')
    \cT_{\mumu{1}{\ell}}(\mp q')
    \, ,
\end{align}
where $\wave{\Phi}^{\io,\mumu{1}{\ell}}_{\Delta,J}$ denotes the integration kernel of the shadow basis, and the step function in the measure has been recast into a covariant form $\theta(q'^0) = \theta(-\hat{q} \co q')$ using the reference vector $\hat{q}$.
Then we smear the lightcone delta function by the identity \cite{Gelfand1}:
\begin{align}
    \label{eq: delta power regularization}
    \delta(P^{2})
    =
    \lim_{\massMellin\to 0}
    \frac{\massMellin}{4}
    \abs{P^{2}}^{\frac{\massMellin}{2}-1}
    \, ,
\end{align}
and obtain
\begin{align}
    \label{eq: CPWF power regularization step 1}
    (\wave{\Phi}^{\io,\ell}_{\Delta,J},\cT)
    =
    \lim_{\massMellin\to 0}
    \frac{\massMellin}{2}
    \intt{
        d^4P
        \abs{P^{2}}^{\frac{\massMellin}{2}-1}
        \theta(-\hat{q} \co P)
    }
    \wave{\Phi}^{\io,\mumu{1}{\ell}}_{\Delta,J}(\qhat,P)
    \cT_{\mumu{1}{\ell}}(\mp P)
    \, .
\end{align}
Now the momentum being integrated is off-shell, and to emphasize this point we have relabeled it as $P$.

The smeared shadow basis \eqref{eq: CPWF power regularization step 1} is inconvenient for computation. Hence we further reduce it by dividing the $P$-integral into two regions: massive $P^{2}\leq0$ and tachyonic $P^{2}\geq0$.
In the first region, by defining $P=p'=m\hat{p}'$ with $\hat{p}'^2=-1$, the contribution to \eqref{eq: CPWF power regularization step 1} is
\begin{equation}
    \lim_{\massMellin\to 0}
    \frac{\massMellin}{2}
    \intrange{dm\, m^{\massMellin-1}}{0}{\oo}
    \intt{[dp'] \theta(-\hat{q} \co p')}
    \wave{\Phi}^{\io,\mumu{1}{\ell}}_{\Delta,J}(\hat{q},p')
    \cT_{\mumu{1}{\ell}}(\mp p')
    \, .
\end{equation}
Similarly in the second region, by defining $P=k'=m\hat{k}'$ with $\hat{k}'^2=1$, the contribution is
\begin{equation}
    \lim_{\massMellin\to 0}
    \frac{\massMellin}{2}
    \intrange{dm\, m^{\massMellin-1}}{0}{\oo}
    \intt{[dk']}\theta(-\hat{q} \co k')
    \wave{\Phi}^{\io,\mumu{1}{\ell}}_{\Delta,J}(\hat{q},k')
    \cT_{\mumu{1}{\ell}}(\mp k')
    \, .
\end{equation}
Then by the relation between conformal bases \eqref{eq: CPWF shadow=massive=tachyonic}, in the above two equations, the parts inside the mass integral can be replaced by the $\io$ massive basis and the $\mathcolor{Tachyonic}{\mp}$ tachyonic basis with extremal spins $\abs{J}=\ell$, respectively.
In other words, \eqref{eq: CPWF power regularization step 1} is equivalent to the following power-regularized conformal bases:
\begin{equation}
    \label{eq: CPWF - power regularization}
    \begin{aligned}
        \wave\Phi^{\io,\ell}_{\Delta,J}
        &=
        \lim_{\massMellin\to 0}
        \frac{\massMellin}{2}
        \intrange{dm\, m^{\massMellin-1}}{0}{\oo}
        \mleft(
            \Phi^{\io,m,\ell}_{\Delta,J}
            +
            \Phi^{\tsign{\mp},im,\ell}_{\Delta,J}
        \mright)
        \, ,
        \\
        \Phi^{\io,\ell}_{\Delta,J}
        &=
        \lim_{\massMellin\to 0}
        \frac{\massMellin}{2}
        \intrange{dm\, m^{\massMellin-1}}{0}{\oo}
        \cpwflimit_{\Delta}^{m,\ell}
        \mleft(
            \Phi^{\io,m,\ell}_{\Delta,J}
            -
            \Phi^{\tsign{\pm},im,\ell}_{\Delta,J}
        \mright)
        \, .
    \end{aligned}
\end{equation}
In the second line, the power-regularized Mellin basis is obtained by performing an inverse shadow transform \eqref{eq: inverse shadow transform} on the first line, and the prefactor $\cpwflimit$ is given in \eqref{eq: massive to Mellin factor}.

In summary, the procedure of power regularization \eqref{eq: CPWF - power regularization} consists of three steps:
\begin{enumerate}
    \item taking the momentum off-shell and foliating it by the mass parameter $m$;
    \item performing a Mellin transform in the mass parameter $m$ with the dual variable $\massMellin$;
    \item isolating the contribution near $m=0$ by extracting the residue at $\massMellin=0$.
\end{enumerate}

\textbf{Mass regularization.}
The mass-regularized conformal bases are
\begin{equation}
    \label{eq: CPWF - mass regularization}
    \begin{aligned}
        \wave\Phi^{\io,\ell}_{\Delta,J}
        &=
        \lim_{m\to 0}
        \frac{1}{2}
        \mleft(
            \Phi^{\io,m,\ell}_{\Delta,J}
            +
            \Phi^{\tsign{\mp},im,\ell}_{\Delta,J}
        \mright)
        \, ,
        \\
        \Phi^{\io,\ell}_{\Delta,J}
        &=
        \lim_{m\to 0}
        \frac{1}{2}
        \cpwflimit_{\Delta}^{m,\ell}
        \mleft(
            \Phi^{\io,m,\ell}_{\Delta,J}
            -
            \Phi^{\tsign{\pm},im,\ell}_{\Delta,J}
        \mright)
        \, .
    \end{aligned}
\end{equation}
This follows directly from the power regularization \eqref{eq: CPWF - power regularization} and the distributional identity \eqref{eq: homogeneous distribution - localization}:
\begin{align}
    \lim_{\massMellin\to 0}
    \massMellin m^{\massMellin-1}
    =
    \delta(m)
    \, .
\end{align}
Intuitively, the mass regularization \eqref{eq: CPWF - mass regularization} corresponds to approaching the lightcone from both the timelike (EAdS) and spacelike (dS) regions of the Minkowski spacetime.

\textbf{Gaussian regularization.}
During the derivation of power regularization, we have used the identity \eqref{eq: delta power regularization} to smear the lightcone delta function, which should be understood in the distributional sense:
\begin{align}
    \lim_{\massMellin\to 0}
    \frac{\massMellin}{4}
    \intt{d^{4}P \abs{P^{2}}^{\frac{\massMellin}{2}-1}}
    f(P)
    =
    f(0)
    \, .
\end{align}
Here $f(P)$ is a test function regular near IR and decays sufficiently fast at UV.
For test functions lacking this decay, we can cure the UV behavior by the Gaussian regularization, which smears $\delta(P^2)=\delta(m^2)$ as
\begin{align}
    \delta(m^2)
    =
    \lim_{\massMellin\to 0}
    \frac{1}{\sqrt{\pi\massMellin}}
    e^{-m^4/\massMellin}
    \, .
\end{align}
Equivalently, the Gaussian-regularized conformal bases are
\begin{equation}
    \label{eq: CPWF - Guassian regularization}
    \begin{aligned}
        \wave{\Phi}^{\io,\ell}_{\Delta,J}
        &=
        \lim_{\massMellin\to 0}
        \frac{1}{\sqrt{\pi\massMellin}}
        \intrange{dm\, m\, e^{-m^4/\massMellin}}{0}{\oo}
        \mleft(
            \Phi^{\io,m,\ell}_{\Delta,J}
            +
            \Phi^{\tsign{\mp},im,\ell}_{\Delta,J}
        \mright)
        \, ,
        \\
        \Phi^{\io,\ell}_{\Delta,J}
        &=
        \lim_{\massMellin\to 0}
        \frac{1}{\sqrt{\pi\massMellin}}
        \intrange{dm\, m\, e^{-m^4/\massMellin}}{0}{\oo}
        \cpwflimit_{\Delta}^{m,\ell}
        \mleft(
            \Phi^{\io,m,\ell}_{\Delta,J}
            -
            \Phi^{\tsign{\pm},im,\ell}_{\Delta,J}
        \mright)
        \, .
    \end{aligned}
\end{equation}

\textbf{Remarks.}
We conclude this section with several important remarks.

First, naively using the results of the massless limit \eqref{eq: massless limit massive shadow}, the mass-regularized conformal basis \eqref{eq: CPWF - mass regularization} seems to be a trivial rewriting of the massless conformal basis, so the regular celestial amplitude would appear to be equal to the conventional celestial amplitude.
However, this is incorrect for the following reasons:
\begin{itemize}
    \item
        As discussed in Section \ref{sec: massless limit}, the mass limit depends on the range of the conformal dimension and the behavior of the test amplitude, therefore it does not necessarily commute with the integrals in the conformal basis.
    \item
        As discussed in Section \ref{sec: discussion on celestial OPE}, regularizing the conformal basis effectively deforms the solution space of momentum conservation, and compared to the conventional celestial amplitude, the regular one contains additional contributions from soft and collinear regions.
    \item
        The conventional celestial amplitude is equivalent to the onshell helicity amplitude. However, as emphasized in Section \ref{sec: massless/massive basis}, apart from the Mellin transform, the regular conformal basis also includes contributions from pure gauge terms present in the massless basis. Since the momentum being integrated is now off-shell, these pure gauges are no longer canceled by the Ward identities, and thus contribute nontrivially to the regular celestial amplitude.
    \item
        Simply using the massless limit does not yield the contributions from the tachyonic part in the mass-regularized conformal basis, nor the relative coefficients of the linear combinations. In later sections, we will see that the tachyonic term is necessary for the self-consistency of OPEs.
\end{itemize}

Moreover, according to the definition of the regular celestial amplitude, in principle, all massless particles need to be regularized, and the final result should not depend on the order of multiple massless limits --- this is technically over-complicated.
We find that, at least for the examples studied in this paper, it is often sufficient to regularize only a subset of massless particles, and the results are not sensitive to the order of multiple massless limits. We leave these issues to future work.

\section{Scalar \MathInTitle{\phi^3}{phi3} theory}
\label{sec: scalar phi3 theory}

We begin by investigating scalar $\phi^3$ theory, evaluating the three-point and four-point regular celestial amplitudes, and verifying their consistency with the celestial OPE \eqref{eq: scalar OPE from collinear} derived from the collinear limit of scattering amplitudes.

\subsection{Three-point regular celestial amplitudes}
\label{sec: scalar three-point}

Taking $\vev{ \wave{\scalar}^{\inn} \scalar^{\out} \scalar^{\out} }$ as an example, we demonstrate that the three-point regular celestial amplitude admits the standard form \eqref{eq: conformal structure}, and is independent of the regularization scheme, the choice of regularized particles, and the order of multiple massless limits.

\subsubsection{Regularizing one massless particle}
\label{sec: regularizing one scalar}

We first apply the three regularization schemes in Section \ref{sec: regular celestial amplitude} to a single massless particle, and verify that they give the same result.

\textbf{Mass regularization.}
Applying the mass regularization \eqref{eq: CPWF - mass regularization} to the first particle gives
\begin{equation}
    \label{eq: scalar 3pt massreg 1}
    \vev{
        \wave{\scalar}^{\inn}_{\Delta_1}
        \scalar^{\out}_{\Delta_2}
        \scalar^{\out}_{\Delta_3}
    }
    =
    \lim_{m \to 0}
    \frac{1}{2}
    \vev{
        \scalar^{\inn,m}_{\Delta_1}
        \scalar^{\out}_{\Delta_2}
        \scalar^{\out}_{\Delta_3}
    }
    \, .
\end{equation}
We have used the fact $\vev{\scalar^{\tsign{-},im}_{\Delta_1}\scalar^{\out}_{\Delta_2}\scalar^{\out}_{\Delta_3}}=0$, since the tachyon decay into massless particles is strictly forbidden by momentum conservation. Using \eqref{eq: 3pt coefficient OOM scalar} we obtain the three-point coefficient
\begin{equation}
    \coThree{
        \wave{\scalar}^{\inn}_{\Delta_1}
        \scalar^{\out}_{\Delta_2}
        \scalar^{\out}_{\Delta_3}
    }
    =
    \lim_{m \to 0}
    2^{\Delta_{1,23}-3} m^{\Delta_{23,1}-2}
    \betaShort{\frac{\Delta_{12,3}}{2},\frac{\Delta_{13,2}}{2}}
    \, .
\end{equation}

For generic conformal dimensions satisfying $\Re\Delta_{23,1}\geq 2$, the factor $m^{\Delta_{23,1}-2}$ under the massless limit behaves as a Kronecker delta function, allowing us to reduce the test function in a manner analogous to the Dirac delta function.\footnote{
    For special conformal dimensions, \eg $\Delta_2 = 1$, $m^{\Delta_{3,1}-1}$ combined with the Gamma function $\gm{\frac{\Delta_{1,3}-1}{2}}$ gets lifted to the complex delta function $\deltaComplex{\Delta_{3,1}-1}$ in Appendix \ref{app: analytic functionals}.
    For simplicity, we only consider generic conformal dimensions here.
}
For convenience, we introduce the notation
\begin{equation}
    \label{eq: Kronecker delta shorthand}
    \deltaK{\Delta}
    \equiv
    \lim_{m \to 0} m^{\Delta}
    \, .
\end{equation}
Then the three-point coefficient can be written as
\begin{equation}
    \label{eq: scalar 3pt massreg result}
    \coThree{
        \wave{\scalar}^{\inn}_{\Delta_1}
        \scalar^{\out}_{\Delta_2}
        \scalar^{\out}_{\Delta_3}
    }
    =
    \frac{1}{32}
    \deltaK{\Delta_{23,1}-2}
    \betaShort{\Delta_2-1,\Delta_3-1}
    \, .
\end{equation}
Alternatively, we can regularize the second particle:
\begin{equation}
    \label{eq: scalar 3pt massreg 2}
    \vev{
        \wave{\scalar}^{\inn}_{\Delta_1}
        \scalar^{\out}_{\Delta_2}
        \scalar^{\out}_{\Delta_3}
    }
    =
    \lim_{m \to 0}
    -\frac{1}{2}\cpwflimit_{\Delta_2}^{m,0}
    \vev{
        \wave{\scalar}^{\inn}_{\Delta_1}
        \scalar^{\tsign{-},im}_{\Delta_2}
        \scalar^{\out}_{\Delta_3}
    }
    \, ,
\end{equation}
in which the tachyonic celestial amplitude is given by \eqref{eq: 3pt coefficient TOO scalar}. This yields the same result as \eqref{eq: scalar 3pt massreg result}.

\textbf{Gaussian and power regularizations}.
For the Gaussian regularization \eqref{eq: CPWF - Guassian regularization}, we have
\begin{equation}
    \coThree{
        \wave{\scalar}^{\inn}_{\Delta_1}
        \scalar^{\out}_{\Delta_2}
        \scalar^{\out}_{\Delta_3}
    }
    =
    \lim_{\massMellin\to 0}
    2^{\Delta_{1,23}-3}
    \massMellin^{
        \frac{1}{4}(\Delta_{23,1}-2)
    }
    \pi^{-\frac{1}{2}}
    \gm{\frac{\Delta_{23,1}}{4}}
    \betaShort{\frac{\Delta_{12,3}}{2},\frac{\Delta_{13,2}}{2}}
    \, .
\end{equation}
The $\massMellin\to 0$ limit gives $\deltaK{\Delta_{23,1}-2}$, reproducing the result of mass regularization \eqref{eq: scalar 3pt massreg result}.
The power regularization \eqref{eq: CPWF - power regularization} is more tricky in this example,
\begin{equation}
    \vev{
        \wave{\scalar}^{\inn}_{\Delta_1}
        \scalar^{\out}_{\Delta_2}
        \scalar^{\out}_{\Delta_3}
    }
    =
    \lim_{\massMellin\to 0}
    \frac{\massMellin}{2}
    \intrange{dm\, m^{\massMellin-1}}{0}{\oo}
    2^{\Delta_{1,23}-2}
    m^{\Delta_{23,1}-2}
    \betaShort{\frac{\Delta_{12,3}}{2},\frac{\Delta_{13,2}}{2}}
    \vevv{\scalar_{1}\scalar_{2}\scalar_{3}}
    \, .
\end{equation}
The integrand does not decay as $m \to\infty$, and as discussed earlier, the contribution from the large mass region must be subtracted. This is accomplished by introducing a hard cutoff $\Lambda$, taking the limit $\massMellin\to 0$ first and then $\Lambda \to \infty$.
Following this prescription we obtain
\begin{equation}
    \coThree{
        \wave{\scalar}^{\inn}_{\Delta_1}
        \scalar^{\out}_{\Delta_2}
        \scalar^{\out}_{\Delta_3}
    }
    =
    \lim_{\Lambda\to\infty}
    \mleft(
        \lim_{\massMellin\to 0}
        \frac{\massMellin\Lambda^{\Delta_{23,1}-2+\massMellin}}{\Delta_{23,1}-2+\massMellin}
        2^{\Delta_{1,23}-3}
        \betaShort{\frac{\Delta_{12,3}}{2},\frac{\Delta_{13,2}}{2}}
    \mright)
    \, .
\end{equation}
For generic conformal dimensions, under the $\massMellin$ limit, it behaves as a Kronecker delta function $\deltaK{\Delta_{23,1}-2}$, then the result agrees with \eqref{eq: scalar 3pt massreg result}.

\subsubsection{Regularizing two or three massless particles}
\label{sec: regularizing two or three scalars}

We now consider the regularization of multiple massless particles. There are three cases:
(1) regularizing one incoming and one outgoing particle;
(2) regularizing two outgoing particles;
(3) regularizing three particles.
For technical simplicity, we adopt the mass regularization scheme.

\textbf{Regularizing two particles}.
We apply the mass regularization to the first two particles. There are four terms, and $\vev{\scalar^{\tsign{-},im_1}_{\Delta_1}\scalar^{\out,m_2}_{\Delta_2}\scalar^{\out}_{\Delta_3}}$ vanishes by momentum conservation, leaving
\begin{equation}
    \label{eq: scalar 3pt massreg 12}
    \vev{\wave{\scalar}^{\inn}_{\Delta_1}\scalar^{\out}_{\Delta_2}\scalar^{\out}_{\Delta_3}}
    =
    \lim_{\substack{m_1\to 0\\m_2\to 0}}
    \frac{1}{4}\cpwflimit_{\Delta_2}^{m_2,0}
    \mleft(
        \vev{\scalar^{\inn,m_1}_{\Delta_1}\scalar^{\out,m_2}_{\Delta_2}\scalar^{\out}_{\Delta_3}}
        -\vev{\scalar^{\inn,m_1}_{\Delta_1}\scalar^{\tsign{-},im_2}_{\Delta_2}\scalar^{\out}_{\Delta_3}}
        -\vev{\scalar^{\tsign{-},im_1}_{\Delta_1}\scalar^{\tsign{-},im_2}_{\Delta_2}\scalar^{\out}_{\Delta_3}}
    \mright)
    \, .
\end{equation}
Physically, if taking $m_1 \to 0$ first, the first term vanishes by momentum conservation, and the remaining two should reduce to the celestial amplitude with only the second particle regularized \eqref{eq: scalar 3pt massreg 2}.
Conversely, if taking $m_2 \to 0$ first, the last term vanishes and the remaining two should reduce to the the celestial amplitude with only the first particle regularized \eqref{eq: scalar 3pt massreg 1}.

Similarly, regularizing the two outgoing particles yields
\begin{equation}
    \label{eq: scalar 3pt massreg 23}
    \vev{\wave{\scalar}^{\inn}_{\Delta_1}\scalar^{\out}_{\Delta_2}\scalar^{\out}_{\Delta_3}}
    =
    \lim_{\substack{m_2\to 0\\ m_3\to 0}}
    \frac{1}{4}\cpwflimit_{\Delta_2}^{m_2,0}\cpwflimit_{\Delta_3}^{m_3,0}
    \mleft(
        -\vev{\wave{\scalar}^{\inn}_{\Delta_1}\scalar^{\tsign{-},im_2}_{\Delta_2}\scalar^{\out,m_3}_{\Delta_3}}
        +\vev{\wave{\scalar}^{\inn}_{\Delta_1}\scalar^{\tsign{-},im_2}_{\Delta_2}\scalar^{\tsign{-},im_3}_{\Delta_3}}
        -\vev{\wave{\scalar}^{\inn}_{\Delta_1}\scalar^{\out,m_2}_{\Delta_2}\scalar^{\tsign{-},im_3}_{\Delta_3}}
    \mright)
    \, .
\end{equation}
Taking $m_2\to 0$ first kills the first term, and the remaining two should reduce to the celestial amplitude with only the third particle regularized.
Conversely, taking $m_3\to 0$ first kills the last term, and the remaining two should reduce to the celestial amplitude with only the second particle regularized.

The validity of the above arguments relies on the compatibility between the massless limit of the conformal basis in Section \ref{sec: massless limit} and the massless limit of the celestial amplitudes, which is not guaranteed a priori as pointed in Section \ref{sec: regular celestial amplitude}.
In Appendix \ref{app: three-point celestial amplitudes - summary}, we compute each term in \eqref{eq: scalar 3pt massreg 12} and \eqref{eq: scalar 3pt massreg 23}, and then in Appendix \ref{app: three-point celestial amplitudes - massless limit of scalars} we confirm that taking either limit indeed reproduces the single-regularized celestial amplitude \eqref{eq: scalar 3pt massreg result}.

\textbf{Regularizing three particles}.
Regularizing all three particles produces eight terms, and momentum conservation eliminates one of them, leaving seven nontrivial contributions:
\begin{align}
    &
    \vev{\wave{\scalar}^{\inn}_{\Delta_1}\scalar^{\out}_{\Delta_2}\scalar^{\out}_{\Delta_3}}
    =
    \raisebox{0.25em}{
        $\displaystyle\lim_{\substack{m_1\to 0\\ m_2\to 0\\ m_3\to 0}}$
    }
    \frac{1}{8}\cpwflimit_{\Delta_2}^{m_2,0}\cpwflimit_{\Delta_3}^{m_3,0}
    \biggl(
        \vev{\scalar^{\inn,m_1}_{\Delta_1}\scalar^{\out,m_2}_{\Delta_2}\scalar^{\out,m_3}_{\Delta_3}}
        -\vev{\scalar^{\inn,m_1}_{\Delta_1}\scalar^{\tsign{-},im_2}_{\Delta_2}\scalar^{\out,m_3}_{\Delta_3}}
        -\vev{\scalar^{\inn,m_1}_{\Delta_1}\scalar^{\out,m_2}_{\Delta_2}\scalar^{\tsign{-},im_3}_{\Delta_3}}
        \nn
        \\[-0.5em]
        &
        \hspace{-0.75em}
        +\vev{\scalar^{\inn,m_1}_{\Delta_1}\scalar^{\tsign{-},im_2}_{\Delta_2}\scalar^{\tsign{-},im_3}_{\Delta_3}}
        +\vev{\scalar^{\tsign{-},im_1}_{\Delta_1}\scalar^{\tsign{-},im_2}_{\Delta_2}\scalar^{\out,m_3}_{\Delta_3}}
        +\vev{\scalar^{\tsign{-},im_1}_{\Delta_1}\scalar^{\out,m_2}_{\Delta_2}\scalar^{\tsign{-},im_3}_{\Delta_3}}
        -\vev{\scalar^{\tsign{-},im_1}_{\Delta_1}\scalar^{\tsign{-},im_2}_{\Delta_2}\scalar^{\tsign{-},im_3}_{\Delta_3}}
    \biggr)
    \, .
\end{align}
While computing the regular celestial amplitude requires knowledge of all seven terms, a closed-form expression is known only for the fully massive case \cite{Liu:2024lbs}. Nonetheless, our physical argument for the limiting procedure is straightforward. Regardless of the order of massless limits, at least two of the three mass parameters will approach zero first, reducing the problem to one of the previously discussed cases with either one or two regularized particles.

\subsubsection{OPE and three-point regular celestial amplitudes}
\label{sec: scalar OPE from three-point}

We extract OPEs from three-point regular celestial amplitudes. In standard CFTs, after inserting the OPE $\op_{1}\op_{2}\sim \op_{\Delta,J}$ into a three-point correlator, the OPE coefficient $\coOPE{\op_{1}\op_{2}|\op_{\Delta,J}}$ is related to the two- and three-point coefficients by
\begin{equation}
    \coThree{\op_{1}\op_{2}\op_{3}}
    =
    \coOPE{\op_{1}\op_{2}|\op_{\Delta,J}}
    \coTwo{\op_{\Delta,J}\op_{3}}
    .
\end{equation}
However, as discussed in Section \ref{sec: discussion on celestial OPE}, this relation holds under the premise that the remaining two-point correlator takes the standard form \eqref{eq: conformal structure}, and in CCFT, the two-point correlators $\vev{\op^{\inn}\wave\op^{\out}}$ and $\vev{\wave\op^{\inn}\op^{\out}}$ in the Mellin-shadow mixed basis satisfy this requirement.

Specifically, for the scalar OPE \eqref{eq: scalar OPE from collinear}, the out-out type can be read off from the regular celestial amplitude $\vev{\scalar^{\out}_{\Delta_1}\scalar^{\out}_{\Delta_2}\wave{\scalar}^{\inn}_{\Delta_3}}$, while the two exchange operators in the in-out OPE can be read off from $\vev{\scalar^{\inn}_{\Delta_1}\scalar^{\out}_{\Delta_2}\wave{\scalar}^{\out}_{\Delta_3}}$ and $\vev{\scalar^{\inn}_{\Delta_1}\scalar^{\out}_{\Delta_2}\wave{\scalar}^{\inn}_{\Delta_3}}$, respectively.
The first amplitude has been discussed in detail in Section \ref{sec: scalar three-point}, and the other two can be obtained similarly using \eqref{eq: 3pt coefficient OOM scalar} and \eqref{eq: 3pt coefficient TOO scalar}:
\begin{align}
    \vev{\scalar^{\inn}_{\Delta_1}\scalar^{\out}_{\Delta_2}\wave{\scalar}^{\out}_{\Delta_3}}
    &=
    \frac{1}{32}
    \deltaK{\Delta_{12,3}-2}
    \betaShort{3-\Delta_{12},\Delta_2-1}
    \vevv{\scalar_{1}\scalar_{2}\scalar_{3}}
    .
    \\
    \vev{\scalar^{\inn}_{\Delta_1}\scalar^{\out}_{\Delta_2}\wave{\scalar}^{\inn}_{\Delta_3}}
    &=
    \frac{1}{32}
    \deltaK{\Delta_{12,3}-2}
    \betaShort{3-\Delta_{12},\Delta_1-1}
    \vevv{\scalar_{1}\scalar_{2}\scalar_{3}}
    \, .
\end{align}
Then we extract the following OPEs:
\begin{align}
    \label{eq: scalar OPE from 3pt}
    &
    \scalar^{\out}_{\Delta_1}\scalar^{\out}_{\Delta_2}
    \sim
    \frac{1}{32|z_{1,2}|^2}
    \frac{\deltaK{\Delta_{12,3}-2}}{\coTwo{\scalar_{\Delta_{12}-2}\wave\scalar_{\Delta_{3}}}}
    \betaShort{\Delta_1-1,\Delta_2-1}
    \scalar^{\out}_{\Delta_{3}}
    \, ,
    \\
    &
    \scalar^{\inn}_{\Delta_1}\scalar^{\out}_{\Delta_2}
    \sim
    \frac{1}{32|z_{1,2}|^2}
    \frac{\deltaK{\Delta_{12,3}-2}}{\coTwo{\scalar_{\Delta_{12}-2}\wave\scalar_{\Delta_{3}}}}
    \Bigl(
        \betaShort{3-\Delta_{12},\Delta_2-1}\scalar^{\inn}_{\Delta_{3}}
        +\betaShort{3-\Delta_{12},\Delta_1-1}\scalar^{\out}_{\Delta_{3}}
    \Bigr)
    \, .
\end{align}
Here $\coTwo{\scalar_{1}\wave\scalar_{2}}$ is the two-point coefficient of $\vev{\scalar_{1}^{\inn}\wave\scalar_{2}^{\out}}$, and it satisfies $\coTwo{\scalar_{1}^{\inn}\wave\scalar_{2}^{\out}} = \coTwo{\wave\scalar_{1}^{\inn}\scalar_{2}^{\out}}$ by shadow transform.
It is worth noting that the self-consistency of CCFT implies that a single three-point amplitude of in-out-out type simultaneously encodes both the out-out OPE and the in-out OPE, which corresponds to the fact that the above three amplitudes can also be derived from $\vev{\scalar^{\inn}_{\Delta_1}\scalar^{\out}_{\Delta_2}{\scalar}^{\out}_{\Delta_3}}$ by applying the shadow transform and using CPT symmetry.

The OPE \eqref{eq: scalar OPE from 3pt} from three-point correlators correctly reproduces the exchange operators and the $\Delta$-dependence of \eqref{eq: scalar OPE from collinear} from collinear limit.
However, two interesting issues arise concerning two-point correlators and propagators, which we briefly discuss below and will revisit in a future work.

\textbf{Two-point normalization.}
Comparing \eqref{eq: scalar OPE from 3pt} with \eqref{eq: scalar OPE from collinear}, the two-point correlator is proportional to a Kronecker delta function:
\begin{equation}
    \label{eq: 2pt with Kronecker delta function}
    \vev{\scalar^{\inn}_{\Delta_1}\wave\scalar^{\out}_{\Delta_2}}
    =
    \frac{1}{8}\deltaK{\Delta_{1,2}}
    \vevv{\scalar_{1}\scalar_{2}}
    \, .
\end{equation}
However, the two-point correlator introduced in \cite{Pasterski:2017kqt} is proportional to the complex delta function $\deltaComplex{\Delta_{1,2}}$.
A possible origin of this discrepancy is that the higher-point correlators we compute come from the interacting part of the scattering matrix, while the two-point in \cite{Pasterski:2017kqt} comes from the identity part, \ie, the inner product of a bulk unitary representation.

For a CFT with a discrete spectrum, two-point correlators are naturally normalized by the Kronecker delta function, and OPEs are discrete sums. With a continuous spectrum, two-point correlators are more likely normalized by the Dirac delta function, and OPEs involve integrals rather than sums, \eg Liouville CFT.

If one considers $\deltaComplex{\Delta_{1,2}}$ to be correct, then three-point regular celestial amplitudes should also contain complex delta functions to match the collinear OPE, suggesting that the regularized conformal basis in Section \ref{sec: regular celestial amplitude} has not been properly normalized.

If instead $\deltaK{\Delta_{1,2}}$ is correct, a possible resolution is as follows. For CCFT, the conformal dimension is usually considered to take continuous values on the principal series $\Delta\in\pseries$, but this is merely an outcome of decomposing bulk unitary representations to the conformal group $\sogroup(3,1)\subset \isogroup(3,1)$ \cite{Chakrabarti1,Chakrabarti2}, and does not necessarily constitute the spectrum of a well-defined CCFT we pursue. For example, \cite{Freidel:2022skz,Cotler:2023qwh,Mitra:2024ugt} introduced discrete basis to understand the soft mode, and $\Delta$ takes discrete values at certain integers. From this perspective, the two-point correlator \eqref{eq: 2pt with Kronecker delta function} with $\deltaK{\Delta_{1,2}}$ is a better candidate to match the discrete collinear OPE \eqref{eq: scalar OPE from collinear}.

\textbf{Sign of in-out OPE.}
Compared with \eqref{eq: scalar OPE from collinear}, \eqref{eq: scalar OPE from 3pt} lacks the relative minus sign in the in-out OPE versus the out-out OPE. This sign arises because the regular three-point celestial amplitudes encode only information of three vertices, whereas the split amplitudes used to compute the collinear OPE include both vertices and propagators. The minus sign in the in-out OPE \eqref{eq: scalar OPE from collinear} is traced back to the propagator $\frac{1}{t}$ with $t<0$.

This mismatch becomes more transparent in higher-derivative theories. For example, in the $\phi^3$ theory with kinetic term $\phi \partial^{4} \phi$, the propagator is quartic and produces collinear singularities $\frac{1}{s^{2}}$ and $\frac{1}{t^{2}}$, then the collinear OPEs are
\begin{align}
    \scalar^{\out}_{\Delta_1}(z_1)
    \scalar^{\out}_{\Delta_2}(z_2)
    &\sim
    \frac{1}{16 \abs{z_{1,2}}^{4}}
    \betaShort{\Delta_1-2,\Delta_2-2}
    \scalar^{\out}_{\Delta_{12}-4}(z_2)
    \, ,
    \\
    \scalar^{\inn}_{\Delta_1}(z_1)
    \scalar^{\out}_{\Delta_2}(z_2)
    &\sim
    \frac{1}{16 \abs{z_{1,2}}^{4}}
    \Bigl(
        \betaShort{\Delta_2-2,5-\Delta_{12}}
        \scalar^{\inn}_{\Delta_{12}-4}(z_2)
        +
        \betaShort{\Delta_1-2,5-\Delta_{12}}
        \scalar^{\out}_{\Delta_{12}-4}(z_2)
    \Bigr)
    \, .
    \nn
\end{align}
In contrast, the regular three-point amplitudes take the same forms as in the usual $\phi^3$ theory, hence the $\Delta$-dependence does not match. This suggests that the regularized conformal basis in Section \ref{sec: regular celestial amplitude} needs further refinement to capture the information of propagators.

\subsection{Four-point regular celestial amplitudes}
\label{sec: scalar four-point}

To bypass the subtleties of two-point celestial amplitudes, we now turn to extracting the OPEs from four-point regular celestial amplitudes.
In standard CFTs, the contribution of an exchange operator $\op_1\op_2\sim \op_{\Delta,J}$ to the four-point correlator takes the form
\begin{equation}
    \label{eq: conformal block coefficient definition}
    \vev{\op_1\op_2\op_3\op_4}
    \sim
    \coBlock{\op_{\Delta,J}}
    \chi^{\frac{\Delta+J}{2}}
    \bar{\chi}^{\frac{\Delta-J}{2}}
    \vevv{\op_1\op_2\op_3\op_4}
    \, ,
\end{equation}
where $\coBlock{\op_{\Delta,J}}$ is called the conformal block coefficient associated with this operator.
If the OPE is consistent, and $\op_{\Delta,J}$ is primary without degeneracy, then $\coBlock{\op_{\Delta,J}}$ must factorize into a product of the OPE coefficient $\coOPE{\op_1\op_2|\op_{\Delta,J}}$ and the three-point coefficient
\begin{equation}
    \label{eq: conformal block coefficient factorization}
    \coBlock{\op_{\Delta,J}}
    =
    \coOPE{\op_1\op_2|\op_{\Delta,J}}\,
    \coThree{\op_{\Delta,J}\op_3\op_4}
    \, .
\end{equation}

If the CCFT is a consistent CFT, we expect that from the following celestial amplitudes:
\begin{equation}
    \vev{\scalar^{\inn}_{\Delta_1}\scalar^{\inn}_{\Delta_2}\scalar^{\out}_{\Delta_3}\scalar^{\out}_{\Delta_4}}
    \, ,
    \quad
    \vev{\scalar^{\inn}_{\Delta_1}\scalar^{\out}_{\Delta_2}\scalar^{\out}_{\Delta_3}\scalar^{\out}_{\Delta_4}}
    \, ,
\end{equation}
the scalar OPE \eqref{eq: scalar OPE from collinear} can be reproduced with the correct factorization of conformal block coefficients.
However, for conventional celestial amplitudes this expectation fails.
The first one  does not admit a \schannel conformal block expansion, and thus the out-out OPE cannot be extracted \cite{Chang:2022jut,Chang:2023ttm,Liu:2024lbs,Liu:2024vmx}; the second one is widely regarded as vanishing due to momentum conservation, again in tension with the scalar OPE.
In this section we employ the regular celestial amplitudes introduced in Section \ref{sec: regular celestial amplitude} to extract the scalar OPE and find exact agreement with \eqref{eq: scalar OPE from collinear}.

\subsubsection{OPE from two-to-two scattering}

In principle, all massless particles should be regularized to obtain the full regular celestial amplitude, but this makes the computation excessively complicated.
If the goal is only to extract specific OPEs, the computation can
be simplified by regularizing only a subset of massless particles, as illustrated in Section \ref{sec: scalar three-point}.

For the four-point correlator $\vev{\scalar^{\inn}_{\Delta_1}\scalar^{\inn}_{\Delta_2}\scalar^{\out}_{\Delta_3}\scalar^{\out}_{\Delta_4}}$, to extract the out-out OPE $\scalar^{\out}_{\Delta_3}\scalar^{\out}_{\Delta_4}\sim \scalar^{\out}_{\Delta_{34}-2}$ in \eqref{eq: scalar OPE from collinear}, it suffices to ensure that, after inserting this OPE into the correlator, the remaining three-point correlator $\vev{\scalar^{\inn}_{\Delta_1}\scalar^{\inn}_{\Delta_2}\scalar^{\out}_{\Delta_{34}-2}}$ admits the standard form \eqref{eq: conformal structure}. This can be achieved by regularizing only the first particle.
Similarly, to extract the in-out OPE $\scalar^{\inn}_{\Delta_2}\scalar^{\out}_{\Delta_4}\sim\scalar^{\inn}_{\Delta_{24}-2}+\scalar^{\out}_{\Delta_{24}-2}$, the relevant three-point correlators $\vev{\scalar^{\inn}_{\Delta_1}\scalar^{\inn}_{\Delta_{24}-2}\scalar^{\out}_{\Delta_3}}$ and $\vev{\scalar^{\inn}_{\Delta_1}\scalar^{\out}_{\Delta_{24}-2}\scalar^{\out}_{\Delta_3}}$ should also take the standard form, which again can be ensured by regularizing only the first particle.
Below we demonstrate this idea through an explicit computation.

Under the mass regularization \eqref{eq: CPWF - mass regularization}, the four-point regular celestial amplitude becomes
\begin{align}
    \vev{\scalar^{\inn}_{\Delta_1}\scalar^{\inn}_{\Delta_2}\scalar^{\out}_{\Delta_3}\scalar^{\out}_{\Delta_4}}
    =
    \lim_{m\to 0}
    \frac{1}{2}\cpwflimit_{\Delta_1}^{m,0}
    \mleft(
        \vev{\scalar^{\inn,m}_{\Delta_1}\scalar^{\inn}_{\Delta_2}\scalar^{\out}_{\Delta_3}\scalar^{\out}_{\Delta_4}}
        -
        \vev{\scalar^{\tsign{+},im}_{\Delta_1}\scalar^{\inn}_{\Delta_2}\scalar^{\out}_{\Delta_3}\scalar^{\out}_{\Delta_4}}
    \mright)
    \, .
\end{align}
We expect the first term to reproduce the in-out OPE $\scalar^{\inn}_{\Delta_2}\scalar^{\out}_{\Delta_4}$ with the outgoing exchange $\scalar^{\out}_{\Delta_{24}-2}$, as indicated by the nonvanishing correlator $\vev{\scalar^{\inn,m}_{\Delta_1}\scalar^{\out}_{\Delta_{24}-2}\scalar^{\out}_{\Delta_3}}$.
In Appendix \ref{app: four-point celestial amplitudes - derivation}, we compute this term as \eqref{eq: 4pt amplitude MOOO S-channel}, and then in Appendix \ref{app: four-point celestial amplitudes - OPE extraction} we extract the collinear OPE by the blow-up method.
For the scattering amplitude $\cT(s,t) \sim \frac{1}{t}$ in the collinear limit $t\to0$, by \eqref{eq: MOOO t-channel collinear conformal block coefficient} the exchange operator is indeed $\scalar^{\out}_{\Delta_{24}-2}$, and the conformal block coefficient factorizes as
\begin{align}
    \coBlock{\scalar^{\out}_{\Delta_{24}-2}}
    &=
    -\frac{1}{128 \pi}
    \deltaK{\Delta_{1234}-6}
    \mgShort{
        \Delta_{13}-3,
        \Delta_{13}-2,
        \Delta_2-1,
        2-\Delta_3
    }{
        \Delta_1-1,
        \Delta_{123}-4
    }
    \\
    &=
    -\frac{1}{4}\betaShort{\Delta_2-1,3-\Delta_{24}}\,
    \coThree{\scalar^{\inn}_{\Delta_1}\scalar^{\out}_{\Delta_{24}-2}\scalar^{\out}_{\Delta_3}}
    \, ,
    \nn
\end{align}
where in the second line we have used the three-point regular coefficient
\begin{equation}
    \coThree{\scalar^{\inn}_{\Delta_1}\scalar^{\out}_{\Delta_{2}}\scalar^{\out}_{\Delta_3}}
    =
    \frac{\Delta_1-1}{32 \pi}
    \deltaK{\Delta_{123}-4}
    \betaShort{2-\Delta_2,2-\Delta_3}
    \, .
\end{equation}
This result is in perfect agreement with the scalar OPE \eqref{eq: scalar OPE from collinear}.

The second term is expected to reproduce both the out-out OPE $\scalar^{\out}_{\Delta_3}\scalar^{\out}_{\Delta_4}\sim \scalar^{\out}_{\Delta_{34}-2}$, and the in-out OPE $\scalar^{\inn}_{\Delta_2}\scalar^{\out}_{\Delta_4}$ with an incoming exchange $\scalar^{\inn}_{\Delta_{24}-2}$, based on analogous reasoning.
This term is given by \eqref{eq: 4pt amplitude TOOO S-channel}.
For the scattering amplitude $\cT(s,t) \sim \frac{1}{s}$ as $s\to0$, by \eqref{eq: TOOO s-channel collinear conformal block coefficient} the exchange operator is $\scalar^{\out}_{\Delta_{34}-2}$, and the conformal block coefficient factorizes as
\begin{equation}
    \label{eq: scalar OPE from TOOO s-channel}
    \coBlock{\scalar^{\out}_{\Delta_{34}-2}}
    =
    \frac{1}{4}\betaShort{\Delta_3-1,\Delta_4-1}\,
    \coThree{\scalar^{\inn}_{\Delta_1}\scalar^{\inn}_{\Delta_2}\scalar^{\out}_{\Delta_{34}-2}}
    \, ,
\end{equation}
For the scattering amplitude $\cT(s,t) \sim \frac{1}{t}$ as $t\to0$, by \eqref{eq: TOOO t-channel collinear conformal block coefficient} the exchange operator is $\scalar^{\inn}_{\Delta_{24}-2}$, and the conformal block coefficient factorizes as
\begin{equation}
    \coBlock{\scalar^{\inn}_{\Delta_{24}-2}}
    =
    -\frac{1}{4}\betaShort{\Delta_4-1,3-\Delta_{24}}\,
    \coThree{\scalar^{\inn}_{\Delta_1}\scalar^{\inn}_{\Delta_{24}-2}\scalar^{\out}_{\Delta_3}}
    \, .
\end{equation}
Both results are in perfect agreement with the scalar OPE \eqref{eq: scalar OPE from collinear}.

\subsubsection{OPE from one-to-three scattering}
\label{sec: scalar OPE from 1to3}

The consistency of CCFT also requires the scalar OPE \eqref{eq: scalar OPE from collinear} can be reproduced from the four-point correlator $\vev{\scalar^{\inn}_{\Delta_1}\scalar^{\out}_{\Delta_2}\scalar^{\out}_{\Delta_3}\scalar^{\out}_{\Delta_4}}$. Due to momentum conservation, the celestial amplitude is usually regarded as vanishing. However, the regular celestial amplitude is in fact nonvanishing, because momentum conservation receives contributions from soft regions.

To extract the out-out OPE, we apply mass regularization to the incoming particle, ensuring the remaining three-point celestial amplitude takes the standard form after inserting the OPE. The regular celestial amplitude is then given by
\begin{align}
    \vev{\scalar^{\inn}_{\Delta_1}\scalar^{\out}_{\Delta_2}\scalar^{\out}_{\Delta_3}\scalar^{\out}_{\Delta_4}}
    =
    \lim_{m\to 0}
    \frac{1}{2}\cpwflimit_{\Delta_1}^{m,0}
    \mleft(
        \vev{\scalar^{\inn,m}_{\Delta_1}\scalar^{\out}_{\Delta_2}\scalar^{\out}_{\Delta_3}\scalar^{\out}_{\Delta_4}}
        -
        \vev{\scalar^{\tsign{+},im}_{\Delta_1}\scalar^{\out}_{\Delta_2}\scalar^{\out}_{\Delta_3}\scalar^{\out}_{\Delta_4}}
    \mright)
    \, .
\end{align}
Here the second term vanishes by momentum conservation, and the first term is given by \eqref{eq: 4pt amplitude MOOO D-channel}.
For the scattering amplitude $\cT(s,t) \sim \frac{1}{s}$ near $s=0$, by \eqref{eq: MOOO d-channel collinear conformal block coefficient} the exchange operator is $\scalar^{\out}_{\Delta_{34}-2}$, and the conformal block coefficient factorizes as
\begin{equation}
    \coBlock{\scalar^{\out}_{\Delta_{34}-2}}
    =
    \frac{1}{4}\betaShort{\Delta_3-1,\Delta_4-1}\,
    \coThree{\scalar^{\inn}_{\Delta_1}\scalar^{\out}_{\Delta_2}\scalar^{\out}_{\Delta_{34}-2}}
    \, ,
\end{equation}
agreeing with the two-to-two case \eqref{eq: scalar OPE from TOOO s-channel} and the scalar OPE \eqref{eq: scalar OPE from collinear}.

According to the permutation symmetry, the other two out-out OPEs can be extracted similarly. Besides, the in-out OPE can be extracted by regularizing one outgoing particle to ensure the corresponding three-point amplitude takes its standard form. The computations are analogous and are therefore omitted.

\subsubsection{Three-OPE from one-to-three scattering}

From the CCFT perspective, scalar $\phi^3$ theory exhibits an obvious deficiency: soft currents do not generate an associative symmetry algebra due to the violation of the double residue condition \cite{Mago:2021wje,Ball:2022bgg,Ren:2022sws,}.
On the amplitude side, this violation originates from the three-particle factorization singularity $\frac{1}{s_{123}}$, which has been confirmed through three-collinear limit analysis \cite{Ball:2023sdz,Guevara:2024ixn} and the split representation method \cite{Liu:2024lbs}.
This singularity provides a new contribution to the three-OPE:
\begin{equation}
    \label{eq: scalar 3-OPE}
    \scalar^{\out}_{\Delta_{1}}\scalar^{\out}_{\Delta_{2}}\scalar^{\out}_{\Delta_{3}}
    \sim
    \LR{
        \coThree_{1}
        \abs{z_{1,3}}^{-2}
        \abs{z_{2,3}}^{-2}
        +
        \coThree_{2}
        \abs{z_{1,3}}^{-2\Delta_1}
        \abs{z_{2,3}}^{2\Delta_1-4}
    }
    \scalar^{\out}_{\Delta_{123}-4}
    \, ,
\end{equation}
where the three-OPE coefficients are given by
\begin{align}
    \label{eq: scalar 3-OPE coefficients}
    &
    \coThree_{1}
    =
    \frac{1}{16}
    \betaShort{\Delta_1-1,\Delta_{23}-3}
    \betaShort{\Delta_2-1,\Delta_3-1}
    \, ,
    \\
    &
    \coThree_{2}
    =
    -\frac{\pi }{16} \csc (\pi  \Delta_1)
    \betaShort{\Delta_{12}-2,\Delta_{13}-2}
    \mleft(
        \frac{\Delta_{12}-2}{\Delta_{13}-3}
        +\frac{\Delta_{13}-2}{\Delta_{12}-3}
        +1
    \mright)
    \, .
\end{align}
Here the first term arises from the leading scalar exchange in the two-OPE \eqref{eq: scalar OPE from collinear}, while the second term is argued to correspond to multi-particle operators in the two-OPE \cite{Ball:2023sdz,Guevara:2024ixn}.

Now consider this phenomenon from the CFT perspective. Whether the OPE algebra satisfies associativity or not, after inserting the three-OPE $\scalar^{\out}_{\Delta_{2}}\scalar^{\out}_{\Delta_{3}}\scalar^{\out}_{\Delta_{4}}$ into the decay process four-point correlator $\vev{\wave\scalar^{\inn}_{\Delta_1}\scalar^{\out}_{\Delta_2}\scalar^{\out}_{\Delta_3}\scalar^{\out}_{\Delta_4}}$, the remaining two-point correlator $\vev{\wave\scalar^{\inn}_{\Delta_{1}}\scalar^{\out}_{\Delta_{123}-4}}$ takes the standard form \eqref{eq: conformal structure}, then the three-OPE coefficients should be extractable from this correlator. This conclusion seems to contradict the fact that the singularity $\frac{1}{s_{123}}$ does not appear in four-point scattering amplitudes.

Surprisingly, from regular celestial amplitudes we can indeed extract the three-OPE \eqref{eq: scalar 3-OPE} with the correct coefficients \eqref{eq: scalar 3-OPE coefficients}.
We consider the scattering amplitude $\cT(s,t)=\frac{1}{s}+\frac{1}{t}+\frac{1}{u}$ and take the mass regularization \eqref{eq: CPWF - mass regularization} on the first particle,
\begin{equation}
    \vev{\wave\scalar^{\inn}_{\Delta_1}\scalar^{\out}_{\Delta_2}\scalar^{\out}_{\Delta_3}\scalar^{\out}_{\Delta_4}}
    =
    \lim_{m\to 0}
    \frac{1}{2}
    \vev{\scalar^{\inn,m}_{\Delta_1}\scalar^{\out}_{\Delta_2}\scalar^{\out}_{\Delta_3}\scalar^{\out}_{\Delta_4}}
    \, ,
\end{equation}
where the right side is given by \eqref{eq: 4pt amplitude MOOO D-channel}.

To utilize the extraction method of two-OPE, we perform a change of variables
\begin{equation}
    z_1 = z_3 + \varepsilon
    \, ,
    \quad
    z_2 = z_3 + \varepsilon \chi
    \, ,
\end{equation}
and take the limit $\varepsilon \to 0$, then $\chi$ becomes precisely the cross-ratio.
Then we apply the blow-up method in Appendix \ref{app: four-point celestial amplitudes - OPE extraction} to determine the small-$\chi$ behavior of the four-point correlator
\begin{equation}
    \vev{\wave\scalar^{\inn}_{\Delta_1}\scalar^{\out}_{\Delta_2}\scalar^{\out}_{\Delta_3}\scalar^{\out}_{\Delta_4}}
    \sim
    \deltaK{\Delta_{234,1}-4}
    \LR{
        \coBlock_{1}
        \abs{\chi}^{\Delta_{34}-2}
        +
        \coBlock_{2}
        \abs{\chi}^{\Delta_{34}+2 \Delta_2-4}
    }
    \vevv{\scalar_{1}\scalar_{2}\scalar_{3}\scalar_{4}}
    \, .
\end{equation}
As discussed in Section \ref{sec: scalar OPE from 1to3}, by \eqref{eq: MOOO d-channel collinear conformal block coefficient} the $\coBlock_{1}$-term comes from the collinear regime $s\sim 0$ of $\cT$, and it corresponds to the first term in the three-OPE \eqref{eq: scalar 3-OPE}. With the two-point coefficient \eqref{eq: 2pt with Kronecker delta function}, the coefficient factorizes into a product of the three-OPE coefficient and the two-point coefficient,
\begin{equation}
    \coBlock_{1}
    =
    \coThree_{1}\,
    \coTwo{\wave\scalar^{\inn}_{\Delta_1}\scalar^{\out}_{\Delta_{123}-4}}
    \, .
\end{equation}
By \eqref{eq: MOOO d-channel double-trace conformal block coefficient}, the $\coBlock_{2}$-term comes from the small mass regime $m\sim 0$ of $\cT$, and it corresponds to the second term in the three-OPE \eqref{eq: scalar 3-OPE} with the factorization
\begin{equation}
    \coBlock_{2}
    =
    \coThree_{2}\,
    \coTwo{\wave\scalar^{\inn}_{\Delta_1}\scalar^{\out}_{\Delta_{123}-4}}
    \, .
\end{equation}

Another interesting observation is that, from the shadow OPE perspective, by $\Delta_{234,1}=4$ the $\coBlock_{2}$-term corresponds to the scalar exchange of double-trace type:
\begin{equation}
    \wave\scalar^{\inn}_{\Delta_1}\scalar^{\out}_{\Delta_2}
    \sim
    \op_{\Delta_{12},0}
    \, .
\end{equation}
As noticed in \cite{Liu:2025dhh}, this type of operator is universally present in massless shadow and massive celestial amplitudes, and in the massless case the conformal block coefficient strikingly captures all the dynamical information of helicity amplitudes.

\section{Yang-Mills and Einstein gravity theories}
\label{sec: Yang-Mills and Gravity}


In this section, we study regular celestial amplitudes in Yang-Mills and Einstein gravity theories, and verify their consistency with the celestial OPEs.
From four-point gluons, the extracted OPEs agree exactly with the collinear OPEs in \cite{Pate:2019lpp}.
From three-point gluons and gravitons, the extracted OPEs agree with \cite{Pate:2019lpp} up to the sign issue of in-out OPE, as discussed in Section \ref{sec: scalar OPE from three-point}.
Additionally, in the four-point gluon case we observe an unexpected ``anomalous'' scalar exchange in the OPE, and clarifying its physical implications is left for future work.

Compared to the scalar $\phi^3$ theory, a new phenomenon arises here: for the regular celestial amplitudes, besides the nontrivial contributions from soft and collinear regions of momentum conservation, Ward identities of gauge symmetry also introduce necessary contributions, as first noted in \cite{Liu:2025dhh}.
The reason is as follows. By the discussion of massless limit in Section \ref{sec: massless limit}, the regularized conformal basis \eqref{eq: CPWF - mass regularization} retains derivatives of Ward identities in the massless basis \eqref{eq: CPWF Mellin}. Since external momenta are taken off-shell at this stage, these derivative terms are not killed by momentum conservation. They therefore survive in regular celestial amplitudes and save the conformal covariance.

\textbf{Conventions.}
We denote a gluon operator as $\gluon^{a}_{\Delta,\pm}$ and a graviton operator as $\graviton_{\Delta,\pm}$, with conformal spins $J=\pm\ell$ abbreviated as $\pm$.
Following the amplitude literature, bulk helicities are defined with respect to outgoing particles, and external momenta are outgoing in Feynman rules.

For the Yang-Mills theory, we set the gauge coupling $g=1$ and choose the covariant $R_{\xi}$ gauge, then the Feynman rules are
\begin{align}
    &
    \TextInMath{propagator: }
    &&
    {-}i\,  \delta^{a_{1} a_{2}} q^{-2}
    \LRa{g^{\mu_1\mu_2}+(1-\xi)q^{\mu_1} q^{\mu_2} q^{-2}}
    \, ,
    \\
    &
    \TextInMath{3-vertex: }
    &&
    f^{a_1a_2a_3}
    \LRa{
        (q_1-q_2)^{\mu_3} g^{\mu_1\mu_2}+(q_2-q_3)^{\mu_1} g^{\mu_2\mu_3}+(q_3-q_1)^{\mu_2} g^{\mu_1\mu_3}
    }
    \, ,
    \nn
    \\
    &
    \TextInMath{4-vertex: }
    &&
    i f^{a_1a_2b} f^{a_3a_4b} (g^{\mu_1\mu_4} g^{\mu_2\mu_3}-g^{\mu_1\mu_3} g^{\mu_2\mu_4})
    +
    i f^{a_1a_3b} f^{a_2a_4b} (g^{\mu_1\mu_4} g^{\mu_2\mu_3}-g^{\mu_1\mu_2} g^{\mu_3\mu_4})
    \nn
    \\
    &
    &&
    +i f^{a_1a_4b} f^{a_2a_3b} (g^{\mu_1\mu_3} g^{\mu_2\mu_4}-g^{\mu_1\mu_2} g^{\mu_3\mu_4})
    \, .
    \nn
\end{align}

For the Einstein gravity theory, we use the following two different graviton vertices to cross-check our results, with expressions stored in \href{\RCARepository}{this GitHub repository}:
\begin{itemize}
    \item
        DeWitt type \cite{DeWitt:1967yk,DeWitt:1967ub,DeWitt:1967uc}: the three-point vertex contains 171 terms.
        For convenience, we set the gravity coupling to $1$.
    \item
        FeynGrav package \cite{Latosh:2022ydd,Latosh:2023zsi,Latosh:2024lhl}: the three-point vertex contains 420 terms and depends on a gauge parameter.
        Compared with the DeWitt type, we take the coupling constant to $4 i$.
\end{itemize}

\subsection{Three-point regular celestial amplitudes}
\label{sec: three-point spinning}

We first sketch the algorithm of computing spinning three-point regular celestial amplitudes, with one particle mass-regularized.
The key observation is that the differential operators in the Mellin basis \eqref{eq: CPWF Mellin} act on all the rest factors.
The algorithm is:
\begin{enumerate}
    \item
        Perform index contraction between interaction vertices and external conformal bases.
    \item
        Use momentum conservation to perform the integrals over one momentum $\phat$/$\khat$ and over two energies $\omega_i$, in direct analogy with the scalar case in Appendix \ref{app: TOO scalar}.
    \item
        Rewrite each term as a scalar celestial amplitude with shifted conformal weights.
    \item
        Finally apply the leftover Mellin-basis differential operators to these scalar seeds, and sum them up to extract the spinning three-point coefficients.
\end{enumerate}
In particular, using momentum conservation and the transversality condition, the relation between the gluon mass-regularized celestial amplitudes and the scalar ones can be neatly written as
\begin{align}
    &
    \vev{
        \gluon^{\out,a_{1}}_{\Delta_1,J_1}
        \gluon^{\out,a_{2}}_{\Delta_2,J_2}
        \gluon^{\inn,m,a_{3}}_{\Delta_3,J_3}
    }
    =
    f^{a_1a_2a_3}
    \frac{1-\Delta_3}{\Delta_3}
    \prod_{i=1}^3
    (\epsilon_{J_i}^{\mu_i}+\frac{1}{\Delta_i-1}\partial_{J_i}\co \qhat_i^{\mu_i})
    \\
    &\quad \xx
    \LRb{
        \LRa{\hat{q}_1^{\mu_3} g^{\mu_1\mu_2}-2\hat{q}_1^{\mu_2} g^{\mu_1\mu_3}}
        \vev{\scalar^{\out}_{\Delta_1+1}\scalar^{\out}_{\Delta_2}\scalar^{\inn,m}_{\Delta_3}}
        -
        \LRa{\hat{q}_2^{\mu_3} g^{\mu_1\mu_2}-2\hat{q}_2^{\mu_1} g^{\mu_2\mu_3}}
        \vev{\scalar^{\out}_{\Delta_1}\scalar^{\out}_{\Delta_2+1}\scalar^{\inn,m}_{\Delta_3}}
    }
    \, ,
    \nn
\end{align}
where the differential operators $\partial_{J_i}\co$ act on all subsequent expressions.

We summarize the mass-regularized celestial amplitudes in Appendix \ref{app: three-point celestial amplitudes - summary}, and now discuss the physical implications.
To extract the out-out gluon OPEs, we apply the mass regularization to the incoming one,
\begin{equation}
    \vev{\gluon^{\out,a_1}_{\Delta_1,J_1}\gluon^{\out,a_2}_{\Delta_2,J_2}\wave{\gluon}^{\inn,a_3}_{\Delta_3,J_3}}
    =
    \lim_{m \to 0}
    \frac{1}{2}
    \vev{\gluon^{\out,a_1}_{\Delta_1,J_1}\gluon^{\out,a_2}_{\Delta_2,J_2}\gluon^{\inn,m,a_3}_{\Delta_3,J_3}}
    \, .
\end{equation}
There are eight spin configurations $(J_1,J_2,J_3)$, among which the $(+,+,-)$ configuration is noteworthy, with three-point coefficient
\begin{align}
    \coThree
    =
    \lim_{m\to 0}
    f^{a_1a_2a_3}
    2^{\Delta_{3,12}-3}
    m^{\Delta_{12,3}-1}
    \frac{(\Delta_{12,3}-1) (\Delta_{12,3}+1)}{(\Delta_1-1) (\Delta_2-1)}
    \betaShort{\frac{\Delta_{13,2}+1}{2},\frac{\Delta_{23,1}+1}{2}}
    \, .
\end{align}
For generic conformal dimensions, the massless limit produces a Kronecker delta function $\deltaK{\Delta_{12,3}-1}$, then the accompanying factor $\Delta_{12,3}-1$ forces this coefficient to vanish. Noticing that the shadow transform flips the conformal spin, all the vanishing three-point correlators are
\begin{equation}
    \label{eq: vanishing 3pt gluon amplitudes}
    \vev{\gluon^{\out,a_{1}}_{\Delta_{1},+}\gluon^{\out,a_{2}}_{\Delta_{2},+}\wave{\gluon}^{\inn,a_{3}}_{\Delta_{3},-}}
    =
    \vev{\gluon^{\out,a_{1}}_{\Delta_{1},-}\gluon^{\out,a_{2}}_{\Delta_{2},-}\wave{\gluon}^{\inn,a_{3}}_{\Delta_{3},+}}
    =
    \vev{\gluon^{\out,a_{1}}_{\Delta_{1},+}\gluon^{\out,a_{2}}_{\Delta_{2},+}{\gluon}^{\inn,a_{3}}_{\Delta_{3},+}}
    =
    \vev{\gluon^{\out,a_{1}}_{\Delta_{1},-}\gluon^{\out,a_{2}}_{\Delta_{2},-}{\gluon}^{\inn,a_{3}}_{\Delta_{3},-}}
    =
    0
    \, .
\end{equation}
On the boundary side, this implies that the OPE between two gluons of the same helicity contains no exchange gluon with opposite helicity.
On the bulk side, the vanishing of these amplitudes reflects the well-known absence of same-helicity gluon scattering amplitudes.

The remaining spin configurations are analyzed similarly, and we summarize the three-point coefficients of gluons and gravitons in Table \ref{tab: three-point coefficients}. As claimed in the beginning, the OPEs extracted from these coefficients agree with \cite{Pate:2019lpp} up to the sign issue of in-out OPE.

\subsection{Four-point regular celestial amplitudes}
\label{sec: four-point spinning}

We now analyze four-point regular celestial amplitudes of gluons at tree-level and their compatibility with the celestial OPEs. The computation of gravitons is time-consuming and will be presented in future work.

Before proceeding to computation, we examine the implications of OPE consistency for four-point gluon correlators. Whether for two-to-two or one-to-three scattering processes, the following three conclusions should hold:
\begin{itemize}
    \item
        The gluon exchanges in the collinear OPEs should contribute to the regular celestial amplitudes as \eqref{eq: conformal block coefficient definition}, and the corresponding conformal block coefficients should factorize into the product of OPE coefficients and three-point coefficients as \eqref{eq: conformal block coefficient factorization}.
    \item
        For same-helicity correlators like $\vev{\gluon_{+}\gluon_{+}\gluon_{+}\gluon_{+}}$ and $\vev{\wave\gluon_{-}\gluon_{+}\gluon_{+}\gluon_{+}}$, gluon exchange should not appear, since same-helicity three-point correlators vanish \eqref{eq: vanishing 3pt gluon amplitudes}.
    \item
        Correlators with single-plus/minus helicity like $\vev{\gluon_{-}\gluon_{+}\gluon_{+}\gluon_{+}}$ and $\vev{\wave\gluon_{+}\gluon_{+}\gluon_{+}\gluon_{+}}$ should not vanish, which can be seen by applying the OPE of last two gluons.

        This novel implication seems to contradict the fact that only MHV amplitudes are nonvanishing at tree level, but as we have explained, compared with the helicity amplitudes, regular celestial amplitudes contain extra contributions from soft/collinear regions and Ward identities.
\end{itemize}
As we will show, while conventional celestial amplitudes satisfy only the second point, regular celestial amplitudes satisfy all the three.

\textbf{Master integrals.}
We now sketch the computation. As in Section \ref{sec: scalar four-point}, it suffices to regularize only one gluon to extract the OPEs.
The difficulty lies in the fact that, unlike the scalar $\cT(s,t)$, off-shell spinning amplitudes do not admit a compact form. Consequently, for mass-regularized spinning celestial amplitudes, we cannot provide universal formulas analogous to the scalar case in Appendix \ref{app: four-point celestial amplitudes - summary}; instead, we need to compute directly from Feynman diagrams.
Similar to the three-point case in Section \ref{sec: three-point spinning}, the algorithm is as follows.
\begin{enumerate}
    \item
        Perform index contraction between Feynman diagrams and external conformal bases.
    \item
        Perform the conformal-basis integrals following the scalar procedure in Appendix \ref{app: four-point celestial amplitudes - derivation}.
    \item
        Apply the Mellin-basis differential operators to the scalar seeds and then sum them up.
    \item
        Strip off conformal structure to obtain the stripped correlator.
        This can be done by fixing to some conformal frame and then check conformal covariance numerically.
    \item
        Check the stripped correlator is independent of gauge choices.
\end{enumerate}
The resulting regular celestial amplitude can be written as a master integral times a polynomial part, $I=I^{\text{master}}\xx I^{\text{poly}}$.
For example, for the decay process we have
\begin{equation}
    \vev{\gluon^{\inn,m,a_{1}}_{\Delta_1,-}\gluon^{\out,a_{2}}_{\Delta_2,+}\gluon^{\out,a_{3}}_{\Delta_3,+}\gluon^{\out,a_{4}}_{\Delta_4,+}}
    =
    \LRa{
        f^{a_{0} a_{1} a_{2}} f^{a_{0} a_{3} a_{4}}
        I_{\texttt{-+++1234}}
        +
        f^{a_{0} a_{1} a_{3}} f^{a_{0} a_{2} a_{4}}
        I_{\texttt{-+++1324}}
    }
    \vevv{\gluon_{1}\gluon_{2}\gluon_{3}\gluon_{4}}
    \, .
\end{equation}
Here the master integral and polynomial part for the first color structure are
\begin{align}
    I_{\texttt{-+++1234}}^{\text{master}}
    &
    =
    \frac{
        -i\,
        2^{\Delta_{1,234}-3}
        m^{\Delta_{234,1}-2}
    }{
        (\Delta_2-1) (\Delta_3-1) (\Delta_4-1)
    }
    \chi^{\frac{\Delta_{12}}{2}-1}
    \bar{\chi}^{\frac{\Delta_{12}}{2}}
    (1-\chi )^{\frac{\Delta_{14,23}}{2}-2}
    (1-\bar{\chi})^{\frac{\Delta_{14,23}}{2}+1}
    \inttt{dS dT}{\region}
    \\
    &\peq\xx
    S^{\frac{\Delta_{134,2}}{2}-1}
    T^{\frac{\Delta_{124,3}}{2}-1}
    (1-S-T)^{\frac{\Delta_{123,4}}{2}-1}
    \mleft(
        S+T \chi  \bar{\chi}-(S+T \chi ) (S+T \bar{\chi})
    \mright)^{-\Delta_1-4}
    \, ,
    \nn
    \\[0.5em]
    I_{\texttt{-+++1234}}^{\text{poly}}
    &
    =
    -2 \Delta_3 (\Delta_{1234}-2) \Delta_{13,24} \chi^7 \bar{\chi}^3 T^7
    +
    \text{(259 more terms in $\chi,\bar{\chi},S,T$)}
    \, ,
\end{align}
where $S,T$ are dimensionless Mandelstam variables $s = m^{2} S$, $t = m^{2} T$ and the integral is over the rescaled physical region $\region = \set{(S,T) \given S\geq 0 \land T\geq 0 \land S+T\leq 1}$.

We have computed four-point mass-regularized celestial amplitudes, for both the two-to-two and one-to-three processes and for all helicity configurations. Each result is expressed in terms of master integrals similarly, and the full data is available in \href{\RCARepository}{this GitHub repository}. Further reduction via IBP relations for these master integrals is expected and left to future work.

\textbf{Closed-form and OPE consistency.}
Among the scattering processes and helicity configurations, there are two special cases that the decay regular celestial amplitudes admit closed-forms.
\begin{equation}
    \vev{\wave\gluon^{\inn, a_{1}}_{\Delta_1,J_1}\gluon^{\out, a_{2}}_{\Delta_2,J_2}\gluon^{\out, a_{3}}_{\Delta_3,J_3}\gluon^{\out, a_{4}}_{\Delta_4,J_4}}
    =
    \lim_{m\to 0}\half
    \vev{\gluon^{\inn,m, a_{1}}_{\Delta_1,J_1}\gluon^{\out, a_{2}}_{\Delta_2,J_2}\gluon^{\out, a_{3}}_{\Delta_3,J_3}\gluon^{\out, a_{4}}_{\Delta_4,J_4}}
    \, ,
\end{equation}
where the master integrals can be performed explicitly by the change of variables
\begin{equation}
    S
    =
    \frac{\chi  \bar{\chi}}{
        \eta_1 (\chi -1) (\bar{\chi}-1)
        +\chi  \bar{\chi}+\eta_2
    }
    \, ,
    \quad
    T
    =
    \frac{\eta_2}{
        \eta_1 (\chi -1) (\bar{\chi}-1)
        +\chi  \bar{\chi}+\eta_2
    }
    \, .
\end{equation}

The first case is the same-helicity configuration, \eg
\begin{equation}
    \vev{\wave\gluon^{\inn, a_{1}}_{\Delta_1,-}\gluon^{\out, a_{2}}_{\Delta_2,+}\gluon^{\out, a_{3}}_{\Delta_3,+}\gluon^{\out, a_{4}}_{\Delta_4,+}}
    =
    0
    \, .
\end{equation}
Notice that the conformal spin $J_{1}$ is opposite to the helicity by the shadow transform. This result is even stronger than the expectation from OPE consistency: not only is there no gluon exchange, but the entire amplitude vanishes.

The single-plus/minus helicity configuration is more interesting, \eg
\begin{equation}
    \label{eq: four-point gluon amplitude - SOOO-decay}
    \vev{\wave\gluon^{\inn, a_{1}}_{\Delta_1,-}\gluon^{\out, a_{2}}_{\Delta_2,-}\gluon^{\out, a_{3}}_{\Delta_3,+}\gluon^{\out, a_{4}}_{\Delta_4,+}}
    =
    \LRb{
        I_{\texttt{n}}
        \LRa{
            f^{a_{0} a_{1} a_{2}} f^{a_{0} a_{3} a_{4}}
            -\chi f^{a_{0} a_{1} a_{3}} f^{a_{0} a_{2} a_{4}}
        }
        +
        I_{\texttt{a}}
        f^{a_{0} a_{1} a_{2}} f^{a_{0} a_{3} a_{4}}
    }
    \vevv{\gluon_{1}\gluon_{2}\gluon_{3}\gluon_{4}}
    \, ,
\end{equation}
where we have separated it into a normal part and an ``anomalous'' part,
\begin{equation}
    \begin{aligned}
        I_{\texttt{n}}
        &
        =
        \frac{i}{4} \deltaK{\Delta_{234,1}-2}
        \mgShort{
            \Delta_2+1,
            \Delta_3-1,
            \Delta_4-1
        }{
            \Delta_{234}-1
        }
        \chi^{\frac{\Delta_{34}}{2}}
        \bar{\chi}^{\frac{\Delta_{34}}{2}-1}
        (1-\chi)^{-1}
        \, ,
        \\[0.5em]
        I_{\texttt{a}}
        &
        =
        -\frac{i}{8} \deltaK{\Delta_{234,1}-2}
        \frac{\Delta_2 \Delta_{3,4}(\Delta_{234}-3) }{(\Delta_3-1) (\Delta_4-1)}
        \mgShort{
            \Delta_2-1,
            \Delta_3-1,
            \Delta_4-1
        }{
            \Delta_{234}-1
        }
        \chi^{\frac{\Delta_{34}}{2}-1}\bar{\chi}^{\frac{\Delta_{34}}{2}-1}
        \, .
    \end{aligned}
\end{equation}
This separation is motivated by consistency with the collinear OPE and the conformally soft Ward identity.

From the OPE perspective, by \eqref{eq: conformal block coefficient definition}, the anomalous part $I_{\texttt{a}}$ corresponds to a colored scalar exchange $\op^{\out,a_{0}}_{\Delta_{34}-2,0}$ in the \schannel OPE $\gluon^{\out,a_{3}}_{\Delta_3,+}\gluon^{\out,a_{4}}_{\Delta_4,+}$, and is more dominant than the expected gluon exchange $\gluon^{\out,a_{0}}_{\Delta_{34}-1,+}$. This scalar has a peculiar property: accompanied by a primary operator $\op$, its descendant operators $\pp^{n}\ppb^{\bar{n}}\op$ also contribute to higher-point correlators. However, through conformal block analysis, we find that due to the restriction $\Delta_{234,1}=2$, all descendants of $\op^{\out,a_{0}}_{\Delta_{34}-2,0}$ do not contribute to the four-point correlator \eqref{eq: four-point gluon amplitude - SOOO-decay}, and hence will not mix with gluon exchanges in the normal part $I_{\texttt{n}}$. For this reason, we introduce the subtracted four-point correlator as
\begin{equation}
    \label{eq: four-point gluon amplitude - SOOO-decay-subtracted}
    \vev{\wave\gluon^{\inn, a_{1}}_{\Delta_1,-}\gluon^{\out, a_{2}}_{\Delta_2,-}\gluon^{\out, a_{3}}_{\Delta_3,+}\gluon^{\out, a_{4}}_{\Delta_4,+}}_{\texttt{n}}
    =
    I_{\texttt{n}}
    \LRa{
        f^{a_{0} a_{1} a_{2}} f^{a_{0} a_{3} a_{4}}
        -\chi f^{a_{0} a_{1} a_{3}} f^{a_{0} a_{2} a_{4}}
    }
    \vevv{\gluon_{1}\gluon_{2}\gluon_{3}\gluon_{4}}
    \, .
\end{equation}
Now from the subtracted correlator, we can extract the exchange gluon $\gluon^{\out,a_{0}}_{\Delta_{34}-1,+}$ and the conformal block coefficient as
\begin{align}
    \coBlock{\gluon^{\out,a_{0}}_{\Delta_{34}-1,+}}
    &=
    \frac{i}{4}
    \deltaK{\Delta_{234,1}-2}
    f^{a_{0} a_{1} a_{2}} f^{a_{0} a_{3} a_{4}}
    \mgShort{
        \Delta_2+1,
        \Delta_3-1,
        \Delta_4-1
    }{
        \Delta_{234}-1
    }
    \\
    &=
    - i
    f^{a_{0} a_{1} a_{2}}
    \betaShort{\Delta_3-1,\Delta_4-1}
    \coThree{\wave\gluon^{\inn,a_{1}}_{\Delta_1,-}\gluon^{\out,a_{2}}_{\Delta_2,-}\gluon^{\out,a_{0}}_{\Delta_{34}-1,+}}
    \, ,
    \nn
\end{align}
which agrees exactly with the factorization \eqref{eq: conformal block coefficient factorization} and the OPE coefficient in \cite{Pate:2019lpp}.

For other four-point regular celestial amplitudes, closed-form expressions are not available, and we analyze their OPE limits using the blow-up technique as the scalar case in Section \ref{sec: scalar four-point}.
We find that there are also anomalous scalars with conformal dimension $\Delta_{34}-2$, and for the gluon exchanges, the conformal block coefficients factorize perfectly into the three-point coefficients in Table \ref{tab: three-point coefficients} and the OPE coefficients in Table \ref{tab: OPE coefficients}, matching the collinear OPEs in \cite{Pate:2019lpp}.

\textbf{Conformally soft theorem and Banerjee-Ghosh equation.}
The soft theorem of scattering amplitudes corresponds to the Ward identity of currents in CCFT, also known as the conformally soft theorem in \eg \cite{Fan:2019emx,Pate:2019mfs}, see also the review \cite{Ball:2024oqa}.
The leading soft gluon is
\begin{equation}
    \gluonS^{a}_{\Delta=1}(z)
    =
    \Res_{\Delta=1}\gluon_{\Delta,+}^{\out,a}(z,\bar{z})
    \, .
\end{equation}
Here to emphasize the holomorphic nature of the current $\ppb \gluonS=0$ we have written the $\zb$-dependence explicitly. Then the conformally soft theorem reads
\begin{equation}
    \label{eq: conformally soft Ward identity}
    \vev{
        \gluonS^{a}(z_{0})
        \gluon^{\io,a_{1}}_{\Delta_1,J_1}
        \cdots
        \gluon^{\io,a_{n}}_{\Delta_n,J_n}
    }
    =
    \sum_{k}
    -i f^{a a_k b}z_{0,k}^{-1}
    \vev{
        \gluon^{\io,a_{1}}_{\Delta_1,J_1}
        \cdots
        \gluon^{\io,b}_{\Delta_k,J_k}
        \cdots
        \gluon^{\io,a_{n}}_{\Delta_n,J_n}
    }
    \, .
\end{equation}
A priori, shadow gluons $\wave\gluon$ do not necessarily satisfy this theorem, because the shadow transform smears the operator over the celestial sphere and thereby delocalizes its insertion point.

However, we find that after subtracting the anomalous scalar, the regular celestial amplitude \eqref{eq: four-point gluon amplitude - SOOO-decay-subtracted} respects the conformally soft Ward identity.
When the second gluon $\gluon^{\out, a_{2}}_{\Delta_2,-}$ becomes soft, the remaining three-point correlator vanishes by \eqref{eq: vanishing 3pt gluon amplitudes}, and this agrees with that the correlator \eqref{eq: four-point gluon amplitude - SOOO-decay-subtracted} has no simple pole at $\Delta_{2}=1$.
When the fourth gluon $\gluon^{\out, a_{4}}_{\Delta_4,+}$ becomes soft, the color structure on the right side of the Ward identity is $i f^{ba_1a_4} f^{ba_2a_3}z_{4,1}^{-1}- i f^{ba_1a_3} f^{ba_2a_4} z_{4,2}^{-1}+ i f^{ba_1a_2} f^{ba_3a_4} z_{4,3}^{-1}$. Then using the Jacobi identity of structure constants and the three-point coefficients in Table \ref{tab: three-point coefficients}, we check that this equals the residue of \eqref{eq: four-point gluon amplitude - SOOO-decay-subtracted} at $\Delta_{4}=1$, thus verifying the Ward identity \eqref{eq: conformally soft Ward identity}.

Another aspect related to the soft theorem is that higher-point MHV celestial amplitudes satisfy the Banerjee-Ghosh difference-differential equations \cite{Banerjee:2020vnt}, which arises from the commutativity of soft limits and collinear limits. However, as noticed in \cite{Fan:2022vbz}, four-point MHV celestial amplitudes do not satisfy these equations due to the missing gluon exchanges in conventional celestial amplitudes.

Interestingly, we find that after subtracting the anomalous scalar, the non-MHV regular celestial amplitude \eqref{eq: four-point gluon amplitude - SOOO-decay-subtracted} also satisfies similar equations, for example, the term with color structure $f^{a_{0} a_{1} a_{2}} f^{a_{0} a_{3} a_{4}}$ is annihilated by the Banerjee-Ghosh operator as in \cite{Fan:2022vbz},
\begin{equation}
    \label{eq: Banerjee-Ghosh operator}
    \pp_{z_{4}}
    -\Delta_{4}z_{3,4}^{-1}
    -z_{1,4}^{-1}
    +z_{3,4}^{-1}
    \LRa{
        \Delta_{3}-2+z_{3,4}\pp_{\zb_{3}}
    }T_{\Delta_{4}}T^{-1}_{\Delta_{3}}
    \, ,
\end{equation}
where $T_{\Delta}$ is the shift operator acting on conformal dimensions as $T_{\Delta}f(\Delta)=f(\Delta+1)$.

These two results provide another physical motivation for subtracting the anomalous scalars, as they violate the conformally soft theorem and the Banerjee-Ghosh equations.

\subsection{Summary of OPE and three-point coefficients}
\label{sec: summary of OPE and three-point coefficients}

In this section, we summarize the results for the three-point coefficients in the Yang-Mills and Einstein gravity theories, as well as the celestial OPEs extracted from four-point regular celestial amplitudes in the Yang-Mills theory.

\begin{table}[H]
    \TableLineWidth{1pt}
    \TableVerticalSpace{2}
    {
        \centering
        \begin{tabular}{c|cccc}
            & $\gluon^{\out,a_{3}}_{\Delta_{3},+}$
            & $\gluon^{\out,a_{3}}_{\Delta_{3},-}$
            & $\gluon^{\inn,a_{3}}_{\Delta_{3},+}$
            & $\gluon^{\inn,a_{3}}_{\Delta_{3},-}$
            \\
            \hline
            $\gluon^{\out,a_{1}}_{\Delta_1,+}\gluon^{\out,a_{2}}_{\Delta_2,+}$
            & $\betaShort{\Delta_{1}{-}1,\Delta_{2}{-}1}$
            & 0
            & 0
            & 0
            \\
            $\gluon^{\out,a_{1}}_{\Delta_1,+}\gluon^{\out,a_{2}}_{\Delta_2,-}$
            & $\betaShort{\Delta_{1}{+}1,\Delta_{2}{-}1}$
            & $\betaShort{\Delta_{1}{-}1,\Delta_{2}{+}1}$
            & 0
            & 0
            \\
            $\gluon^{\inn,a_{1}}_{\Delta_1,+}\gluon^{\out,a_{2}}_{\Delta_2,+}$
            & $-\betaShort{\Delta_{1}{-}1,3{-}\Delta_{12}}$
            & 0
            & $\betaShort{\Delta_{2}-1,3-\Delta_{12}}$
            & 0
            \\
            $\gluon^{\inn,a_{1}}_{\Delta_1,+}\gluon^{\out,a_{2}}_{\Delta_2,-}$
            & $-\betaShort{\Delta_{1}{+}1,1{-}\Delta_{12}}$
            & $-\betaShort{\Delta_{1}{-}1,1{-}\Delta_{12}}$
            & $\betaShort{\Delta_{2}-1,1-\Delta_{12}}$
            & $\betaShort{\Delta_{2}+1,1-\Delta_{12}}$
        \end{tabular}
        \caption{OPE coefficients extracted from four-point regular celestial amplitudes.}
        \label{tab: OPE coefficients}
    }
    \smallskip
    {
        \small
        Rows correspond to operator products, and columns correspond to exchange operators.
        Each OPE coefficient is proportional to a common factor: $ -i f^{a_1a_2a_3} \deltaK{\Delta_{12,3}-1}$.
    }
\end{table}

\begin{table}[H]
    {
        \centering
        \TableLineWidth{1pt}
        \TableVerticalSpace{2}
        \begin{tabular}{c|cccc}
            & $(+,+,+)$
            & $(+,+,-)$
            & $(+,-,+)$
            & $(+,-,-)$
            \\
            \hline
            $\vev{\gluon^{\out}_{1}\gluon^{\out}_{2}\wave{\gluon}^{\inn}_{3}}$
            & $\betaShort{\Delta_1{-}1,\Delta_2{-}1}$
            & $0$
            & $\betaShort{\Delta_1{+}1,\Delta_2{-}1}$
            & $\betaShort{\Delta_1{-}1,\Delta_2{+}1}$
            \\
            $\vev{\gluon^{\inn}_{1}\gluon^{\out}_{2}\wave{\gluon}^{\inn}_{3}}$
            & $\betaShort{\Delta_1{-}1,3{-}\Delta_{12}}$
            & $0$
            & $\betaShort{\Delta_1{+}1,1{-}\Delta_{12}}$
            & $\betaShort{\Delta_1{-}1,1{-}\Delta_{12}}$
            \\
            $\vev{\gluon^{\inn}_{1}\gluon^{\out}_{2}\wave{\gluon}^{\out}_{3}}$
            & $-\betaShort{\Delta_2{-}1,3{-}\Delta_{12}}$
            & $0$
            & $-\betaShort{\Delta_2{-}1,1{-}\Delta_{12}}$
            & $-\betaShort{\Delta_2{+}1,1{-}\Delta_{12}}$
            \\[0.5em]
            \hline
            $\vev{\graviton^{\out}_{1}\graviton^{\out}_{2}\wave{\graviton}^{\inn}_{3}}$
            & $\betaShort{\Delta_1{-}1,\Delta_2{-}1}$
            & $0$
            & $\betaShort{\Delta_1{+}3,\Delta_2{-}1}$
            & $\betaShort{\Delta_1{-}1,\Delta_2{+}3}$
            \\
            $\vev{\graviton^{\inn}_{1}{\graviton}^{\out}_{2}\wave{\graviton}^{\inn}_{3}}$
            & $\betaShort{\Delta_{1}{-}1,3{-}\Delta_{12}}$
            & $0$
            & $\betaShort{\Delta_{1}{+}3,-1{-}\Delta_{12}}$
            & $\betaShort{\Delta_{1}{-}1,-1{-}\Delta_{12}}$
            \\
            $\vev{\graviton^{\inn}_{1}\graviton^{\out}_{2}\wave\graviton^{\out}_{3}}$
            & $\betaShort{\Delta_{2}{-}1,3{-}\Delta_{12}}$
            & $0$
            & $\betaShort{\Delta_{2}{-}1,-1{-}\Delta_{12}}$
            & $\betaShort{\Delta_{2}{+}3,-1{-}\Delta_{12}}$
        \end{tabular}
        \caption{Three-point coefficients of regular celestial amplitudes.}
        \label{tab: three-point coefficients}
    }
    \smallskip
    {
        \small
        Rows correspond to three-point correlators with color indices suppressed.
        Columns correspond to spin configurations, and the remaining four spin configurations not listed here can be obtained by symmetry.
        Each three-point coefficient is proportional to a common factor: $-\frac{1}{4}f^{a_1a_2a_3}\deltaK{\Delta_{12,3}-1}$ for gluons and $-2\deltaK{\Delta_{12,3}}$ for gravitons.
    }
\end{table}


\section{Discussions}

This work resolves two fundamental issues in celestial holography: the distributional nature of conventional celestial amplitudes, which obstructs the application of standard CFT techniques, and their inconsistency with the known celestial OPEs.
We find that both are symptoms of a single underlying problem: massless celestial amplitudes defined by the Mellin transform are not suitable as CCFT correlators.

Consequently, we introduce regular celestial amplitudes, computed using the regularized conformal bases in \eqref{eq: CPWF - power regularization}, \eqref{eq: CPWF - mass regularization}, and \eqref{eq: CPWF - Guassian regularization}.
Compared to conventional celestial amplitudes, our regular amplitudes take the standard form of conformal correlators and, more importantly, are compatible with the known celestial OPEs, up to the sign issue in the in-out OPE.
We confirm this claim through explicit computations in massless $\phi^3$, Yang-Mills, and Einstein gravity theories.
Particularly, the non-MHV amplitude $\vev{\gluon_{-}\gluon_{+}\gluon_{+}\gluon_{+}}$ now exhibits a nonvanishing contribution in the required OPE channel, thereby ensuring the consistency of the OPE algebra.

Our approach reveals a crucial physical picture in celestial holography: a consistent CCFT requires off-shell data from the bulk theory.
The reason is that regular celestial amplitudes contain at least the following three types of contributions:
\begin{equation*}
    \text{regular celestial amplitude}
    \supset
    \text{helicity amplitude}
    +
    \text{soft/collinear}
    +
    \text{Ward identity}
    \, .
\end{equation*}
Therefore, although computing regular celestial amplitudes via Feynman rules is more involved, this is not a mere technical drawback, but a necessary condition to ensure the consistency of the OPE algebra.

There are several future directions:
\begin{itemize}
    \item
        Order of massless limits: prove that regular celestial amplitudes are independent of the order of limits when regularizing multiple particles, as discussed in Section \ref{sec: regular celestial amplitude}.
    \item
        Normalization of two-point correlators and sign of in-out OPE: resolve the tension between Kronecker and complex delta-function normalizations, and fix the sign discrepancy of in-out OPEs extracted from three- and four-point regular celestial amplitudes, as discussed in Section \ref{sec: scalar OPE from three-point}.
    \item
        Anomalous scalars in the celestial OPE of Yang-Mills theory: clarify the physical origin of anomalous scalar exchanges, their properties, and implications for the consistency of CCFT.
    \item
        Reduction of spinning regular celestial amplitudes: apply IBP relations to simplify the master integrals appearing in gluon and graviton regular celestial amplitudes.
    \item
        Regular celestial amplitudes in Klein space: extend the formalism of regular celestial amplitudes to Klein space to facilitate the study of celestial OPEs.
\end{itemize}


\section*{Acknowledgements}

The authors would like to thank Yangrui Hu, Sabrina Pasterski and Ellis Ye Yuan for useful discussions.
WJM is supported by the National Natural Science Foundation of China No. 12405082 and Shanghai Pujiang Program No. 24PJA118.


\clearpage

\appendix

\section{Useful identities}
\label{app: useful identities}

We list some useful identities here.

The Gamma symbol denotes the product of Gamma functions,
\begin{align}
    &\gm{a_{1},\dotsc,a_{n}}
    \equiv
    \prod_{i=1}^{n}\gm{a_{i}}
    \, ,
    \\
    &\mg{a_{1},\dotsc,a_{n} }{b_{1},\dotsc,b_{m}}
    \equiv
    \left.
    \prod_{i=1}^{n}\gm{a_{i}}
    \middle/
    \prod_{j=1}^{m}\gm{b_{j}}
    \right.
    \, .
\end{align}

The Mellin-Barnes relation is
\begin{equation}
    \label{eq: Mellin-Barnes relation}
    (A+B)^{-\Delta}=
    \intrange{\frac{ds}{2\pi i}}{-i\oo}{+i\oo}
    \mg{
        \Delta+s,-s
    }{
        \Delta
    }
    A^{s}B^{-s-\Delta}
    \, .
\end{equation}

The Feynman-Schwinger parameterization for $\Re(\Delta_i)>0$ is
\begin{align}
    \label{eq: Feynman-Schwinger parameterization}
    \prod_{i=1}^{n} A_i^{-\Delta_i}
    &=
    \mg{
        \sum_{i=1}^{n} \Delta_i
    }{
        \Delta_{1},\dotsc,\Delta_{n}
    }
    \LRb{\prod_{i=2}^n \intrange{d\alpha_{i}}{0}{\oo}\alpha_i^{\Delta_i-1}}
    \LRb{A_1+\sum_{i=2}^n \alpha_i A_i}^{-\sum_{i=1}^{n} \Delta_i}
    \, .
\end{align}

The hypergeometric function near $z=0$ is related to that near $z=\oo$ by
\begin{equation}
    \label{eq: hypergeometric 0->infinity}
    \Fpq{2}{1}{a,b}{c}{z}
    =
    (-z)^{-a}
    \mg{
        b-a,
        c
    }{
        b,
        c-a
    }
    \Fpq{2}{1}{a,a-c+1}{a-b+1}{\frac{1}{z}}
    +
    (a\leftrightarrow b)
    \, .
\end{equation}
The discontinuity across the branch cut $z\in(1,\oo)$ is
\begin{equation}
    \label{eq: hypergeometric discontinuity}
    \Disc\Fpq{2}{1}{a,b}{c}{z}
    =
    2 \pi i\,
    (z-1)^{-a-b+c}
    \mg{
        c
    }{
        a,
        b,
        1-a-b+c
    }
    \Fpq{2}{1}{c-a,c-b}{1-a-b+c}{1-z}
    \, .
\end{equation}

\subsection{Identities of conformal basis}

After performing the derivatives, the massless conformal basis \eqref{eq: CPWF Mellin} can be rewritten as
\begin{equation}
    \label{eq: CPWF Mellin form2}
    (\Phi_{\Delta,J}^{\io,\ell},\cT)
    =
    \intt{d\omega} \omega^{\Delta-1}
    \sum_{n=0}^{\ell}
    \frac{(\Delta+\ell-1) (\Delta)_{n}}{\Delta-1}
    \binom{\ell}{n}
    \qhat_{\symL\mumu{1}{n}}\epsilon_{J,\mumu{n+1}{\ell}\symR}
    \LRa{
        \pp_{J}^{n}\cT^{\mumu{1}{\ell}}(\mp q)
    }
    \, .
\end{equation}

The shadow, massive and tachyonic conformal bases \eqref{eq: CPWF shadow}, \eqref{eq: CPWF massive}, \eqref{eq: CPWF tachyonic} satisfy the following identity:
\begin{align}
    \label{eq: CPWF shadow=massive=tachyonic}
    &\intt{[dp']}
    \wave\Phi^{\io,\mumu{1}{\ell}}_{\Delta,J}(\hat{q},p')
    \cT_{\mumu{1}{\ell}}(\mp p')
    =
    \intt{[dp']}
    \Phi_{m,\Delta,J}^{\io,\mumu{1}{\ell}}(\hat{q},p')
    \cT_{\mumu{1}{\ell}}(\mp p')
    \, ,
    \\
    &\intt{[dk']}
    \wave\Phi^{\io,\mumu{1}{\ell}}_{\Delta,J}(\hat{q},k')
    \cT_{\mumu{1}{\ell}}(\mp k')
    =
    \intt{[dk]}
    \Phi_{im,\Delta,J}^{\tsign{\mp},\mumu{1}{\ell}}(\hat{q},k')
    \cT_{\mumu{1}{\ell}}(k')
    \, ,
\end{align}
where $\wave\Phi^{\io,\mumu{1}{\ell}}_{\Delta,J}(\hat{q},p'|k')$ denotes the integration kernel of the shadow basis with the massless momentum $q'$ replaced by the massive/tachyonic momentum $p'$/$k'$.
This identity can be verified using \eqref{eq: CPWF massive j=l}, \eqref{eq: CPWF tachyonic j=l} together with the symmetric, traceless, and transverse properties of the polarization tensor.

\section{Distributions}

We review some basic aspects of distributions used in this work.

\subsection{Analytic continuation of distributions}
\label{app: regularization of distributions}

The analytic continuation of distributions was initiated by Gelfand and his collaborators in \cite{Gelfand1,Gelfand2}, and this motivated Bernstein to develop the theory of D-modules \cite{Bernstein1,Bernstein2}.
In this work, we only need some simple cases. Following \cite{Gelfand1,Gelfand2}, we provide a brief and physics-oriented review.

\textbf{Regularization and normalization.}
For simplicity, we focus on tempered distributions on $\RR$.
A tempered distribution $\phi\in S'(\RR)$ acting on the rapidly decreasing test function $f\in S(\RR)$ can be formally written as an integral
\begin{equation}
    \label{eq: functional}
    (\phi,f)=\intt{d x}\phi(x) f(x)
    \, ,
\end{equation}
with the kernel $\phi(x)$.
It is useful to conceptualize $f$ as a Gaussian wave-packet and $\phi$ as a sharp classical observable.

When the kernel $\phi(x)$ contains singularities, the integral \eqref{eq: functional} is convergent only for a subspace $V_{\phi}\subset S(\RR)$ of test functions.
To extend the domain of $\phi$ from $V_{\phi}$ to $S(\RR)$, we need to subtract off the divergences in \eqref{eq: functional} in a systematic way, and this procedure is called regularization of distributions.
The extension is nonunique, and we are interested in the case where a family of distributions $\phi_{\lambda}(x)$ depends on the parameter $\lambda$ analytically. Then the analyticity of $\lambda$ can help choose a unique regularization of $\phi_{\lambda}(x)$.

If $\phi_{\lambda}(x)$ is meromorphic in $\lambda\in U\subset\CC$, we can cancel the poles of $\phi_{\lambda}(x)$ by another meromorphic function $N(\lambda)$ such that the normalized distribution $N(\lambda)\phi_{\lambda}(x)$ is holomorphic in $U$, called normalization of distributions.

\textbf{Homogeneous distributions.}
For $x\in\RR$, there are three bases of homogeneous distributions: the plus/minus basis is
\begin{equation}
    x_{\tsign{+}}^\lambda \eqq x^\lambda \theta(x)
    \, ,
    \quad
    x_{\tsign{-}}^\lambda \eqq (-x)^\lambda \theta(-x)
    \, ,
\end{equation}
and the even/odd basis is
\begin{equation}
    x_{\tsign{0}}^\lambda \eqq |x|^\lambda= x_{\tsign{+}}^\lambda+ x_{\tsign{-}}^\lambda
    \, ,
    \quad
    x_{\tsign{1}}^\lambda \eqq |x|^\lambda \sign(x)=x_{\tsign{+}}^\lambda- x_{\tsign{-}}^\lambda
    \, .
\end{equation}
These four distributions are meromorphic functions in $\lambda\in\CC$, and the normalized versions are
\begin{equation}
    \frac{1}{\gm{\lambda+1}}x_{\tsign{+}}^\lambda
    \, , \quad
    \frac{1}{\gm{\lambda+1}}x_{\tsign{-}}^\lambda
    \, ,  \quad
    \frac{1}{\gm{\frac{\lambda+1}{2}}}x_{\tsign{0}}^\lambda
    \, ,\quad
    \frac{1}{\gm{\frac{\lambda+2}{2}}}x_{\tsign{1}}^\lambda
    \, .
\end{equation}
The third imaginary basis consists of boundary values of holomorphic functions,
\begin{equation}
    \label{eq: homogeneous distribution - imaginary basis}
    \begin{aligned}
        &x_{\tsign{+i}}^\lambda \eqq  \lim_{\varepsilon\to 0}(x+i\varepsilon )^\lambda=x_{\tsign{+}}^\lambda+e^{i\pi \lambda}x_{\tsign{-}}^\lambda
        \, ,
        \\
        &x_{\tsign{-i}}^\lambda \eqq  \lim_{\varepsilon\to 0}(x-i\varepsilon )^\lambda=x_{\tsign{+}}^\lambda+e^{-i\pi \lambda}x_{\tsign{-}}^\lambda
        \, .
    \end{aligned}
\end{equation}
The $\lambda$-poles get canceled due to the phase factor, and $x_{\tsign{\pm i}}^\lambda$ is holomorphic in $\lambda\in\CC$.

At the removed poles, these six distributions localize to the Dirac delta function and its derivatives. For $n\in\NN$,
\begin{align}
    \label{eq: homogeneous distribution - localization}
    &
    \frac{x_{\tsign{\pm}}^\lambda}{\gm{\lambda+1}}
    =
    \delta^{(n)}(\pm x)
    \, ,
    \TextInMath{for} \lambda=-n-1
    \, ,
    \\
    &
    \frac{x_{\tsign{0}}^{\lambda}}{\gm{\frac{\lambda+1}{2}}}
    =
    \frac{(-1)^{n} n!}{(2n)!}\delta^{(2n)}(x)
    \, ,
    \TextInMath{for} \lambda=-2n-1
    \, ,
    \\
    &
    \frac{x_{\tsign{1}}^{\lambda}}{\gm{\frac{\lambda+2}{2}}}
    =
    \frac{(-1)^{n+1} n!}{(2n+1)!}\delta^{(2n+1)}(x)
    \, ,
    \TextInMath{for} \lambda=-2n-2
    \, ,
    \\
    &
    x_{\tsign{\pm i}}^{\lambda}
    =
    x^{-n-1}\mp i\pi \frac{(-1)^{n}}{n!} \delta^{(n)}(x)
    \, ,
    \TextInMath{for} \lambda=-n-1
    \, .
\end{align}

\textbf{Parity symmetry.}
Under the parity change $x\to -x$, the three bases transform as
\begin{equation}
    \label{eq: homogeneous distribution - parity symmetry}
    (-x)_{\tsign{\pm}}^{\lambda}=x_{\tsign{\mp}}^{\lambda}
    \, ,
    \quad
    (-x)_{\tsign{0|1}}^{\lambda}=(-1)^{0|1}x_{\tsign{0|1}}^{\lambda}
    \, ,
    \quad
    (-x)_{\tsign{\pm i}}^{\lambda}=e^{\pm i\pi \lambda} x_{\tsign{\mp i}}^{\lambda}
    \, .
\end{equation}
The last one is useful in computation and is consistent with the branch cut $x\in (-\infty,0)$ of $x^{\lambda}$.

\textbf{Analytic structure of $x_{\tsign{+}}^\lambda$.}
The integral $(x_{\tsign{+}}^\lambda,f(x))=\intrange{dx}{0}{\oo}x^\lambda f(x)$ is convergent and hence is holomorphic for $\Re \lambda> -1$.
For $\Re \lambda\leq -1$ the integral can be divergent and acquires regularization near $x=0$. The easiest way to see $\lambda$-poles in this region is to choose the test function as $e^{-x}$, then $(x_{\tsign{+}}^\lambda,e^{-x})=\gm{\lambda+1}$ manifests the analytic structure of $x_{\tsign{+}}^\lambda$ and provides the normalization.

A finer argument to read off the residues at $\lambda=-n-1$ is as follows.
Given a real-analytic test function, inserting its Taylor expansion and dividing the integration region into $[0,1]$ and $[1,\oo)$, we obtain
\begin{equation}
    (x_{\tsign{+}}^\lambda,f(x))
    =
    \sum_{n=0}^{\oo} \frac{f^{(n)}(0)}{n!}\frac{1}{\lambda+n+1}
    +
    \intrange{dx}{1}{\oo}x^\lambda f(x)
    \, .
\end{equation}
The second term is holomorphic in $\lambda$ by the fast decay. The first term shows the simple poles at $\lambda=-n-1$, and after chopping off the test function, the residues are
\begin{equation}
    \Res_{\lambda=-n-1} x_{\tsign{+}}^\lambda = \frac{(-1)^{n}}{n!}\delta^{(n)}(x)
    \, .
\end{equation}

\textbf{Singularity at infinity.}
In practice, the test functions do not decay sufficiently fast as $x\to \oo$, and the singularity at infinity plays an important role.
For example, choosing the test function as $(1+x)^{-1}$, there are additional poles at nonnegative integers:
\begin{equation}
    (x_{\tsign{+}}^\lambda,(1+x)^{-1})=\gm{-\lambda}\,  \gm{\lambda+1}
    \, .
\end{equation}
In this case the test function is not in $S(\RR)$ and should be divided into two parts: (1) one is in $S(\RR)$ and detects the singularity of $x_{\tsign{+}}^\lambda$ at origin, providing the factor $\gm{\lambda+1}$; (2) another is well-behaved as $x\to \oo$ and decays sufficiently fast as $x\to 0$, then detects the singularity of $x_{\tsign{+}}^\lambda$ at infinity, providing the factor $\gm{-\lambda}$, \eg $(x_{\tsign{+}}^\lambda,e^{-1/x})=-\gm{-\lambda-1}$.

\textbf{Two-dimensional case.}
In higher dimensions, homogeneous distributions are proportional to the spherical harmonics in the representations of the rotation group.
In $2d$, with the complex coordinate $z=r e^{i\theta}$, the homogeneous distributions can be written as
\begin{equation}
    z^{\frac{\delta+j}{2}}\zb^{\frac{\delta-j}{2}}
    \eqq
    r^{\delta}e^{i j \theta}
    \, ,
    \TextInMath{for}
    \delta\in\CC
    \, ,
    j\in\ZZ
    \, .
\end{equation}
Similar to the $1d$ case, to read off the analytic structure, we divide the integration region into the disk $r\in[0,1]$ and $r\in[1,\oo)$, then
\begin{equation}
    (z^{\frac{\delta+j}{2}}\zb^{\frac{\delta-j}{2}},f(z,\zb))
    =
    \sum_{n,m=0}^{\oo}
    \frac{f^{(n,m)}(0,0)}{n! m!}
    \frac{2\pi \deltaK{j+n-m}}{\delta+n+m+2}
    +
    \inttt{d^2 z}{r\geq 1}\rest
    \, .
\end{equation}
For fixed $j=m-n$, this exhibits the simple poles at $\delta=-n-m-2$ with residues proportional to $\delta^{(n,m)}(z,\zb)$.
The normalized version and its values at the removed poles are
\begin{equation}
    \label{eq: homogeneous distribution - localization 2d}
    \frac{1}{\gm{\frac{\delta+|j|+2}{2}}}
    z^{\lambda}\zb^{\bar\lambda}
    =
    \frac{\pi(-1)^{\max(n,m)}}{\max(n,m)!}
    \delta^{(n,m)}(z,\bar{z})
    \, .
\end{equation}

\subsection{Analytic functionals}
\label{app: analytic functionals}

The complex delta function $\delta_{\CC}$ is an analytic functional belonging to the Gelfand-Shilov space $Z'$, which is the dual of the space $Z$ of entire functions of at most exponential growth, see \cite{Gelfand1,Gelfand2}.
This distribution is employed in the celestial literature to understand the analytic continuation of $\Delta$, see \eg \cite{Donnay:2020guq}, and it also appears in the method of brackets for evaluating Feynman integrals, see \eg \cite{Gonzalez:2008xm,Gonzalez:2010nm,Gonzalez:2021vqh,Ananthanarayan:2021not}.
Following \cite{Donnay:2020guq}, we provide a brief and self-contained introduction here.

The complex delta function is defined as the identity of the Mellin transform,
\begin{equation}
    \label{eq: complex delta function and Mellin transform}
    \intrange{\frac{d\Delta}{2\pi i}}{a-i\infty}{a+i\infty}
    \deltaComplex{\Delta-\Delta_{0}}f(\Delta)
    =
    f(\Delta_{0})
    \, ,
\end{equation}
and formally can be written as
\begin{equation}
    \label{eq: complex delta function}
    \deltaComplex{\Delta-\Delta_{0}}
    =
    \intrange{d\omega}{0}{\oo}\omega^{\Delta_{0}-\Delta-1}
    \, .
\end{equation}

As a generalization of the conventional Dirac delta function, when $\Delta_{0}$ locates on the integration contour $\Delta\in a + i\RR$ in \eqref{eq: complex delta function and Mellin transform}, $\delta_{\CC}$ reduces to
\begin{equation}
    \deltaComplex{\Delta-\Delta_{0}}
    =
    2\pi \delta\left(\Im(\Delta-\Delta_{0})\right)
    \, ,
    \TextInMath{for} \Re\Delta_{0}=a
    \, .
\end{equation}
When $\Delta_{0}$ leaves off the integration contour, $\delta_{\CC}$ admits the following approximations:
\begin{equation}
    \label{eq: complex Delta function approximation}
    \deltaComplex{\Delta-\Delta_{0}}
    =
    \begin{dcases}
        \lim_{\varepsilon\to0}
        \gm{\Delta_{0}-\Delta}\varepsilon^{\Delta-\Delta_{0}}
        =
        \lim_{\varepsilon\to0}
        \intrange{d\omega}{0}{\oo}\omega^{\Delta_{0}-\Delta-1}e^{-\varepsilon \omega}
        \, ,
        &
        \TextInMath{for} \Re\Delta_{0}>a
        \, ,
        \\
        \lim_{\varepsilon\to0}
        \gm{\Delta-\Delta_{0}}\varepsilon^{\Delta_{0}-\Delta}
        =
        \lim_{\varepsilon\to0}
        \intrange{d\omega}{0}{\oo}\omega^{\Delta_{0}-\Delta-1}e^{-\varepsilon/\omega}
        \, ,
        &
        \TextInMath{for} \Re\Delta_{0}<a
        \, .
    \end{dcases}
\end{equation}

\textbf{Derivation of \eqref{eq: complex Delta function approximation}.}
This can be shown by a contour deformation argument.
For $\Re\Delta_{0}>a$, we consider a test function $f(\Delta)$ that is analytic in the region $a<\Re \Delta$ and decays sufficiently fast as $\Delta\to +\oo$.
As $\varepsilon^{\Delta-\Delta_{0}}\to 0$, we can enclose the contour to the right and pick up the poles at $\Delta=\Delta_{0}+n$, $n\in\NN$, leading to
\begin{equation}
    \lim_{\varepsilon\to 0}
    \intrange{\frac{d\Delta}{2\pi i}}{a-i\oo}{a+i\oo}
    \gm{\Delta_{0}-\Delta}\varepsilon^{\Delta-\Delta_{0}}
    f(\Delta)
    =
    \lim_{\varepsilon\to 0}\sum_{n=0}^{+\oo}\frac{(-1)^{n}}{n!}\varepsilon^{n}f(\Delta_{0}+n)
    =
    f(\Delta_{0})
    \, ,
\end{equation}
which justifies the defining property \eqref{eq: complex delta function and Mellin transform}.

From the analysis above, we observe that only the term $n = 0$ needs to be dominant. Consequently, the requirement on test functions can be relaxed: it suffices for $f(\Delta)$ to be holomorphic in the strip $a < \Re \Delta < \Re \Delta_{0}$, to be meromorphic or contain branch cuts in the region $\Re \Delta_{0} < \Re \Delta$, and to decay sufficiently fast as $\Delta \to +\infty$.

\textbf{Principal value.}
Notice that if naively applying the approximation \eqref{eq: complex Delta function approximation} to the case $\Re\Delta_{0}=a$, there would appear an extra factor $\half$.
The reason is as follows: in this case the leading pole $\Delta=\Delta_{0}$ lies on the contour, hence the integral is divergent and should be understood as the principal value. When deforming the contour to the right, it is necessary to consider the contribution of a small semicircle $C_{\Delta_{0},\delta} = \set{\Delta \given \Delta=\Delta_{0}+\delta e^{i\theta}, \frac{\pi}{2}<\theta<\frac{3\pi}{2}}$ surrounding the leading pole:
\begin{equation}
    \lim_{\delta\to 0}
    \inttt{\frac{d\Delta}{2\pi i}}{C_{\Delta_{0},\delta}}
    \gm{\Delta_{0}-\Delta}\varepsilon^{\Delta-\Delta_{0}}
    f(\Delta)
    =
    \lim_{\delta\to 0}
    f(\Delta_{0})
    \inttt{\frac{d\Delta}{2\pi i}}{C_{\Delta_{0},\delta}}
    \frac{\varepsilon^{\Delta-\Delta_{0}}}{\Delta_{0}-\Delta}
    =
    -\half f(\Delta_{0})
    \, .
\end{equation}

\section{EAdS, dS and conformal integrals}

We compute several integrals related to $\eads{d+1}$, $\ds{d+1}$ and the conformal boundary $\RR^{d}$.
We summarize the results in Appendix \ref{sec: conformal integrals - summary} and provide the derivations in the subsequent appendices.

The bulk momenta are parametrized similar to the $2d$ case \eqref{eq: momentum parametrization} as
\begin{equation}
    \label{eq: momentum parametrization higher dim}
    \begin{aligned}
        &
        q=\omega\hat{q}=\omega(1+ |x|^2,2 x,1-|x|^2)
        \, ,
        \TextInMath{for} \omega\geq 0
        \, ,
        \\
        &
        p=m\phat=\frac{m}{2y}(1+|x|^2+y^2,2x,1-|x|^2-y^2)
        \, ,
        \TextInMath{for} y>0
        \, ,
        \\
        &
        k=m \khat
        =
        \frac{m}{2y}(1+|x|^2-y^2,2x,1-|x|^2+y^2)
        \, ,
        \TextInMath{for} y\in\RR
        \, ,
    \end{aligned}
\end{equation}
where $x\in \RR^{d}$, and the mass-shell integrals are
\begin{equation}
    \label{eq: phase space integral higher dim}
    \begin{aligned}
        &
        \intt{[dq]}
        =
        \intt{d^{d}x}\intrange{2^d \omega^{d-1} d\omega}{0}{\oo}
        \, ,
        \\
        &
        \intt{[d\phat]}
        =
        \intt{d^{d}x}\intrange{y^{-d-1} dy}{0}{\oo}
        \, ,
        \\
        &
        \intt{[d\khat]}
        =
        \intt{d^{d}x}\intrange{|y|^{-d-1} dy}{-\oo}{+\oo}
        \, .
    \end{aligned}
\end{equation}

\subsection{Summary}
\label{sec: conformal integrals - summary}

\textbf{One-point.}
The one-point integrals are
\begin{align}
    \label{eq: EAdS one-point integral}
    &
    I_{p,p}
    =
    \int [d\hat{p}_0] (-\hat{p}_0 \co p_1)^{-\Delta}
    =
    \pi^{\frac{d+1}{2}}
    \mg{
        \frac{\Delta -d}{2}
    }{
        \frac{\Delta +1}{2}
    }
    m_1^{-\Delta}
    \, ,
    \\
    \label{eq: dS one-point integral kp}
    &
    I_{k,p}^{\tsign{\pm i}}
    =
    \int [d\hat{k}_0]
    (-\hat{k}_0 \co p_1)^{-\Delta}_{\tsign{\pm i}}
    =
    2 \pi^{\frac{d+3}{2}}
    \mg{
        \frac{\Delta -d}{2}
    }{
        \frac{1-d}{2},
        \frac{d+1}{2},
        \frac{\Delta +1}{2}
    }
    e^{\mp \half i \pi \Delta}
    m_1^{-\Delta}
    \, ,
    \\
    \label{eq: dS one-point integral kk}
    &
    I_{k,k}^{\tsign{\pm i}}
    =
    \int [d\hat{k}_0]
    (-\hat{k}_0 \co k_1)^{-\Delta}_{\tsign{\pm i}}
    =
    2 \pi^{\frac{d+3}{2}}
    \mg{
        \frac{\Delta -d}{2}
    }{
        \frac{d-\Delta +1}{2},
        \frac{\Delta +1}{2},
        \frac{-d+\Delta +1}{2}
    }
    e^{\mp \half i \pi \Delta}
    m_1^{-\Delta}
    \, ,
    \\
    \label{eq: cft one-point integral qp}
    &
    I_{q,p}
    =
    \intt{d^{d}x_0}
    (-\hat{q}_0 \co p_1 )^{-d}
    =
    2^{1-d} \pi^{\frac{d+1}{2}}
    \frac{1}{\gm{\frac{d+1}{2}}}
    m_1^{-d}
    \, ,
    \\
    \label{eq: cft one-point integral qk}
    &
    I_{q,k}^{\tsign{\pm i}}
    =
    \intt{d^{d}x_0}
    (-\hat{q}_0 \co k_1 )^{-d}_{\tsign{\pm i}}
    =
    2^{1-d}
    \pi^{\frac{d+1}{2}}
    \frac{1}{\gm{\frac{d+1}{2}}}
    e^{\mp \half i \pi d}
    m_1^{-d}
    \, .
\end{align}

\textbf{Two-point.}
The two-point integrals are
\begin{align}
    \label{eq: cft two-point integral qqp}
    &
    I_{q,qp}
    =
    \intt{d^{d}x_{0}}
    (-\hat{q}_0 \co \hat{q}_1)^{-\Delta}
    (-\hat{q}_0 \co p_2)^{-\wave{\Delta}}
    =
    2^{-\Delta}
    \pi^{d/2}
    \mg{
        \frac{d}{2}-\Delta
    }{
        d-\Delta
    }
    m_2^{\Delta -d}
    (-\hat{q}_1 \co \hat{p}_2)^{-\Delta}
    \, ,
    \\
    \label{eq: cft two-point integral qqk}
    &
    I_{q,qk}^{\tsign{\pm i}}
    =
    \intt{d^{d}x_{0}}
    (-\hat{q}_0 \co \hat{q}_1)^{-\Delta}
    (-\hat{q}_0 \co k_2)_{\tsign{\pm i}}^{-\wave{\Delta}}
    =
    2^{-\Delta}
    \pi^{d/2}
    \mg{
        \frac{d}{2}-\Delta
    }{
        d-\Delta
    }
    e^{\pm i \pi (\Delta-\frac{d}{2} )}
    m_2^{\Delta -d}
    (-\hat{q}_1 \co k_2)_{\tsign{\pm i}}^{-\Delta}
    \, .
\end{align}

\textbf{Three-point.}
The EAdS three-point integral is
\begin{equation}
    \label{eq: EAdS three-point integral}
    \begin{aligned}
        I_{\Delta_{1},\Delta_{2},\Delta_{3}}
        &=
        \int [d\hat{p}_0]
        (-\hat{p}_0 \co \hat{q}_1)^{-\Delta_1}
        (-\hat{p}_0 \co \hat{q}_2)^{-\Delta_2}
        (-\hat{p}_0 \co \hat{q}_3)^{-\Delta_3}
        \\
        &=
        \frac{1}{2} \pi^{\frac{d}{2}}
        \mg{
            \frac{\Delta_{12,3}}{2},
            \frac{\Delta_{13,2}}{2},
            \frac{\Delta_{23,1}}{2},
            \frac{\Delta_{123}-d}{2}
        }{
            \Delta_1,\Delta_2,\Delta_3
        }
        \vevv{\op_{\Delta_1}\op_{\Delta_2}\op_{\Delta_3}}
        \, .
    \end{aligned}
\end{equation}

The dS three-point integral is
\begin{equation}
    \label{eq: dS three-point integral}
    I^{\ts{1},\ts{2},\ts{3}}_{\Delta_{1},\Delta_{2},\Delta_{3}}
    =
    \int [d\hat{k}_0]
    (-\hat{k}_0 \co \hat{q}_1)_{\ts{1}}^{-\Delta_1}
    (-\hat{k}_0 \co \hat{q}_2)_{\ts{2}}^{-\Delta_2}
    (-\hat{k}_0 \co \hat{q}_3)_{\ts{3}}^{-\Delta_3}
    \, .
\end{equation}
For the imaginary type $\ts{n}=\pm i$, $n=1,2,3$, besides the permutation symmetry among $\set{\Delta_n,\ts{n}}$, the integral satisfies the parity symmetry by the change of variable $\khat_{0}\to -\khat_{0}$ and \eqref{eq: homogeneous distribution - parity symmetry},
\begin{equation}
    I^{\ts{1},\ts{2},\ts{3}}_{\Delta_{1},\Delta_{2},\Delta_{3}}
    =
    I^{\tsign{-s_1},\tsign{-s_2},\tsign{-s_3}}_{\Delta_{1},\Delta_{2},\Delta_{3}}
    e^{-\ts{1}\Delta_{1}-\ts{2}\Delta_{2}-\ts{3}\Delta_{3}}
    \, .
\end{equation}
Hence only two configurations of $\ts{n}$ are independent, and are related to the EAdS one \eqref{eq: EAdS three-point integral} as
\begin{equation}
    \label{eq: dS three-point integral - imaginary basis}
    \begin{aligned}
        &
        I^{\tsign{+i},\tsign{+i},\tsign{+i}}_{\Delta_{1},\Delta_{2},\Delta_{3}}
        =
        2
        e^{-\frac{1}{2} i \pi  \Delta_{123}}
        \cos\LRa{\frac{\pi}{2} d}
        I_{\Delta_{1},\Delta_{2},\Delta_{3}}
        \, ,
        \\
        &
        I^{\tsign{+i},\tsign{+i},\tsign{-i}}_{\Delta_{1},\Delta_{2},\Delta_{3}}
        =
        2
        e^{-\frac{1}{2} i \pi  \Delta_{12,3}}
        \cos\LRa{\frac{\pi }{2}(d-2\Delta_{3})}
        I_{\Delta_{1},\Delta_{2},\Delta_{3}}
        \, .
    \end{aligned}
\end{equation}
Later we will need this integral for $\ts{n}=\pm$ in even dimensions, which can be obtained by the change of basis \eqref{eq: tachyon basis change}.
For $d=0\mod 2$,
\begin{equation}
    \label{eq: dS three-point integral - plus/minus basis 2d}
    \begin{aligned}
        &
        I^{\tsign{+},\tsign{+},\tsign{+}}_{\Delta_{1},\Delta_{2},\Delta_{3}}
        =
        -\frac{1}{2} (-1)^{\frac{d}{2}} \pi^{\frac{d}{2}}
        \mg{
            1-\Delta_1,
            1-\Delta_2,
            1-\Delta_3,
            \frac{\Delta_{123}-d}{2},
        }{
            \frac{\Delta_{1,23}+2}{2},
            \frac{\Delta_{2,13}+2}{2},
            \frac{\Delta_{3,12}+2}{2}
        }
        \vevv{\op_{\Delta_1}\op_{\Delta_2}\op_{\Delta_3}}
        \, ,
        \\
        &
        I^{\tsign{+},\tsign{+},\tsign{-}}_{\Delta_{1},\Delta_{2},\Delta_{3}}
        =
        \frac{1}{2} \pi^{\frac{d}{2}}
        \mg{
            1-\Delta_1,
            1-\Delta_2,
            1-\Delta_3,
            \frac{\Delta_{12,3}}{2}
        }{
            \frac{\Delta_{1,23}+2}{2},
            \frac{\Delta_{2,13}+2}{2},
            \frac{-\Delta_{123}+d+2}{2},
        }
        \vevv{\op_{\Delta_1}\op_{\Delta_2}\op_{\Delta_3}}
        \, .
    \end{aligned}
\end{equation}
The other configurations are related by the permutation symmetry and the parity symmetry,
\begin{equation}
    I^{\ts{1},\ts{2},\ts{3}}_{\Delta_{1},\Delta_{2},\Delta_{3}}
    =
    I^{\tsign{-s_1},\tsign{-s_2},\tsign{-s_3}}_{\Delta_{1},\Delta_{2},\Delta_{3}}
    \, ,
\end{equation}

\subsection{One-point integrals}

We start from the EAdS one-point integral $I_{p,p}$ in \eqref{eq: EAdS one-point integral}. By the symmetry and scaling behavior, $I_{p,p}$ is proportional to $m_1^{-\Delta}=(-p_1^2)^{-\Delta/2}$.
To fix the coefficient, we set $x_{1}=0$, $y_{1}=1$ and perform the integrals in the radial coordinates $r=\abs{x}$,
\begin{align}
    I_{p,p}
    &=
    2^{\Delta} m_1^{-\Delta}
    \intt{d^{d}x_0}
    \intrange{dy_{0}}{0}{\oo}
    y_0^{\Delta-d-1}
    (|x_0|^2+y_0^2+1)^{-\Delta}
    \\
    &=
    2^{\Delta+d-1}
    \pi^{\frac{d-1}{2}}
    m_1^{-\Delta}
    \mg{\frac{d-1}{2}}{d-1}
    \intrange{dr}{0}{\oo}
    \intrange{dy_{0}}{0}{\oo}
    r^{d-1}
    y_0^{\Delta-d-1}
    (r^2+y_0^2+1)^{-\Delta}
    \, ,
\end{align}
which leads to the result \eqref{eq: EAdS one-point integral}.

Similar to the EAdS case, the dS one-point integral $I_{k,p}$ in \eqref{eq: dS one-point integral kp} reduces to
\begin{equation}
    I_{k,p}^{\tsign{\pm i}}=
    \intt{d^{d}x_{0}}\intrange{dy_{0}}{-\oo}{+\oo}
    |y_{0}|^{-d-1}
    \mleft(\frac{
            m_1
            (|x_0|^2-y_0^2+1)
        }{2 y_0}\pm i \varepsilon
    \mright)^{-\Delta}
    \, .
\end{equation}
Then we separate the integral into two regions $y_{0}>0$ and $y_{0}<0$, change the variable $y_{0}\to -y_{0}$ in the second region, and use the Mellin-Barnes representation \eqref{eq: Mellin-Barnes relation} to rewrite the integral as
\begin{equation}
    \begin{aligned}
        I_{k,p}^{\tsign{\pm i}}
        &=
        2^{\Delta+2} \pi^{\frac{d}{2}+1}
        (-m_{1}^2\pm i \varepsilon)^{-\frac{\Delta}{2}}
        \intrange{dr}{0}{\oo}
        \intrange{dy_{0}}{0}{\oo}
        \intt{\frac{dt}{2\pi i}}
        r^{d-1}
        (r^2+1)^{-\Delta-t}
        y_0^{\Delta-d+2 t-1}
        \\
        &\peq\xx
        \mg{-t,t+\Delta}{
            \Delta,\frac{d}{2},
            \pm\frac{1}{2}(\Delta+2 t)+\frac{1}{2},
            \mp\frac{1}{2}(\Delta+2 t)+\frac{1}{2}
        }
        \, .
    \end{aligned}
\end{equation}
The $y_{0}$- and $t$-integrals can be performed by the complex delta function \eqref{eq: complex delta function}, and the $r$-integral is a Beta function.
The rest integrals $I_{k,k}$ in \eqref{eq: dS one-point integral kk}, $I_{q,p}$ in \eqref{eq: cft one-point integral qp} and $I_{q,k}$ in \eqref{eq: cft one-point integral qk} can be computed similarly.

\subsection{Two-point integrals}

For the conformal two-point integral $I_{q,qp}$ in \eqref{eq: cft two-point integral qqp}, we use the Feynman-Schwinger parameterization \eqref{eq: Feynman-Schwinger parameterization} to rewrite the integral as
\begin{equation}
    I_{q,qp}
    =
    \mg{
        d
    }{
        d-\Delta ,
        \Delta
    }
    \intt{d^{d}x_{0}}
    \intrange{d\alpha}{0}{\oo}
    \alpha^{d-\Delta -1}
    (-\hat{q}_0 \co (\alpha  p_2+\hat{q}_1))^{-d}
    \, .
\end{equation}
Here the $\hat{q}_0$-integral can be performed by the one-point integral \eqref{eq: cft one-point integral qp} since $\alpha  p_2+\hat{q}_1$ is timelike, then the $\alpha$-integral is a Beta function.

The integral $I_{q,qk}$ in \eqref{eq: cft two-point integral qqk} needs more careful treatment after the Feynman-Schwinger parameterization,
\begin{equation}
    I_{q,qk}^{\tsign{\pm i}}
    =
    \mg{
        d
    }{
        d-\Delta ,
        \Delta
    }
    \intt{d^{d}x_{0}}
    \intrange{d\alpha}{0}{\oo}
    \alpha^{\Delta -1}
    (\hat{q}_0 \co (-k_2-\alpha  \hat{q}_1) \pm i \varepsilon)^{-d}
    \, .
\end{equation}
If $\hat{q}_1 \co \hat{k}_2>0$, then $-k_2-\alpha  \hat{q}_1$ is spacelike for all $\alpha>0$ and the $x_{0}$-integral can be preformed by the one-point integral \eqref{eq: cft one-point integral qk}, leading to
\begin{equation}
    I_{>0}^{\tsign{\pm i}}
    =
    \pi^{d/2} 2^{-\Delta}
    \mg{
        \frac{d}{2}-\Delta
    }{
        d-\Delta
    }
    e^{\mp\frac{1}{2} i \pi  d }
    m_2^{\Delta -d}
    (\hat{q}_1 \co \hat{k}_2)^{-\Delta}
    \, .
\end{equation}

If $\hat{q}_1 \co \hat{k}_2<0$, there are two cases: (1) $-k_2-\alpha  \hat{q}_1$ is spacelike for $0<\alpha<-\frac{m_2}{2 \hat{q}_1 \co \hat{k}_2}$, and we perform the $x_{0}$-integral by \eqref{eq: cft one-point integral qk} as before; (2) $-k_2-\alpha  \hat{q}_1$ is timelike for $\alpha>-\frac{m_2}{2 \hat{q}_1 \co \hat{k}_2}$, and we need to use \eqref{eq: cft one-point integral qp} to perform the $x_{0}$-integral.
After performing the $\alpha$-integrals in both cases, we obtain
\begin{equation}
    I_{<0}^{\tsign{\pm i}}
    =
    2^{-\Delta}
    \pi^{d/2}
    m_2^{\Delta -d}
    (-\hat{q}_1 \co \hat{k}_2)^{-\Delta}
    \LR{
        e^{\mp\frac{1}{2} i \pi  d}
        \mg{
            1-\frac{d}{2},
            \frac{d}{2}
        }{
            d-\Delta ,
            -\frac{d}{2}+\Delta +1
        }
        +
        \mg{
            1-\frac{d}{2},
            \frac{d}{2},
            \frac{d}{2}-\Delta
        }{
            1-\Delta ,
            d-\Delta ,
            \Delta
        }
    }
    \, .
\end{equation}

Summing the two parts together, we arrive at
\begin{equation}
    I_{q,qk}^{\tsign{\pm i}}
    =
    I_{>0}^{\tsign{\pm i}}\theta(\hat{q}_1 \co \hat{k}_2)
    +
    I_{<0}^{\tsign{\pm i}}\theta(-\hat{q}_1 \co \hat{k}_2)
    \, .
\end{equation}
Then we expand the distributions $(\pm\hat{q}_1 \co \hat{k}_2)^{-\Delta}\theta(\pm\hat{q}_1 \co \hat{k}_2)=(-\hat{q}_1 \co \hat{k}_2)_{\tsign{\mp}}^{-\Delta}$ back into $(-\hat{q}_1 \co \hat{k}_2)_{\tsign{\pm i}}^{-\Delta}$ by \eqref{eq: homogeneous distribution - imaginary basis}, yielding the result \eqref{eq: cft two-point integral qqk}.

\subsection{Three-point integrals}

We first revisit the EAdS three-point integral \eqref{eq: EAdS three-point integral}. Using the Feynman-Schwinger parameterization \eqref{eq: Feynman-Schwinger parameterization}, we rewrite the integral as
\begin{equation}
    I_{\Delta_{1},\Delta_{2},\Delta_{3}}
    =
    \mg{
        \Delta_{123}
    }{
        \Delta_1,
        \Delta_2,
        \Delta_3
    }
    \intrange{d\alpha}{0}{\oo}
    \intrange{d\beta}{0}{\oo}
    \intt{[d\hat{p}_0]}
    \alpha^{\Delta_2-1} \beta^{\Delta_3-1} (-\hat{p}_0 \co (\hat{q}_1+\alpha  \hat{q}_2+\beta  \hat{q}_3))^{-\Delta_{123}}
    \, .
\end{equation}
The $\hat{p}_0$-integral can be performed by the EAdS one-point integral \eqref{eq: EAdS one-point integral},
\begin{equation}
    I_{\Delta_{1},\Delta_{2},\Delta_{3}}
    =
    2^{\frac{\Delta_{123}-2}{2}}
    \pi^{\frac{d}{2}}
    \mg{
        \frac{\Delta_{123}}{2},
        \frac{\Delta_{123}-d}{2}
    }{
        \Delta_1,
        \Delta_2,
        \Delta_3
    }
    \intrange{d\alpha}{0}{\oo}
    \intrange{d\beta}{0}{\oo}
    \alpha^{\Delta_2-1} \beta^{\Delta_3-1}
    (-\alpha  \beta  \hat{q}_{23}-\alpha  \hat{q}_{12}-\beta  \hat{q}_{13})^{-\frac{\Delta_{123}}{2}}
    \, ,
\end{equation}
then the $\alpha$- and $\beta$-integrals are Beta functions, leading to the result \eqref{eq: EAdS three-point integral}.

For the dS three-point integral \eqref{eq: dS three-point integral} with imaginary type $\ts{n}=\pm i$, as discussed before, it suffices to compute the two independent configurations $(+i,+i,+i)$ and $(+i,+i,-i)$.
To utilize the Feynman-Schwinger parameterization \eqref{eq: Feynman-Schwinger parameterization}, we need to rewrite the integrand by \eqref{eq: homogeneous distribution - parity symmetry} as
\begin{align}
    \nn
    I^{\ts{1},\ts{2},\ts{3}}_{\Delta_{1},\Delta_{2},\Delta_{3}}
    &=
    e^{
        -\frac{1}{2} \pi \sum_{n=1}^{3}\Delta_n (\ts{n}-i)
    }
    \intt{[d\hat{k}_0]}
    (i \varepsilon +i \ts{1} \hat{k}_0 \co \hat{q}_1)^{-\Delta_1}
    (i \varepsilon +i \ts{2} \hat{k}_0 \co \hat{q}_2)^{-\Delta_2}
    (i \varepsilon +i \ts{3} \hat{k}_0 \co \hat{q}_3)^{-\Delta_3}
    \, ,
    \\
    &=
    e^{
        -\frac{1}{2} \pi \sum_{n=1}^{3}\Delta_n (\ts{n}-i)
    }
    \mg{
        \Delta_{123}
    }{
        \Delta_1,
        \Delta_2,
        \Delta_3
    }
    \intt{[d\hat{k}_0]}
    \intrange{d\alpha}{0}{\oo}
    \intrange{d\beta}{0}{\oo}
    \alpha^{\Delta_2-1} \beta^{\Delta_3-1}
    \\
    \nn
    &\peq\xx
    (i \varepsilon +\hat{k}_0 \co (i \ts{1} \hat{q}_1+i \alpha  \ts{2} \hat{q}_2+i \beta  \ts{3} \hat{q}_3))^{-\Delta_{123}}
    \, ,
\end{align}
where we have shrinked the infinitesimal parameter $i \varepsilon  (\alpha +\beta +1)$ to $i \varepsilon$.

For the $(+i,+i,+i)$ configuration, $\hat{q}_1+\alpha \hat{q}_2+\beta \hat{q}_3$ is timelike, and the $\hat{k}_0$-integral can be performed by the dS one-point integral \eqref{eq: dS one-point integral kp}, then the $\alpha$- and $\beta$-integrals are Beta functions as well, leading to the first line of \eqref{eq: dS three-point integral - imaginary basis}.

For the $(+i,+i,-i)$ configuration, $\hat{q}_1+\alpha \hat{q}_2-\beta \hat{q}_3$ can be either timelike or spacelike, and we separate the $\alpha$- and $\beta$-integrals into two regions accordingly,
\begin{equation}
    \begin{aligned}
        &R_{p}=\set{(\alpha,\beta)\given \alpha>0,0<\beta <\frac{\alpha  \hat{q}_{12}}{\alpha  \hat{q}_{23}+\hat{q}_{13}}}
        \, ,
        \\
        &R_{k}=\set{(\alpha,\beta)\given \alpha>0,\beta >\frac{\alpha  \hat{q}_{12}}{\alpha  \hat{q}_{23}+\hat{q}_{13}}}
        \, .
    \end{aligned}
\end{equation}
In $R_{p}$, the $\hat{k}_0$-integral is performed by the dS one-point integral \eqref{eq: dS one-point integral kp}, while in $R_{k}$, it is performed by \eqref{eq: dS one-point integral kk}, leading to
\begin{equation}
    \begin{aligned}
        \peq
        I^{\tsign{+i},\tsign{+i},\tsign{-i}}_{\Delta_{1},\Delta_{2},\Delta_{3}}
        &
        =
        2^{\frac{\Delta_{123}}{2}}
        \pi^{\frac{d}{2}}
        \mg{
            \frac{\Delta_{123}}{2},
            \frac{\Delta_{123}-d}{2}
        }{
            \Delta_1,
            \Delta_2,
            \Delta_3
        }
        \Biggl(
            \cos\LRa{\frac{\pi}{2} d}
            \inttt{d\alpha d\beta}{R_{p}}
            \alpha^{\Delta_2-1} \beta^{\Delta_3-1}
            (\alpha  \beta  \hat{q}_{23}-\alpha  \hat{q}_{12}+\beta  \hat{q}_{13})^{-\frac{\Delta_{123}}{2}}
            \\
            &\peq
            +
            \cos\LRa{\frac{\pi}{2} (\Delta_{123}-d)}
            \inttt{d\alpha d\beta}{R_{k}}
            \alpha^{\Delta_2-1} \beta^{\Delta_3-1}
            (-\alpha  \beta  \hat{q}_{23}+\alpha  \hat{q}_{12}-\beta  \hat{q}_{13})^{-\frac{\Delta_{123}}{2}}
        \Biggr)
        \, .
    \end{aligned}
\end{equation}
Then performing the $\beta$- and $\alpha$-integrals successively, we obtain the second line of \eqref{eq: dS three-point integral - imaginary basis}.

\section{Three-point celestial amplitudes}
\label{app: three-point celestial amplitudes}

We compute several three-point celestial amplitudes of scalars, gluons, and gravitons.
For convenience, we use the following abbreviations to denote the types of conformal bases involved in the celestial amplitudes:\\
\centerline{
    \texttt{O} for massless Mellin, \quad
    \texttt{\~{O}} for massless shadow, \quad
    \texttt{M} for massive, \quad
    \texttt{T} for tachyonic.
}
Then for example, \oom denotes the celestial amplitude with two outgoing massless particles and one incoming massive particle.

We summarize the results in Appendix \ref{app: three-point celestial amplitudes - summary} and provide detailed derivations in the subsequent appendices.
The \oom scalar celestial amplitude was computed in \cite{Lam:2017ofc,Liu:2024lbs}, \mom in \cite{Liu:2024lbs}, and \too in \cite{Chang:2023ttm}. We list them here with the new normalization of the conformal basis in Section \ref{sec: conformal basis}.
Then we study the massless limit of three-point celestial amplitudes involving massive and tachyonic scalars in Appendices \ref{app: three-point celestial amplitudes - massless limit of scalars}.

\subsection{Summary}
\label{app: three-point celestial amplitudes - summary}

\subsubsection{Scalar}

\begin{itemize}
    \item
        \oom and \Sout{}\Oout{}\Minn (see \cite{Lam:2017ofc,Liu:2024lbs}):
        \begin{align}
            \label{eq: 3pt coefficient OOM scalar}
            \coThree{
                \scalar^{\out}_{\Delta_{1}}
                \scalar^{\out}_{\Delta_{2}}
                \scalar^{\inn,m_{3}}_{\Delta_{3}}
            }
            &=
            2^{\Delta_{3,12}-2}
            m_3^{\Delta_{12,3}-2}
            \mg{
                \frac{\Delta_{13,2}}{2},
                \frac{\Delta_{23,1}}{2}
            }{
                \Delta_3
            }
            \, ,
            \\[0.5em]
            \label{eq: 3pt coefficient S0M scalar}
            \coThree{
                \wave\scalar^{\out}_{\Delta_{1}}
                \scalar^{\out}_{\Delta_{2}}
                \scalar^{\inn,m_{3}}_{\Delta_{3}}
            }
            &=
            \pi  2^{\Delta_{13,2}-4} m_3^{\Delta_{2,13}}
            \mg{
                1-\Delta_1,
                \frac{\Delta_{12,3}}{2},
                \frac{\Delta_{123}-2}{2},
                \frac{\Delta_{13,2}}{2}
            }{
                \Delta_1,
                \Delta_3,
                \frac{\Delta_{2,13}+2}{2}
            }
            \, .
        \end{align}
        The three-point coefficient in shadow basis \eqref{eq: 3pt coefficient S0M scalar} is obtained from that in Mellin basis \eqref{eq: 3pt coefficient OOM scalar} by the massless shadow relation \eqref{eq: massless basis shadow relation} and the star-triangle relation \eqref{eq: star-triangle relation}.
    \item
        \too and \Tall{}\Oinn{}\Sout (see \cite{Chang:2023ttm} and Appendix \ref{app: TOO scalar}):
        \begin{align}
            \label{eq: 3pt coefficient TOO scalar}
            \coThree{
                \scalar^{\tsign{+},im_{1}}_{\Delta_{1}}
                \scalar^{\inn}_{\Delta_{2}}
                \scalar^{\out}_{\Delta_{3}}
            }
            &=
            2^{\Delta_{1,23}-2}
            m_1^{\Delta_{23,1}-2}
            \mg{
                1-\Delta_1,
                \frac{\Delta_{13,2}}{2}
            }{
                \frac{\Delta_{3,12}+2}{2}
            }
            \, ,
            \\[0.5em]
            \coThree{
                \scalar^{\tsign{-},im_{1}}_{\Delta_{1}}
                \scalar^{\inn}_{\Delta_{2}}
                \scalar^{\out}_{\Delta_{3}}
            }
            &=
            \coThree{
                \scalar^{\tsign{+},im_{1}}_{\Delta_{1}}
                \scalar^{\inn}_{\Delta_{3}}
                \scalar^{\out}_{\Delta_{2}}
            }
            \, ,
            \\[0.5em]
            \label{eq: 3pt coefficient T0S scalar}
            \coThree{
                \scalar^{\tsign{+},im_{1}}_{\Delta_{1}}
                \scalar^{\inn}_{\Delta_{2}}
                \wave\scalar^{\out}_{\Delta_{3}}
            }
            &=
            \pi  2^{\Delta_{13,2}-4} m_1^{\Delta_{2,13}}
            \mg{
                1-\Delta_1,
                1-\Delta_3,
                \frac{\Delta_{13,2}}{2},
                \frac{\Delta_{23,1}}{2}
            }{
                \Delta_3,
                \frac{4-\Delta_{123}}{2},
                \frac{\Delta_{2,13}+2}{2}
            }
            \, ,
            \\[0.5em]
            \coThree{
                \scalar^{\tsign{-},im_{1}}_{\Delta_{1}}
                \scalar^{\inn}_{\Delta_{2}}
                \wave\scalar^{\out}_{\Delta_{3}}
            }
            &=
            \pi  2^{\Delta_{13,2}-4} m_1^{\Delta_{2,13}}
            \mg{
                1-\Delta_1,
                1-\Delta_3,
                \frac{\Delta_{123}-2}{2},
                \frac{\Delta_{13,2}}{2}
            }{
                \Delta_3,
                \frac{\Delta_{1,23}+2}{2},
                \frac{\Delta_{2,13}+2}{2}
            }
            \, .
        \end{align}
    \item
        \mom (see \cite{Liu:2024lbs}):
        \begin{align}
            \label{eq: 3pt coefficient M0M scalar}
            \coThree{
                \scalar^{\out,m_{1}}_{\Delta_{1}}
                \scalar^{\out}_{\Delta_{2}}
                \scalar^{\inn,m_{3}}_{\Delta_{3}}
            }
            &=
            \pi  2^{\Delta_{13,2}-4}
            m_1^{2-\Delta_{123}}
            (m_3^2-m_1^2)^{\Delta_2-1}
            \theta(m_3-m_1)
            \\
            &\peq\xx
            \mg{
                \frac{\Delta_{12,3}}{2},
                \frac{\Delta_{13,2}}{2},
                \frac{\Delta_{23,1}}{2},
                \frac{\Delta_{123}-2}{2}
            }{
                \Delta_1,
                \Delta_2,
                \Delta_3
            }
            \Fpq{2}{1}{\frac{\Delta_{123}-2}{2},\frac{\Delta_{23,1}}{2}}{\Delta_2}{\frac{
                    m_1^2-m_3^2
            }{m_1^2}}
            \, .
            \nn
        \end{align}
    \item
        \tom (see Appendix \ref{app: TOM scalar}):
        \begin{align}
            \label{eq: 3pt coefficient TOM scalar}
            \coThree{
                \scalar^{\tsign{+},im_{1}}_{\Delta_{1}}
                \scalar^{\inn}_{\Delta_{2}}
                \scalar^{\out,m_{3}}_{\Delta_{3}}
            }
            &=
            \frac{\pi 2^{\Delta_{13,2}-4}}{1-\Delta_{1}}
            m_3^{\Delta_{1,23}}
            m_1^{2-2 \Delta_1}
            (m_1^2+m_3^2)^{\Delta_2-1}
            \\
            &\peq\xx
            \mg{
                \frac{\Delta_{13,2}}{2},
                \frac{\Delta_{23,1}}{2}
            }{
                \Delta_3
            }
            \Fpq{2}{1}{\frac{\Delta_{2,13}+2}{2},\frac{\Delta_{23,1}}{2}}{2-\Delta_1}{-\frac{m_1^2}{m_3^2}}
            \, ,
            \nn
            \\[0.5em]
            \coThree{
                \scalar^{\tsign{-},im_{1}}_{\Delta_{1}}
                \scalar^{\inn}_{\Delta_{2}}
                \scalar^{\out,m_{3}}_{\Delta_{3}}
            }
            &=
            \pi  2^{\Delta_{13,2}-4}
            m_3^{2-\Delta_{123}}
            (m_1^2+m_3^2)^{\Delta_2-1}
            \\
            &\peq\xx
            \mg{
                1-\Delta_1,
                \frac{\Delta_{12,3}}{2},
                \frac{\Delta_{123}-2}{2},
                \frac{\Delta_{13,2}}{2}
            }{
                \Delta_1,
                \Delta_3,
                \frac{\Delta_{2,13}+2}{2}
            }
            \Fpq{2}{1}{\frac{\Delta_{12,3}}{2},\frac{\Delta_{123}-2}{2}}{\Delta_1}{-\frac{m_1^2}{m_3^2}}
            \, .
            \nn
        \end{align}
    \item
        \tot (see Appendix \ref{app: TOT scalar}):
        \begin{align}
            \label{eq: 3pt coefficient TOT scalar}
            \coThree{
                \scalar^{\tsign{+},im_{1}}_{\Delta_{1}}
                \scalar^{\inn}_{\Delta_{2}}
                \scalar^{\tsign{+},im_{3}}_{\Delta_{3}}
            }
            &=
            (1\leftrightarrow 3)
            +
            \pi  2^{\Delta_{13,2}-4}
            (m_1^2-m_3^2)^{\frac{\Delta_{2,13}}{2}}
            \theta (m_{1}-m_{3})
            \\
            &\peq\xx
            \mg{
                1-\Delta_1,
                1-\Delta_3,
                \frac{\Delta_{13,2}}{2},
                \frac{\Delta_{23,1}}{2}
            }{
                \Delta_3,
                \frac{4-\Delta_{123}}{2},
                \frac{\Delta_{2,13}+2}{2}
            }
            \Fpq{2}{1}{\frac{\Delta_{123}-2}{2},\frac{\Delta_{13,2}}{2}}{\Delta_3}{\frac{m_3^2}{
                    m_3^2-m_1^2
            }}
            \, ,
            \nn
            \\[0.5em]
            \coThree{
                \scalar^{\tsign{+},im_{1}}_{\Delta_{1}}
                \scalar^{\inn}_{\Delta_{2}}
                \scalar^{\tsign{-},im_{3}}_{\Delta_{3}}
            }
            &=
            (1\leftrightarrow 3)
            +
            \frac{\pi  2^{\Delta_{13,2}-4}}{1-\Delta_3}
            m_3^{2-2 \Delta_3}
            (m_1^2-m_3^2)^{\frac{\Delta_{23,1}-2}{2}}
            \theta (m_{1}-m_{3})
            \\
            &\peq\xx
            \mg{
                1-\Delta_1,
                \frac{\Delta_{13,2}}{2}
            }{
                \frac{\Delta_{3,12}+2}{2}
            }
            \Fpq{2}{1}{\frac{\Delta_{1,23}+2}{2},\frac{\Delta_{12,3}}{2}}{2-\Delta_3}{\frac{m_3^2}{
                    m_3^2-m_1^2
            }}
            \, ,
            \nn
            \\[0.5em]
            \coThree{
                \scalar^{\tsign{-},im_{1}}_{\Delta_{1}}
                \scalar^{\inn}_{\Delta_{2}}
                \scalar^{\tsign{+},im_{3}}_{\Delta_{3}}
            }
            &=
            (1\leftrightarrow 3)
            +
            \pi  2^{\Delta_{13,2}-4}
            (m_1^2-m_3^2)^{\frac{\Delta_{2,13}}{2}}
            \theta (m_{1}-m_{3})
            \\
            &\peq\xx
            \mg{
                1-\Delta_1,
                1-\Delta_3,
                \frac{\Delta_{123}-2}{2},
                \frac{\Delta_{13,2}}{2}
            }{
                \Delta_3,
                \frac{\Delta_{1,23}+2}{2},
                \frac{\Delta_{2,13}+2}{2}
            }
            \Fpq{2}{1}{\frac{\Delta_{123}-2}{2},\frac{\Delta_{13,2}}{2}}{\Delta_3}{\frac{m_3^2}{
                    m_3^2-m_1^2
            }}
            \, ,
            \nn
            \\[0.5em]
            \coThree{
                \scalar^{\tsign{-},im_{1}}_{\Delta_{1}}
                \scalar^{\inn}_{\Delta_{2}}
                \scalar^{\tsign{-},im_{3}}_{\Delta_{3}}
            }
            &=
            (1\leftrightarrow 3)
            +
            \frac{\pi  2^{\Delta_{13,2}-4}}{1-\Delta_3}
            m_3^{2-2 \Delta_3}
            (m_1^2-m_3^2)^{\frac{\Delta_{23,1}-2}{2}}
            \theta (m_{1}-m_{3})
            \\
            &\peq\xx
            \mg{
                1-\Delta_1,
                \frac{\Delta_{12,3}}{2}
            }{
                \frac{\Delta_{2,13}+2}{2}
            }
            \Fpq{2}{1}{\frac{\Delta_{1,23}+2}{2},\frac{\Delta_{12,3}}{2}}{2-\Delta_3}{\frac{m_3^2}{
                    m_3^2-m_1^2
            }}
            \, .
            \nn
        \end{align}
        Notice that for each of the three-point celestial amplitudes, there are two terms symmetric under the permutation $(13)$.

\end{itemize}

\subsubsection{Gluon}

\begin{itemize}
    \item
        \oom (see Section \ref{sec: three-point spinning}):
        \begin{equation}
            \label{eq: 3pt coefficient OOM gluon}
            \coThree{
                \gluon^{\out,a_{1}}_{\Delta_1,J_1}
                \gluon^{\out,a_{2}}_{\Delta_2,J_2}
                \gluon^{\inn,m_{3},a_{3}}_{\Delta_3,J_3}
            }
            =
            f^{a_{1}a_{2}a_{3}}
            \frac{-2^{\Delta_{3,12}-2}m_3^{\Delta_{12,3}-1}}{(\Delta_1-1) (\Delta_2-1)}
            \mg{
                \frac{| J_{13,2}| +\Delta_{13,2}}{2}
                ,
                \frac{| J_{23,1}| +\Delta_{23,1}}{2}
            }{
                \Delta_3+1
            }
            \cC^{\gluon}_{J_{1},J_{2},J_{3}}
            \, .
        \end{equation}
        Here the reduced coefficient $\cC^{\gluon}_{J_{1},J_{2},J_{3}}$ satisfies the parity symmetry
        \begin{equation}
            \cC^{\gluon}_{J_{1},J_{2},J_{3}} = \cC^{\gluon}_{-J_{1},-J_{2},-J_{3}}
            \, ,
        \end{equation}
        and depends on the spin configurations as follows:
        \begin{equation}
            \label{eq: 3pt subcoefficient OOM gluon}
            \begin{aligned}
                &
                \cC^{\gluon}_{1,1,1}
                =
                (\Delta_{123}-3) (\Delta_{123}-1)
                \, ,
                \\
                &
                \cC^{\gluon}_{1,-1,1}
                =
                2 (\Delta_{13,2}-1)
                \, ,
                \\
                &
                \cC^{\gluon}_{-1,1,1}
                =
                2 (\Delta_{23,1}-1)
                \, ,
                \\
                &
                \cC^{\gluon}_{-1,-1,1}
                =
                (\Delta_{12,3}-1) (\Delta_{12,3}+1)
                \, .
            \end{aligned}
        \end{equation}
    \item
        \too (see Section \ref{sec: three-point spinning}):
        \begin{align}
            \label{eq: 3pt coefficient TOO gluon}
            &
            \coThree{
                \gluon^{\tsign{+},im_{1},a_{1}}_{\Delta_1,J_1}
                \gluon^{\inn,a_{2}}_{\Delta_2,J_2}
                \gluon^{\out,a_{3}}_{\Delta_3,J_3}
            }
            =
            f^{a_{1}a_{2}a_{3}}
            \frac{2^{\Delta_{1,23}-2}
                m_1^{\Delta_{23,1}-1}
            }{(\Delta_2-1) (\Delta_3-1)}
            \mg{
                -\Delta_1,
                \frac{| J_{13,2}| +\Delta_{13,2}}{2}
            }{
                \frac{-| J_{12,3}| +\Delta_{3,12}+2}{2}
            }
            \cC^{\gluon,\tsign{+}}_{J_{1},J_{2},J_{3}}
            \, ,
            \\[0.5em]
            &
            \coThree{
                \gluon^{\tsign{-},im_{1},a_{1}}_{\Delta_1,J_1}
                \gluon^{\inn,a_{2}}_{\Delta_2,J_2}
                \gluon^{\out,a_{3}}_{\Delta_3,J_3}
            }
            =
            \coThree{
                \gluon^{\tsign{+},im_{1},a_{1}}_{\Delta_1,J_1}
                \gluon^{\inn,a_{3}}_{\Delta_3,J_3}
                \gluon^{\out,a_{2}}_{\Delta_2,J_2}
            }
            \, .
        \end{align}
        Here the reduced coefficient $\cC^{\gluon,\tsign{+}}_{J_{1},J_{2},J_{3}}$ satisfies the parity symmetry
        \begin{equation}
            \cC^{\gluon,\tsign{+}}_{J_{1},J_{2},J_{3}} = \cC^{\gluon,\tsign{+}}_{-J_{1},-J_{2},-J_{3}}
            \, ,
        \end{equation}
        and depends on the spin configurations as follows:
        \begin{equation}
            \label{eq: 3pt subcoefficient TOO gluon}
            \begin{aligned}
                &
                \cC^{\gluon,\tsign{+}}_{1,1,1}
                =
                -(\Delta_{123}-3) (\Delta_{123}-1)
                \, ,
                \\
                &
                \cC^{\gluon,\tsign{+}}_{1,1,-1}
                =
                2 (\Delta_{12,3}-1)
                \, ,
                \\
                &
                \cC^{\gluon,\tsign{+}}_{1,-1,1}
                =
                -2 (\Delta_{13,2}-1)
                \, ,
                \\
                &
                \cC^{\gluon,\tsign{+}}_{1,-1,-1}
                =
                -(\Delta_{23,1}-1) (\Delta_{23,1}+1)
                \, .
            \end{aligned}
        \end{equation}

\end{itemize}

\subsubsection{Graviton}

\begin{itemize}
    \item
        \oom (see Section \ref{sec: three-point spinning}):
        \begin{equation}
            \label{eq: 3pt coefficient OOM graviton}
            \coThree{
                \graviton^{\out}_{\Delta_1,J_1}
                \graviton^{\out}_{\Delta_2,J_2}
                \graviton^{\inn,m_{3}}_{\Delta_3,J_3}
            }
            =
            \frac{2^{\Delta_{3,12}-2} m_3^{\Delta_{12,3}}}{\Delta_1 \Delta_2 (\Delta_1-1) (\Delta_2-1)}
            \mg{
                \frac{| J_{13,2}| +\Delta_{13,2}}{2}
                ,
                \frac{| J_{23,1}| +\Delta_{23,1}}{2}
            }{
                \Delta_3+2
            }
            \cC^{\graviton}_{J_{1},J_{2},J_{3}}
            \, .
        \end{equation}
        Here the reduced coefficient $\cC^{\graviton}_{J_{1},J_{2},J_{3}}$ satisfies the parity symmetry
        \begin{equation}
            \cC^{\graviton}_{J_{1},J_{2},J_{3}} = \cC^{\graviton}_{-J_{1},-J_{2},-J_{3}}
            \, ,
        \end{equation}
        and depends on the spin configurations as follows:
        \begin{align*}
            \nnn
            \label{eq: 3pt subcoefficient OOM graviton}
            &
            \cC^{\graviton}_{2,2,2}
            =
            (\Delta_{123}-2) \Delta_{123} (\Delta_{123}+2)
            (3 \Delta_{12}-\Delta_3-4)
            \, ,
            \\
            &
            \cC^{\graviton}_{2,-2,2}
            =
            -4 \Delta_3^2
            +8 \Delta_3 (\Delta_{1,2}+1)
            +4
            (
                3 \Delta_1^2-6 \Delta_1+3 \Delta_2^2-2 \Delta_2
                +2 \Delta_1 \Delta_2
            )
            \, ,
            \\
            &
            \cC^{\graviton}_{-2,2,2}
            =
            -4 \Delta_3^2
            -8 \Delta_3 (\Delta_{1,2}-1)
            +4
            (
                3 \Delta_1^2-2 \Delta_1+3 \Delta_2^2-6 \Delta_2
                +2 \Delta_1 \Delta_2
            )
            \, ,
            \\
            &
            \cC^{\graviton}_{-2,-2,2}
            =
            \Delta_{12,3} (\Delta_{12,3}+2) (\Delta_{12,3}+4)
            (3 \Delta_{12}+\Delta_3-6)
            \, .
        \end{align*}
    \item
        \too (see Section \ref{sec: three-point spinning}):
        \begin{align}
            \label{eq: 3pt coefficient TOO graviton}
            &
            \coThree{
                \graviton^{\tsign{+},im_{1}}_{\Delta_1,J_1}
                \graviton^{\inn}_{\Delta_2,J_2}
                \graviton^{\out}_{\Delta_3,J_3}
            }
            =
            \frac{2^{\Delta_{1,23}-2}
                m_1^{\Delta_{23,1}}
            }{\Delta_2 (\Delta_2-1) \Delta_3 (\Delta_3-1)}
            \mg{
                -\Delta_1,
                \frac{| J_{13,2}| +\Delta_{13,2}}{2}
            }{
                \frac{-| J_{12,3}| +\Delta_{3,12}+4}{2}
            }
            \cC^{\graviton,\tsign{+}}_{J_{1},J_{2},J_{3}}
            \, ,
            \\[0.5em]
            &
            \coThree{
                \graviton^{\tsign{-},im_{1}}_{\Delta_1,J_1}
                \graviton^{\inn}_{\Delta_2,J_2}
                \graviton^{\out}_{\Delta_3,J_3}
            }
            =
            \coThree{
                \graviton^{\tsign{+},im_{1}}_{\Delta_1,J_1}
                \graviton^{\inn}_{\Delta_3,J_3}
                \graviton^{\out}_{\Delta_2,J_2}
            }
            \, .
        \end{align}
        Here the reduced coefficient $\cC^{\graviton,\tsign{+}}_{J_{1},J_{2},J_{3}}$ satisfies the parity symmetry
        \begin{equation}
            \cC^{\graviton,\tsign{+}}_{J_{1},J_{2},J_{3}} = \cC^{\graviton,\tsign{+}}_{-J_{1},-J_{2},-J_{3}}
            \, ,
        \end{equation}
        and depends on the spin configurations as follows:
        \begin{align*}
            \nnn
            \label{eq: 3pt subcoefficient TOO graviton}
            &
            \cC^{\graviton,\tsign{+}}_{2,2,2}
            =
            -(\Delta_{123}-2) \Delta_{123} (\Delta_{123}+2)
            (\Delta_1-3 \Delta_{23}+4)
            \, ,
            \\
            &
            \cC^{\graviton,\tsign{+}}_{2,2,-2}
            =
            -4 \Delta_1^2
            +8 \Delta_1 (\Delta_{2,3}+1)
            +4
            (
                3 \Delta_2^2-6 \Delta_2+3 \Delta_3^2-2 \Delta_3
                +2 \Delta_2 \Delta_3
            )
            \, ,
            \\
            &
            \cC^{\graviton,\tsign{+}}_{2,-2,2}
            =
            -4 \Delta_1^2
            -8 \Delta_1 (\Delta_{2,3}-1)
            +4
            (
                3 \Delta_2^2-2 \Delta_2+3 \Delta_3^2-6 \Delta_3
                +2 \Delta_2 \Delta_3
            )
            \, ,
            \\
            &
            \cC^{\graviton,\tsign{+}}_{2,-2,-2}
            =
            -(\Delta_{1,23}-4) (\Delta_{1,23}-2) \Delta_{1,23}
            (\Delta_1+3 \Delta_{23}-6)
            \, .
        \end{align*}

\end{itemize}

\subsection{\too scalars}
\label{app: TOO scalar}

We warm up by rederiving the \too scalar celestial amplitude. For $\ts{1}=\pm i$,
\begin{equation}
    \vev{
        \scalar^{\ts{1},im_1}_{\Delta_1}
        \scalar^{\inn}_{\Delta_2}
        \scalar^{\out}_{\Delta_3}
    }
    =
    2^{\Delta_1-1}
    m_1^{2-\Delta_1}
    \intt{[d\hat{k}_{1'}] d\omega_2 d\omega_3}
    \omega_2^{\Delta_2-1}
    \omega_3^{\Delta_3-1}
    (-\hat{q}_1 \co \hat{k}_{1'})_{\ts{1}}^{-\Delta_1}
    \deltaMC{k_{1'}-q_2+q_3}
    \, .
\end{equation}
By \eqref{eq: mass-shell measure}, we write the $\hat{k}_{1'}$-integral as $\intt{d^{4}\khat_{1'}}2\delta(\khat_{1'} \co \khat_{1'}-1)$, and perform this integral using the momentum conservation delta function,
\begin{equation}
    \vev{
        \scalar^{\ts{1},im_1}_{\Delta_1}
        \scalar^{\inn}_{\Delta_2}
        \scalar^{\out}_{\Delta_3}
    }
    =
    2^{\Delta_1-1}
    \intt{d\omega_2 d\omega_3}
    \omega_2^{\Delta_2-1}
    \omega_3^{\Delta_3-1}
    \delta (
        -2 \omega_2 \omega_3 \hat{q}_{23}-m_1^2
    )
    (\omega_3 \hat{q}_{13}-\omega_2 \hat{q}_{12})^{-\Delta_1}_{\ts{1}}
    \, ,
\end{equation}
then the $\omega_{2}$- and $\omega_{3}$-integrals can be performed successively, leading to
\begin{equation}
    \vev{
        \scalar^{\tsign{\pm i},im_1}_{\Delta_1}
        \scalar^{\inn}_{\Delta_2}
        \scalar^{\out}_{\Delta_3}
    }
    =
    2^{\Delta_{1,23}-2}
    e^{\mp\frac{1}{2} i \pi  \Delta_{13,2}}
    m_1^{\Delta_{23,1}-2}
    \mg{
        \frac{\Delta_{12,3}}{2},
        \frac{\Delta_{13,2}}{2}
    }{
        \Delta_1
    }
    \, .
\end{equation}
By the change of basis \eqref{eq: tachyon basis change}, we obtain the result \eqref{eq: 3pt coefficient TOO scalar}.

\subsection{\tom scalars}
\label{app: TOM scalar}

Next we compute the \tom scalar celestial amplitude. For $\ts{1}=\pm i$,
\begin{equation}
    \begin{aligned}
        \vev{
            \scalar^{\ts{1},im_{1}}_{\Delta_{1}}
            \scalar^{\inn}_{\Delta_{2}}
            \scalar^{\out,m_{3}}_{\Delta_{3}}
        }
        &=
        2^{\Delta_{13}-3}
        m_3^{2-\Delta_3}
        \intt{[d\hat{k}_{1'}] [d\hat{p}_{3'}] d\omega_2}
        \omega_2^{\Delta_2-1}
        \\
        &\peq\xx
        (-\hat{q}_1 \co \hat{k}_{1'})^{-\Delta_1}_{\ts{1}}
        (-\hat{p}_{3'} \co \hat{q}_3)^{-\Delta_3}
        \deltaMC{k_{1'}-q_2+p_{3'}}
        \, .
    \end{aligned}
\end{equation}
Similar to the \too scalar, we perform the $\hat{k}_{1'}$- and $\omega_{2}$-integrals using the momentum conservation delta function and mass-shell delta function, leading to
\begin{equation}
    \begin{aligned}
        \vev{
            \scalar^{\ts{1},im_{1}}_{\Delta_{1}}
            \scalar^{\inn}_{\Delta_{2}}
            \scalar^{\out,m_{3}}_{\Delta_{3}}
        }
        &=
        2^{\Delta_{3,2}+2 \Delta_1-3}
        m_3^{\Delta_{1,23}+2}
        (m_1^2+m_3^2)^{\Delta_2-1}
        \intt{d\hat{p}_{3'}}
        (-\hat{p}_{3'} \co \hat{q}_2)^{\Delta_{1,2}}
        (-\hat{p}_{3'} \co \hat{q}_3)^{-\Delta_3}
        \\
        &\peq\xx
        (
            -2 m_3^2 \hat{p}_{3'} \co \hat{q}_1 \hat{p}_{3'} \co \hat{q}_2
            -(m_1^2+m_3^2) \hat{q}_{12}
        )^{-\Delta_1}_{\ts{1}}
        \, .
    \end{aligned}
\end{equation}
Then we separate the term $-(m_1^2+m_3^2) \hat{q}_{12}> 0$ from the rest in $\rest^{-\Delta_1}_{\ts{1}}$ by the Mellin-Barnes representation \eqref{eq: Mellin-Barnes relation},
\begin{equation}
    \begin{aligned}
        \vev{
            \scalar^{\ts{1},im_{1}}_{\Delta_{1}}
            \scalar^{\inn}_{\Delta_{2}}
            \scalar^{\out,m_{3}}_{\Delta_{3}}
        }
        &=
        \intt{[d\hat{p}_{3'}]\frac{ds}{2 \pi i}}
        2^{\Delta_{3,2}+2 \Delta_1+s-3}
        m_3^{\Delta_{1,23}+2 s+2}
        (m_1^2+m_3^2)^{\Delta_{2,1}-s-1}
        e^{\ts{1}\pi s}
        \\
        &\peq\xx
        \mg{
            -s,
            \Delta_1+s
        }{
            \Delta_1
        }
        (-\hat{q}_{12})^{-\Delta_1-s}
        (-\hat{p}_{3'} \co \hat{q}_1)^s
        (-\hat{p}_{3'} \co \hat{q}_2)^{\Delta_{1,2}+s}
        (-\hat{p}_{3'} \co \hat{q}_3)^{-\Delta_3}
        \, .
    \end{aligned}
\end{equation}
Performing the $\hat{p}_{3'}$-integral by the EAdS three-point integral \eqref{eq: EAdS three-point integral}, we obtain the three-point correlator with coefficient
\begin{equation}
    \begin{aligned}
        \coThree{
            \scalar^{\ts{1},im_{1}}_{\Delta_{1}}
            \scalar^{\inn}_{\Delta_{2}}
            \scalar^{\out,m_{3}}_{\Delta_{3}}
        }
        &=
        \intt{\frac{ds}{2 \pi i}}
        2^{\Delta_{13,2}-3}
        m_3^{\Delta_{1,23}+2 s+2}
        (m_1^2+m_3^2)^{\Delta_{2,1}-s-1}
        e^{\ts{1}\pi s}
        \\
        &\peq\xx
        \mg{
            \Delta_2,
            \Delta_1+s,
            \frac{\Delta_{2,13}-2 s}{2},
            \frac{\Delta_{23,1}-2 s-2}{2}
        }{
            \frac{\Delta_{12,3}}{2},
            \frac{\Delta_{123}-2}{2},
            \Delta_{2,1}-s
        }
        \, .
    \end{aligned}
\end{equation}
We close the $s$-contour to the left and sum over the residues at $s= -\Delta_1-n$, $n\in\NN$ leading to
\begin{equation}
    \begin{aligned}
        \coThree{
            \scalar^{\tsign{\pm},im_{1}}_{\Delta_{1}}
            \scalar^{\inn}_{\Delta_{2}}
            \scalar^{\out,m_{3}}_{\Delta_{3}}
        }
        &=
        2^{\Delta_{13,2}-3}
        e^{\mp i\pi  \Delta_1}
        m_3^{2-\Delta_{123}}
        (m_1^2+m_3^2)^{\Delta_2-1}
        \Fpq{2}{1}{
            \frac{\Delta_{12,3}}{2},\frac{\Delta_{123}-2}{2}
        }{\Delta_2}{\frac{
                m_1^2+m_3^2
        }{m_3^2 (1 \mp i \varepsilon )}}
        \, .
    \end{aligned}
\end{equation}
The two are boundary values along the branch cut of the hypergeometric function. By the change of basis \eqref{eq: tachyon basis change} and the discontinuity of the hypergeometric function \eqref{eq: hypergeometric discontinuity}, we obtain the result \eqref{eq: 3pt coefficient TOM scalar}.

\subsection{\tot scalars}
\label{app: TOT scalar}

Now we compute the \tot scalar celestial amplitude. For $\ts{1},\ts{3}=\pm i$,
\begin{equation}
    \begin{aligned}
        \vev{
            \scalar^{\ts{1},im_{1}}_{\Delta_{1}}
            \scalar^{\inn}_{\Delta_{2}}
            \scalar^{\ts{3},im_{3}}_{\Delta_{3}}
        }
        &=
        2^{\Delta_{13}-3}
        m_1^{2-\Delta_1}
        m_3^{2-\Delta_3}
        \intt{[d\hat{k}_{1'}] [d\hat{k}_{3'}] d\omega_2}
        \omega_2^{\Delta_2-1}
        \\
        &\peq\xx
        (-\hat{q}_1 \co \hat{k}_{1'})^{-\Delta_1}_{\ts{1}}
        (-\hat{q}_3 \co \hat{k}_{3'})^{-\Delta_3}_{\ts{3}}
        \deltaMC{k_{1'}-q_2+k_{3'}}
        \, .
    \end{aligned}
\end{equation}
Due to the mass threshold and the permutation symmetry $(13)$, the result contains two symmetric terms proportional to $\theta(m_1-m_3)$ and $\theta(m_3-m_1)$ respectively
\begin{equation}
    \vev{
        \scalar^{\ts{1},im_{1}}_{\Delta_{1}}
        \scalar^{\inn}_{\Delta_{2}}
        \scalar^{\ts{3},im_{3}}_{\Delta_{3}}
    }
    =I_{13} \theta(m_1-m_3)+I_{31} \theta(m_3-m_1)
    \, .
\end{equation}

Without loss of generality, we focus on the first term $I_{13}$ with $m_1\geq m_3$.
We perform the $\hat{k}_{1'}$-integral using the momentum conservation delta function, then the $\omega_{2}$-integral is localized at
\begin{equation}
    \omega_2
    =
    \frac{
        m_3^2-m_1^2
    }{2 m_3 \hat{q}_2 \co \hat{k}_{3'}}\geq 0
    \, .
\end{equation}
which implies $-\hat{q}_2 \co \hat{k}_{3'}>0$, and
\begin{equation}
    \begin{aligned}
        I_{13}
        &=
        2^{\Delta_{13,2}-3}
        m_3^{\Delta_{1,23}+2}
        (m_1^2-m_3^2)^{\Delta_2-1}
        \intt{[d\hat{k}_{3'}]}
        (-\hat{q}_2 \co \hat{k}_{3'})^{\Delta_{1,2}}_{\tsign{+}}
        (-\hat{q}_3 \co \hat{k}_{3'})^{-\Delta_3}_{\ts{3}}
        \\
        &\peq\xx
        (
            -2 m_3^2 \hat{q}_1 \co \hat{k}_{3'} \hat{q}_2 \co \hat{k}_{3'}
            -(m_1^2-m_3^2) \hat{q}_{12}
        )^{-\Delta_1}_{\ts{1}}
        \, .
    \end{aligned}
\end{equation}
Similar to the \tom scalar, we separate the term $-(m_1^2-m_3^2) \hat{q}_{12}> 0$ from the rest in $\rest^{-\Delta_1}_{\ts{1}}$ by the Mellin-Barnes representation \eqref{eq: Mellin-Barnes relation},
\begin{equation}
    \begin{aligned}
        I_{13}
        &=
        \intt{[d\hat{k}_{3'}] \frac{ds}{2 \pi i}}
        2^{\Delta_{3,2}+2 \Delta_1+s-3}
        e^{\ts{1}\pi s}
        m_3^{\Delta_{1,23}+2 s+2}
        (m_1^2-m_3^2)^{\Delta_{2,1}-s-1}
        \\
        &\peq\xx
        \mg{
            -s,
            \Delta_1+s
        }{
            \Delta_1
        }
        (-\hat{q}_{12})^{-\Delta_1-s}
        (-\hat{q}_1 \co \hat{k}_{3'})^s_{\tsign{-s_1}}
        (-\hat{q}_2 \co \hat{k}_{3'})^{\Delta_{1,2}+s}_{\tsign{+}}
        (-\hat{q}_3 \co \hat{k}_{3'})^{-\Delta_3}_{\ts{3}}
        \, ,
    \end{aligned}
\end{equation}
where $\tsign{-s_1}$ is due to the identity \eqref{eq: homogeneous distribution - parity symmetry}.
Performing the $\hat{k}_{3'}$-integral by the dS three-point integral \eqref{eq: dS three-point integral - plus/minus basis 2d}, performing the $s$-integral by contour deformation, and change the basis by \eqref{eq: tachyon basis change}, we obtain the result \eqref{eq: 3pt coefficient TOT scalar}.

\subsection{Massless limit}
\label{app: three-point celestial amplitudes - massless limit of scalars}

We check various massless limits of scalar celestial amplitudes listed in Appendix \eqref{app: three-point celestial amplitudes - summary}.
Recall that by \eqref{eq: massless limit massive shadow} and \eqref{eq: massless limit massive Mellin}, for massive particles, the pattern of massless limits is \Mout \become \Oout/\Sout and \Minn \become \Oinn/\Sinn.
By \eqref{eq: massless limit tachyonic shadow} and \eqref{eq: massless limit tachyonic Mellin}, for tachyonic particles, the pattern is \Tplus \become \Sout/\Oinn and \Tminus \become \Sinn/\Oout.

\mom.
For the $m_{3}\to0$ limit, \mom \become \Mout{}\Oout{}\Oinn vanishes due to the step function $\theta(m_3-m_1)$, and this is consistent with the physical intuition that a massless particle cannot decay into massive particles.
For the $m_{1}\to0$ limit, we rewrite it by \eqref{eq: hypergeometric 0->infinity} to manifest the mass dependence,
\begin{equation}
    \coThree{
        \scalar^{\out,m_{1}}_{\Delta_{1}}
        \scalar^{\out}_{\Delta_{2}}
        \scalar^{\inn,m_{3}}_{\Delta_{3}}
    }
    =
    m_1^{2-2 \Delta_1}
    \Fba{\cdots,\frac{m_1^2}{m_1^2-m_3^2}}
    \rest
    +
    \Fba{\cdots,\frac{m_1^2}{m_1^2-m_3^2}}
    \rest
    \, .
\end{equation}
For the Mellin limit under $\Re\Delta_1>1$, with the factor $m_1^{2\Delta_1-2}$ from \eqref{eq: massive to Mellin factor}, the second term vanishes and the first term becomes \oom.
Conversely, for the shadow limit with $\Re\Delta_1<1$, the first term vanishes and the second term becomes \Sout{}\Oout{}\Minn.

\tom.
For the $m_{1}\to0$ limit, \Tplus{}\Oinn{}\Mout is proportional to $m_1^{2-2\Delta_1}$, and \Tminus{}\Oinn{}\Mout is of order $m_1^{0}$.
For the Mellin limit under $\Re\Delta_1>1$, with the factor $m_1^{2\Delta_1-2}$ from \eqref{eq: massive to Mellin factor}, \Tplus{}\Oinn{}\Mout \become \Oinn{}\Oinn{}\Mout, whereas \Tminus{}\Oinn{}\Mout \become \Oout{}\Oinn{}\Mout vanishes and agrees with momentum conservation.
For the shadow limit under $\Re\Delta_1<1$, \Tplus{}\Oinn{}\Mout \become \Sout{}\Oinn{}\Mout vanishes and agrees with momentum conservation, whereas \Tminus{}\Oinn{}\Mout \become \Sinn{}\Oinn{}\Mout.

For the $m_{3}\to0$ limit, we need to rewrite \tom by \eqref{eq: hypergeometric 0->infinity},
\begin{equation}
    \coThree{
        \scalar^{\ts{1},im_{1}}_{\Delta_{1}}
        \scalar^{\inn}_{\Delta_{2}}
        \scalar^{\out,m_{3}}_{\Delta_{3}}
    }
    =
    m_3^{2-2 \Delta_3}
    \Fba{\cdots,-\frac{m_3^2}{m_1^2}}
    \rest
    +
    \Fba{\cdots,-\frac{m_3^2}{m_1^2}}
    \rest
    \, ,
\end{equation}
Then the analysis is similar to the case of \mom. Only one term survives in each limit, leading to \Tall{}\Oinn{}\Mout \become \Tall{}\Oinn{}\Oout and \Tall{}\Oinn{}\Mout \become \Tall{}\Oinn{}\Sout.

\tot.
Now we have the permutation symmetry $(13)$, so it suffices to consider the $m_{3}\to0$ limit and the term proportional to $\theta(m_1-m_3)$.
Notice that \Tall{}\Oinn{}\Tplus is of order $m_3^{0}$, whereas \Tall{}\Oinn{}\Tminus is proportional to $m_3^{2-2\Delta_3}$, hence the analysis is similar to the $m_{1}\to0$ limit of \tom, and the nonvanishing ones are \Tall{}\Oinn{}\Tplus \become \Tall{}\Oinn{}\Sout and \Tall{}\Oinn{}\Tminus \become \Tall{}\Oinn{}\Oout.

\section{Four-point celestial amplitudes}
\label{app: four-point celestial amplitudes}

In this section, we compute several four-point scalar celestial amplitudes.
We summarize the results in Appendix \ref{app: four-point celestial amplitudes - summary} and provide detailed derivations in Appendix \ref{app: four-point celestial amplitudes - derivation}.
Then we discuss the blow-up method of OPE extraction in Appendix \ref{app: four-point celestial amplitudes - OPE extraction}, and analyze the massless limit in Appendix \ref{app: four-point celestial amplitudes - massless limit}.

\subsection{Summary}
\label{app: four-point celestial amplitudes - summary}

For convenience we define the dimensionless positive Mandelstam variables in any scattering process involving one massive/tachyonic particle:
\begin{equation}
    \label{eq: dimensionless Mandelstam variables}
    \quad
    S = m_{1}^{-2} |s| \geq 0
    \, ,
    \quad
    T = m_{1}^{-2} |t| \geq 0
    \, ,
    \quad
    U = m_{1}^{-2} |u| \geq 0
    \, .
\end{equation}
We also introduce the kinematical factor
\begin{equation}
    K(\chi,\chib)
    =
    2^{\Delta_{1,234}-2}
    \chi^{\frac{\Delta_{12}}{2}}
    \bar{\chi}^{\frac{\Delta_{12}}{2}}
    (1-\chi )^{\frac{\Delta_{14,23}}{2}}
    (1-\bar{\chi})^{\frac{\Delta_{14,23}}{2}}
    \, .
\end{equation}

\begin{itemize}
    \item
        \soooS (\cf \cite{Liu:2025dhh}):
        \begin{align}
            \label{eq: 4pt amplitude SOOO S-channel}
            \vev{\wave\scalar^{\inn}_{\Delta_1}\scalar^{\out}_{\Delta_2}\scalar^{\out}_{\Delta_3}\scalar^{\out}_{\Delta_4}}
            &=
            K(\chi,\chib)
            \vevv{\scalar_{1}\scalar_{2}\scalar_{3}\scalar_{4}}
            \inttt{ds dt}{\region}
            s^{\frac{\Delta_{134,2}-2}{2}}
            (-t)^{\frac{\Delta_{124,3}-2}{2}}
            (-u)^{\frac{\Delta_{123,4}-2}{2}}
            \\
            &\peq\xx
            (s+t \chi )^{-\Delta_1}
            (s+t \bar{\chi})^{-\Delta_1}
            \cT(s,t)
            \, ,
            \nn
        \end{align}
        where the physical region is
        \begin{align}
            &\region
            =
            \set{(s,t) \given t\leq 0 \land s+t\geq 0}
            \, .
        \end{align}
    \item
        \moooS:
        \begin{align}
            \label{eq: 4pt amplitude MOOO S-channel}
            \vev{\scalar^{\inn,m_{1}}_{\Delta_1}\scalar^{\inn}_{\Delta_2}\scalar^{\out}_{\Delta_3}\scalar^{\out}_{\Delta_4}}
            &=
            K(\chi,\chib)
            \vevv{\scalar_{1}\scalar_{2}\scalar_{3}\scalar_{4}}
            \inttt{ds dt}{\region}
            s^{\frac{\Delta_{134,2}-2}{2}}
            (-t)^{\frac{\Delta_{124,3}-2}{2}}
            (-u)^{\frac{\Delta_{123,4}-2}{2}}
            \\
            &\peq\xx
            \mleft(
                (s+t \chi ) (s+t \bar{\chi})
                -m_1^2 (s+t \chi  \bar{\chi})
            \mright)^{-\Delta_1}
            \cT(s,t)
            \, ,
            \nn
        \end{align}
        where the physical region is
        \begin{equation}
            \region
            =
            \set{(s,t) \given t\leq 0 \land s+t\geq m_1^2}
            =
            \set{(S,T) \given T\geq 0 \land S \geq T+1}
            \, .
        \end{equation}
    \item
        \toooS:
        \begin{align}
            \label{eq: 4pt amplitude TOOO S-channel}
            \vev{\scalar^{\tsign{\pm},im_{1}}_{\Delta_1}\scalar^{\inn}_{\Delta_2}\scalar^{\out}_{\Delta_3}\scalar^{\out}_{\Delta_4}}
            &=
            K(\chi,\chib)
            \vevv{\scalar_{1}\scalar_{2}\scalar_{3}\scalar_{4}}
            \inttt{ds dt}{\region}
            s^{\frac{\Delta_{134,2}-2}{2}}
            (-t)^{\frac{\Delta_{124,3}-2}{2}}
            (-u)^{\frac{\Delta_{123,4}-2}{2}}
            \\
            &\peq\xx
            \mleft(
                (s+t \chi ) (s+t \bar{\chi})
                +m_1^2 (s+t \chi  \bar{\chi})
            \mright)^{-\Delta_1}_{\tsign{\mp}}
            \cT(s,t)
            \, ,
            \nn
        \end{align}
        where the physical region is
        \begin{equation}
            \region
            =
            \set{(s,t) \given t \leq0 \land s\geq 0 \land s+t+m_1^2 \geq 0}
            =
            \set{(S,T) \given T\geq 0 \land S\geq 0 \land S+1\geq T}
            \, .
        \end{equation}
    \item
        \moooD:
        \begin{align}
            \label{eq: 4pt amplitude MOOO D-channel}
            \vev{\scalar^{\inn,m_{1}}_{\Delta_1}\scalar^{\out}_{\Delta_2}\scalar^{\out}_{\Delta_3}\scalar^{\out}_{\Delta_4}}
            &=
            K(\chi,\chib)
            \vevv{\scalar_{1}\scalar_{2}\scalar_{3}\scalar_{4}}
            \inttt{ds dt}{\region}
            s^{\frac{\Delta_{134,2}-2}{2}}
            t^{\frac{\Delta_{124,3}-2}{2}}
            u^{\frac{\Delta_{123,4}-2}{2}}
            \\
            &\peq\xx
            \mleft(
                m_1^2 (s+t \chi  \bar{\chi})
                -(s+t \chi ) (s+t \bar{\chi})
            \mright)^{-\Delta_1}
            \cT(s,t)
            \, ,
            \nn
        \end{align}
        where the physical region is
        \begin{align}
            \region
            =
            \set{(s,t) \given s\geq 0 \land t\geq 0 \land s+t\leq m_1^2}
            =
            \set{(S,T) \given S\geq 0 \land T\geq 0 \land S+T\leq 1}
            \, .
        \end{align}
\end{itemize}

\subsection{Derivation}
\label{app: four-point celestial amplitudes - derivation}

We first compute the \moooS scalar celestial amplitude. For an arbitrary scalar amplitude $\mathcal{T}$, we have
\begin{align}
    \vev{\scalar^{\inn,m_{1}}_{\Delta_1}\scalar^{\inn}_{\Delta_2}\scalar^{\out}_{\Delta_3}\scalar^{\out}_{\Delta_4}}
    &=
    2^{\Delta_1-2}
    m_1^{2-\Delta_1}
    \intt{[d\hat{p}_{1'}] d\omega_2 d\omega_3 d\omega_4}
    \omega_2^{\Delta_2-1}
    \omega_3^{\Delta_3-1}
    \omega_4^{\Delta_4-1}
    (-\hat{p}_{1'} \co \hat{q}_1)^{-\Delta_1}
    \\
    &\peq\xx
    \deltaMC{p_{1'}+q_2-q_3-q_4}
    \cT\mleft(-(q_3{+}q_4)^{2},-(q_2{-}q_4)^{2}\mright)
    \, .
    \nn
\end{align}
We write the amplitude in this way to separate the dependence on $\hat{p}_{1'}$, and perform the $\hat{p}_{1'}$-integral using the momentum conservation delta function, leading to
\begin{align}
    \rhs
    &=
    2^{\Delta_1-1}
    \intt{d\omega_2 d\omega_3 d\omega_4}
    \omega_2^{\Delta_2-1}
    \omega_3^{\Delta_3-1}
    \omega_4^{\Delta_4-1}
    (\hat{q}_1 \co \hat{q}_2 \omega_2-\hat{q}_1 \co \hat{q}_3 \omega_3-\hat{q}_1 \co \hat{q}_4 \omega_4)^{-\Delta_1}
    \\
    &\peq\xx
    \delta (
        m_1^2
        -2 \hat{q}_2 \co \hat{q}_3 \omega_2 \omega_3
        -2 \hat{q}_2 \co \hat{q}_4 \omega_2 \omega_4
        +2 \hat{q}_3 \co \hat{q}_4 \omega_3 \omega_4
    )
    \cT\mleft(-(q_3{+}q_4)^{2},-(q_2{-}q_4)^{2}\mright)
    \, .
    \nn
\end{align}
Here the step function in the mass-shell measure \eqref{eq: mass-shell measure} has been dropped because $q_3^0+q_4^0-q_2^0\geq0$ when $(q_3+q_4-q_2)^2=-m_{1}^2$.
Then we change $(\omega_2,\omega_3)$ to the Mandelstam variables
\begin{equation}
    \label{eq: MOOO omega to st}
    s = 4 \omega_3 \omega_4 z_{3,4} \bar{z}_{3,4}>0
    \, ,
    \quad
    t = -4 \omega_2 \omega_4 z_{2,4} \bar{z}_{2,4}<0
    \, ,
\end{equation}
and perform the $\omega_4$-integral using the delta function, leading to
\begin{align}
    \rhs
    &=
    2^{
        \frac{3 \Delta_1-\Delta_{234}-4}{2}
    }
    (-\hat{q}_3 \co \hat{q}_4)^{\frac{\Delta_{12,34}}{2}}
    (-\hat{q}_2 \co \hat{q}_4)^{\frac{\Delta_{13,24}}{2}}
    (-\hat{q}_2 \co \hat{q}_3)^{\frac{\Delta_{14,23}}{2}}
    \inttt{ds dt}{\region}
    \cT(s,t)
    \\
    &\peq\xx
    s^{\frac{\Delta_{134,2}-2}{2}}
    (-t)^{\frac{\Delta_{124,3}-2}{2}}
    (-u)^{\frac{\Delta_{123,4}-2}{2}}
    (
        -\hat{q}_1 \co \hat{q}_4 \hat{q}_2 \co \hat{q}_3 s t
        -\hat{q}_1 \co \hat{q}_3 \hat{q}_2 \co \hat{q}_4 s u
        -\hat{q}_1 \co \hat{q}_2 \hat{q}_3 \co \hat{q}_4 t u
    )^{-\Delta_1}
    \, .
    \nn
\end{align}
Here the integration region of $(s,t)$ has shrunk from \eqref{eq: MOOO omega to st} to the physical region $\region$ of two-to-two scattering.
The reason is that, for any $\qhat_{2},\, \qhat_{3},\, \qhat_{4}$ and $\omega_4 \geq 0$, the delta function has a solution at
\begin{equation}
    \omega_{4}^{2}=
    -\frac{\hat{q}_{23} s t}{2 \hat{q}_{24} \hat{q}_{34} u}
    \, ,
\end{equation}
if and only if the Mandelstam variables $(s,t)$ belong to the physical region $\region$.
Finally we strip off the conformal structure in \eqref{eq: conformal structure} and write the remaining in terms of cross ratios, leading to \eqref{eq: 4pt amplitude MOOO S-channel}.

With the same procedure, we obtain the other celestial amplitudes listed in Appendix \ref{app: four-point celestial amplitudes - summary}.

\subsection{Massless limit}
\label{app: four-point celestial amplitudes - massless limit}

The massless limit from \moooS to the conventional shadow celestial amplitude \soooS is straightforward: both the integrand and the physical region of \moooS reduce directly to those of \soooS.
However, according to the discussion of regular celestial amplitudes, even in the massless limit, \moooS contains more contributions from soft/collinear regions than \soooS.
This is manifest when rewriting \moooS in terms of the dimensionless Mandelstam variables \eqref{eq: dimensionless Mandelstam variables}, which produces an overall prefactor $m_1^{\Delta_{234,1}-2}$ in front of the integral.
Consequently,
\begin{equation}
    \vev{\scalar^{\inn,m_{1}}_{\Delta_1}\scalar^{\inn}_{\Delta_2}\scalar^{\out}_{\Delta_3}\scalar^{\out}_{\Delta_4}}
    \sim
    m_1^{\Delta_{234,1}-2} \vevv{\scalar_{1}\scalar_{2}\scalar_{3}\scalar_{4}} \rest
    +
    m_{1}^{0}
    \vev{\wave\scalar^{\inn}_{\Delta_1}\scalar^{\inn}_{\Delta_2}\scalar^{\out}_{\Delta_3}\scalar^{\out}_{\Delta_4}}
    +
    \cdots
    \, ,
\end{equation}
and it is the first term that contributes to the collinear OPE as discussed later \eqref{eq: MOOO t-channel collinear conformal block coefficient}, which is missing in \soooS.
With the asymptotic amplitude $\cT\sim m_{1}^{-n}\cT_{0}(S,T)$ near $m_{1}\sim 0$, the contribution of the first term is
\begin{align}
    &
    \deltaK{\Delta_{234,1}-n-2}
    K(\chi,\chib)
    \inttt{dS dT}{\region}
    T^{\Delta_{24}-\frac{n}{2}-2}
    S^{\Delta_{34}-\frac{n}{2}-2}
    (S-T-1)^{\Delta_{23}-\frac{n}{2}-2}
    \\
    &\qquad \xx
    \mleft(
        (S-T \chi ) (S-T \bar{\chi})
        -S+T \chi  \bar{\chi}
    \mright)^{-\Delta_{234}+n+2}
    \cT_{0}(S,T)
    \, .
    \nn
\end{align}
This phenomenon can be illustrated by the toy example: the following integral is a hypergeometric function, and the asymptotic behavior near $m_{1}\sim 0$ is given by
\begin{equation}
    \intrange{ds}{m^{2}}{\infty}
    \frac{s^{\alpha} (s-m^2)^{\beta}}{s+1}
    \sim
    -\pi \csc(\pi(\alpha{+}\beta))
    +
    m^{2 (\alpha +\beta +1)} \betaShort{\beta+1,-\alpha -\beta -1}
    \, .
\end{equation}
Taking the limit directly on both the integrand and the region yields only the first term, whereas changing $s = m^{2} S$ before taking the limit yields only the second term.

This effect is more striking in the decay process \moooD: the conventional shadow celestial amplitude vanishes by momentum conservation, but the massless limit of \moooD yields a nontrivial contribution that captures the collinear OPE \eqref{eq: MOOO d-channel collinear conformal block coefficient}.
While the physical region contracts to zero measure in the massless limit, it expands to a finite region when rewritten in terms of the dimensionless Mandelstam variables. For the asymptotic scattering amplitude $\cT\sim m_{1}^{-n}\cT_{0}(S,T)$ near $m_{1}\sim 0$, the contribution reads
\begin{align}
    &
    \deltaK{\Delta_{234,1}-n-2}
    K(\chi,\chib)
    \inttt{dS dT}{\region}
    T^{\Delta_{24}-\frac{n}{2}-2}
    S^{\Delta_{34}-\frac{n}{2}-2}
    (1-S-T)^{\Delta_{23}-\frac{n}{2}-2}
    \\
    &\qquad \times
    \mleft(
        S+T \chi  \bar{\chi}
        -(S+T \chi ) (S+T \bar{\chi})
    \mright)^{-\Delta_{234}+n+2}
    \cT_{0}(S,T)
    \, .
    \nn
\end{align}

\subsection{OPE extraction}
\label{app: four-point celestial amplitudes - OPE extraction}

We now extract OPEs from four-point celestial amplitudes.

\moooS.
We extract the leading massless operators in the \tchannel collinear OPE $\scalar^{\inn}_{\Delta_2}\scalar^{\out}_{\Delta_4}$. Physically, they come from the scattering amplitude $\cT$ in the $t \sim 0$ regime.
To separate these contributions from those in other regimes, we first introduce the blow-up variable $t=t'|\chi|^{-2}$, and then take the \tchannel OPE limit $\chi=\frac{z_{1,2}z_{3,4}}{z_{1,3}z_{2,4}}\to \oo$. This compresses all finite $t'$ to the vicinity of $t\sim 0$, and accordingly, the integration region changes from $R$ to $\set{(s,t') \given s\geq m_{1}^{2} \land t'\leq 0}$. In this process, we replace the scattering amplitude $\cT(s,t)$ with its asymptotic behavior $\cT(s,t)\sim f(t)\cT_{0}(s)$ in the $t\to 0$ limit.

In practical calculations, we first rewrite the celestial amplitude \eqref{eq: 4pt amplitude MOOO S-channel} in terms of the \tchannel cross ratios $\chit = \chi^{-1}$ and the dimensionless Mandelstam variables $(S,T)$ in \eqref{eq: dimensionless Mandelstam variables}, and then perform the blow-up $T=T' |\chit|^{2}$.
If the amplitude has asymptotic behavior $\cT\sim (-t)^{-n}\cT_{0}(s)$, then the celestial amplitude behaves as
\begin{equation}
    \vev{\scalar^{\inn,m_{1}}_{\Delta_1}\scalar^{\inn}_{\Delta_2}\scalar^{\out}_{\Delta_3}\scalar^{\out}_{\Delta_4}}
    \sim
    \vevv{\scalar_{1}\scalar_{3}\scalar_{2}\scalar_{4}}
    \chit^{\frac{1}{2}\Delta_{24}-n}
    \chibt^{\frac{1}{2}\Delta_{24}-n}
    \, ,
\end{equation}
which indicates the presence of the exchange operator
\begin{equation}
    \scalar^{\inn}_{\Delta_2}\scalar^{\out}_{\Delta_4}
    \sim
    \op_{\Delta_{24}-2n,0}
    \, .
\end{equation}
The corresponding conformal block coefficient can be read off as
\begin{align}
    \label{eq: MOOO t-channel collinear conformal block coefficient}
    \coBlock
    &=
    2^{\Delta_{1,234}-2}
    m_1^{\Delta_{234,1}-2 n-2}
    \mg{
        \frac{\Delta_{13,24}+2 n}{2}
        ,
        \frac{\Delta_{124,3}-2 n}{2}
    }{
        \Delta_1
    }
    \intrange{dS}{1}{\infty}
    S^{\Delta_4-n-1}
    (S-1)^{\Delta_2-n-1}
    \cT_{0}(m_1^2 S)
    \, .
\end{align}

\Tplus{}\Oinn{}\Oout{}\Oout.
For the \tchannel OPE $\scalar^{\inn}_{\Delta_2}\scalar^{\out}_{\Delta_4}$, the analysis is similar to the \moooS case. The exchange operator is
\begin{equation}
    \scalar^{\inn}_{\Delta_2}\scalar^{\out}_{\Delta_4}
    \sim
    \op_{\Delta_{24}-2n,0}
    \, ,
\end{equation}
with the conformal block coefficient
\begin{align}
    \label{eq: TOOO t-channel collinear conformal block coefficient}
    \coBlock
    &=
    2^{\Delta_{1,234}-2}
    m_1^{\Delta_{234,1}-2 n-2}
    \mg{
        1-\Delta_1,
        \frac{\Delta_{13,24}+2 n}{2}
    }{
        \frac{\Delta_{3,124}+2 n+2}{2}
    }
    \intrange{dS}{0}{\infty}
    S^{\Delta_4-n-1}
    (S+1)^{\Delta_2-n-1}
    \cT_{0}(m_1^2 S)
    \, .
\end{align}

For the \schannel OPE $\scalar^{\out}_{\Delta_3}\scalar^{\out}_{\Delta_4}$, we change the blow-up variable $S=S'|\chi|^{-2}$ and take the OPE limit $\chi\to 0$, replacing the scattering amplitude $\cT(s,t)$ with the asymptotic behavior $\cT\sim s^{-n}\cT_{0}(t)$ near $s\sim 0$. Then the exchange operator is
\begin{equation}
    \scalar^{\out}_{\Delta_3}\scalar^{\out}_{\Delta_4}
    \sim
    \op_{\Delta_{34}-2n,0}
    \, ,
\end{equation}
with the conformal block coefficient
\begin{align}
    \label{eq: TOOO s-channel collinear conformal block coefficient}
    \coBlock
    &=
    2^{\Delta_{1,234}-2}
    m_1^{\Delta_{234,1}-2 n-2}
    \mg{
        1-\Delta_1,
        \frac{\Delta_{134,2}-2 n}{2}
    }{
        \frac{\Delta_{34,12}-2 n+2}{2}
    }
    \intrange{dT}{0}{1}
    T^{\Delta_4-n-1}
    (1-T)^{\Delta_3-n-1}
    \cT_{0}(-m_1^2 T)
    \, .
\end{align}

\moooD.
For the \schannel OPE $\scalar^{\out}_{\Delta_3}\scalar^{\out}_{\Delta_4}$, with the asymptotic behavior $\cT\sim s^{-n}\cT_{0}(t)$ near $s\sim 0$, the exchange operator is
\begin{equation}
    \scalar^{\out}_{\Delta_3}\scalar^{\out}_{\Delta_4}
    \sim
    \op_{\Delta_{34}-2n,0}
    \, .
\end{equation}
with the conformal block coefficient
\begin{align}
    \label{eq: MOOO d-channel collinear conformal block coefficient}
    \coBlock
    &=
    2^{\Delta_{1,234}-2}
    m_1^{\Delta_{234,1}-2 n-2}
    \mg{
        \frac{\Delta_{12,34}+2 n}{2}
        ,
        \frac{\Delta_{134,2}-2 n}{2}
    }{
        \Delta_1
    }
    \intrange{dT}{0}{1}
    T^{\Delta_4-n-1}
    (1-T)^{\Delta_3-n-1}
    \cT_{0}(m_1^2 T)
    \, .
\end{align}

\textbf{Double-trace exchange.}
For all the massive celestial amplitudes listed in Appendix \ref{app: four-point celestial amplitudes - summary}, due to the factor $\chi^{\frac{\Delta_{12}}{2}}\bar{\chi}^{\frac{\Delta_{12}}{2}}$ in the kinematical prefactor $K(\chi,\chib)$, there is an scalar exchange $\op_{\Delta_{12},0}$ of double-trace type in the \schannel OPE.

Particularly for \moooD, with the asymptotic scattering amplitude  $\cT\sim m_{1}^{-n}\cT_{0}(S,T)$ near $m_{1}\sim 0$, the leading \schannel OPE is
\begin{equation}
    \scalar^{\inn,m_{1}}_{\Delta_1}\scalar^{\out}_{\Delta_2}
    \sim
    \op_{\Delta_{12},0}
    \, ,
\end{equation}
with the conformal block coefficient
\begin{equation}
    \label{eq: MOOO d-channel double-trace conformal block coefficient}
    \coBlock
    =
    2^{\Delta_{1,234}-2}
    m_1^{\Delta_{234,1}-n-2}
    \inttt{dS dT}{\region}
    S^{\frac{\Delta_{34,12}}{2}-1}
    T^{\frac{\Delta_{124,3}}{2}-1}
    (1-S)^{-\Delta_1}
    (1-S-T)^{\frac{\Delta_{123,4}}{2}-1}
    \cT_{0}(S,T)
    \, .
\end{equation}

\section{Miscellaneous}

\subsection{Comparison with coordinate-space Mellin basis}
\label{app: CPWF lower spin comparison}

In this section we compare the massless Mellin basis in Section \ref{sec: conformal basis} with the coordinate-space representation in \eg \cite{Pasterski:2017kqt,Donnay:2020guq}.

\textbf{Spin-1.}
In the coordinate space, the conformal basis is
\begin{equation}
    A_{a,\mu}^{\oi}=
    \epsilon_{a,\mu}(-\qhat \co X_{\pm})^{-\Delta}
    +
    \qhat_{\mu}
    (\epsilon_{a} \co X)
    (-\qhat \co X_{\pm})^{-\Delta-1}
    \, ,
\end{equation}
where $X_{\pm}\eqq X \pm i \varepsilon (-1,0,0,0)$.
Without loss of generality, we consider the outgoing case and omit the subscript $\pm$. Using the IBP relation
\begin{equation}
    \qhat_{\mu}
    (\epsilon_{a} \co X)
    (-\qhat \co X)^{-\Delta-1}
    =
    \frac{1}{\Delta}\pp_{\mu}\bigl(\epsilon_{a} \co X (-\qhat \co X)^{-\Delta}\bigr)
    -
    \frac{1}{\Delta} \epsilon_{a,\mu}(-\qhat \co X)^{-\Delta}
    \, ,
\end{equation}
we rewrite $\epsilon_{a} \co X$ into total derivatives
\begin{equation}
    A_{a,\mu}
    =
    \frac{\Delta-1}{\Delta}
    \epsilon_{a,\mu}(-\qhat \co X)^{-\Delta}
    +
    \frac{1}{\Delta}\pp_{\mu}\bigl(\epsilon_{a} \co X (-\qhat \co X)^{-\Delta}\bigr)
    \, .
\end{equation}
Then using the regularized integral
\begin{equation}
    (-\qhat \co X_{\pm})^{-\Delta}
    =
    c_{\pm}(\Delta)
    \intt{d\omega}
    \omega^{\Delta-1}
    e^{\pm i q X_{\pm}}
    \, ,
    \TextInMath{where}
    c_{\pm}(\Delta)=\frac{1}{\gm{\Delta}}e^{\pm i\pi \Delta/2}
    \, ,
\end{equation}
we obtain
\begin{align}
    A_{a,\mu}
    &=
    c(\Delta)
    \intt{d\omega}
    \omega^{\Delta-1}
    \LR{
        \frac{\Delta-1}{\Delta}
        \epsilon_{a,\mu}
        e^{i q X}
        +
        \frac{1}{\Delta}
        \pp_{\mu}(
            \epsilon_{a} \co X
            e^{i q X}
        )
    }
    \\
    &=
    c(\Delta)
    \intt{d\omega}
    \omega^{\Delta-1}
    \LR{
        \frac{\Delta-1}{\Delta}
        \epsilon_{a,\mu}
        e^{i q X}
        +
        \frac{1}{\Delta}
        \pp_{a}(\qhat_{\mu}e^{i q X})
    }
    \, ,
    \nn
\end{align}
where in the second line we have used
\begin{equation}
    \label{eq: derivative intertwiner}
    \pp_{\mu_{1}}\cdots \pp_{\mu_{n}}
    \left(
        (\epsilon_{a} \co X)^{n} e^{\pm i q X}
    \right)
    =
    \pp^{n}_{a}
    \left(
        \qhat_{\mu_{1}}\cdots \qhat_{\mu_{n}}
        e^{\pm i q X}
    \right)
    \, .
\end{equation}
Integrated with the scattering amplitude $T(q)=\intt{d^{d+2}X}T(X) e^{iq X}$ we obtain the relative normalization
\begin{equation}
    A^{\oi}_{a,\mu}=
    \frac{e^{\pm i\pi \Delta/2}}{\Delta\, \gm{\Delta -1}}
    \Phi_{\Delta,a}^{\oi,1}
    \, .
\end{equation}

\textbf{Spin-2.}
The coordinate-space representation is
\begin{equation}
    h_{a,\mu\nu}^{\oi}=
    \half
    \LRa{
        -\qhat \co X\, \epsilon_{a,\mu}
        +
        \epsilon \co X\, \qhat_{\mu}
    }
    \LRa{
        -\qhat \co X\, \epsilon_{a,\nu}
        +
        \epsilon \co X\, \qhat_{\nu}
    }
    (-\qhat \co X_{\pm})^{-\Delta-2}
    \, .
\end{equation}
Similar to the spin-1 case, we first rewrite $\epsilon_{a} \co X$ into total derivatives
\begin{align}
    h_{a,\mu\nu}
    &=
    \frac{\Delta-1}{2(\Delta+1)}
    \epsilon_{a,\mu}\epsilon_{a,\nu}
    (-\qhat \co X)^{-\Delta}
    +
    \frac{\Delta-1}{2\Delta(\Delta+1)}
    (\epsilon_{a,\nu}\pp_{\mu}+\epsilon_{a,\mu}\pp_{\nu})
    \LRa{(\epsilon_{a} \co X)(-\qhat \co X)^{-\Delta}}
    \nn
    \\
    &\peq
    +
    \frac{1}{2\Delta(\Delta+1)}
    \pp_{\mu}\pp_{\nu}\LRa{
        (\epsilon_{a} \co X)^{2}
        (-\qhat \co X)^{-\Delta}
    }
    \, ,
    \nn
\end{align}
by the following IBP relations
\begin{align}
    \qhat_{\mu}
    (\epsilon_{a} \co X)
    (-\qhat \co X)^{-\Delta-1}
    &=
    \frac{1}{\Delta}\pp_{\mu}\LRa{\epsilon_{a} \co X (-\qhat \co X)^{-\Delta}}
    -
    \frac{1}{\Delta} \epsilon_{a,\mu}(-\qhat \co X)^{-\Delta}
    \, ,
    \\
    \qhat_{\mu}\qhat_{\nu}
    (\epsilon_{a} \co X)^{2}
    (-\qhat \co X)^{-\Delta-2}
    &=
    \frac{1}{\Delta(\Delta+1)}
    \pp_{\mu}\pp_{\nu}\LRa{
        (\epsilon_{a} \co X)^{2}
        (-\qhat \co X)^{-\Delta}
    }
    -
    \frac{2}{\Delta(\Delta+1)}
    \epsilon_{a,\mu}\epsilon_{a,\nu}
    (-\qhat \co X)^{-\Delta}
    \nn
    \\
    &\peq
    -
    \frac{2}{\Delta+1}
    (\qhat_{\mu}\epsilon_{a,\nu}+\qhat_{\nu}\epsilon_{a,\mu})
    (\epsilon_{a} \co X)(-\qhat \co X)^{-\Delta-1}
    \, ,
\end{align}
Then performing the Fourier transform and using the identity \eqref{eq: derivative intertwiner} again, we obtain
\begin{equation}
    h_{a,\mu\nu}^{\oi}
    =
    \frac{e^{\pm i\pi \Delta/2}}{2 (\Delta +1) \gm{\Delta -1}}
    \Phi_{\Delta,a}^{\oi,2}
    \, .
\end{equation}

\subsection{Shadow transform of massive/tachyonic basis with extremal spins}
\label{app: shadow transform of massive and tachyonic bases}

In this appendix, we compute the shadow transform of massive and tachyonic conformal bases with extremal spins $J=\pm \ell$.

\textbf{Massive case.}
We begin with the massive case. For $J=\pm\ell$, the massive basis kernel in \eqref{eq: CPWF massive} simplifies to
\begin{equation}\label{eq: CPWF massive j=l}
    \begin{aligned}
        \Phi_{m,\Delta,\pm\ell}^{\io,\mumu{1}{\ell}}(\hat{q},p)
        &=
        2^{\Delta-2}(-1)^{\ell}
        (-\qhat \co p)^{-\Delta}
        \cP_{\nu_{1}}{\!}^{\mu_{1}}(\qhat,\phat)\cdots\cP_{\nu_{\ell}}{\!}^{\mu_{\ell}}(\qhat,\phat)\,
        \epsilon_{\pm\ell}^{\nunu{1}{\ell}}
        \, .
    \end{aligned}
\end{equation}
This simplification follows from the symmetric, traceless, and transverse properties of the polarization tensor $\epsilon_{\pm\ell}^{\nunu{1}{\ell}}(\qhat)$, together with the identity
\begin{align}
    \cP_{\nunu{1}{\ell}}{\!}^{\rhorho{1}{\ell}}(\phat)\cP_{\rhorho{1}{\ell}}{\!}^{\mumu{1}{\ell}}(\qhat,\phat)=\cP_{\nunu{1}{\ell}}{\!}^{\mumu{1}{\ell}}(\qhat,\phat)
    \, ,
\end{align}
which can be verified using the tracelessness and transversality properties of the projectors.

Following \cite{Simmons-Duffin:2012juh,Pasterski:2017kqt}, the shadow transform of the massive basis kernel $\shadow[\Phi_{m,\Delta,\pm\ell}^{\io,\mumu{1}{\ell}}]$ takes the form
\begin{align}
    \shadow[\Phi_{m,\Delta,\pm\ell}^{\io,\mumu{1}{\ell}}]=2^{\Delta-2}\epsilon_{\mp\ell}^{\nunu{1}{\ell}}(\qhat)\int d^2z'\frac{\cP_{\nu_1}{\!}^{\rho_1}(\qhat,\qhat')\cdots\cP_{\nu_\ell}{\!}^{\rho_\ell}(\qhat,\qhat')}{(-\frac{1}{2}\qhat\co \qhat')^{2-\Delta}(-\qhat' \co p)^{\Delta}}\cP_{\rho_{1}}{\!}^{\mu_{1}}(\qhat',\phat)\cdots\cP_{\rho_{\ell}}{\!}^{\mu_{\ell}}(\qhat',\phat)
    \, .
\end{align}
Then by the following identity of the projector \eqref{eq: projector qp}:
\begin{align}
    &\peq
    \epsilon_{\mp\ell}^{\nunu{1}{\ell}}(\qhat)
    \cP_{\nu_1}{\!}^{\rho_1}(\qhat,\qhat')
    \cdots
    \cP_{\nu_\ell}{\!}^{\rho_\ell}(\qhat,\qhat')
    \cP_{\rho_{1}}{\!}^{\mu_{1}}(\qhat',\phat)
    \cdots
    \cP_{\rho_{\ell}}{\!}^{\mu_{\ell}}(\qhat',\phat)
    \\
    &
    =
    \epsilon_{\mp\ell}^{\nunu{1}{\ell}}
    \sum_{i,j=0}^{\ell}
    \binom{\ell}{i}\binom{\ell}{j}
    \frac{
        \qhat'_{\nu_1}\qhat^{\symL\rho_1}\cdots\qhat'_{\nu_j}\qhat^{\rho_j}g_{\nu_{j+1}}{\!}^{\rho_{j+1}}\cdots g_{\nu_{\ell}}{\!}^{\rho_{\ell}\symR}
    }{
        (-\qhat'\co \qhat)^j
    }
    \frac{
        p_{\rho_1}\qhat'^{\symL\mu_1}\cdots p_{\rho_i}\qhat'^{\mu_i}g_{\rho_{i+1}}{\!}^{\mu_{i+1}}\cdots g_{\rho_{\ell}}{\!}^{\mu_{\ell}\symR}
    }{
        (-\qhat' \co p)^i
    }
    \, ,
    \nn
\end{align}
we can rewrite the kernel as
\begin{align}
    \shadow[\Phi_{m,\Delta,\pm\ell}^{\io,\mumu{1}{\ell}}]&
    =
    \epsilon_{\mp\ell}^{\nunu{1}{\ell}}
    \sum_{i,j=0}^{\ell}
    \binom{\ell}{i}\binom{\ell}{j}
    \frac{
        p_{\rho_1}\cdots p_{\rho_i}g_{\rho_{i+1}}{\!}^{\mu_{i+1}}\cdots g_{\rho_{\ell}}{\!}^{\mu_{\ell}}\qhat^{\symL\rho_1}\cdots\qhat^{\rho_j}g_{\nu_{j+1}}{\!}^{\rho_{j+1}}\cdots g_{\nu_{\ell}}{\!}^{\rho_{\ell}\symR}
    }{
        (\Delta)_i(\Delta-j)_j
    }
    \nn
    \\
    &\peq\xx
    \frac{\partial}{\partial p_{\mu_1}}\cdots\frac{\partial}{\partial p_{\mu_i}}
    \frac{\partial}{\partial p^{\nu_1}}\cdots\frac{\partial}{\partial p^{\nu_j}}
    \intt{d^2z'}
    (-\qhat\co \qhat')^{\Delta-2-j}(-\qhat' \co p)^{-\Delta+j}
    \, .
\end{align}
Here all the $\mu$-indices are symmetrized, and the mass of the momentum $p$ is treated as an independent variable so that the derivative $\partial_{p_{\mu}}$ is well-defined.

The $z'$-integral is the conformal two-point integral \eqref{eq: cft two-point integral qqp}, then we have
\begin{align}
    \shadow[\Phi_{m,\Delta,\pm\ell}^{\io,\mumu{1}{\ell}}]&=\sum_{i,j=0}^{\ell}\frac{\pi}{2^{2-\Delta+j}}\binom{\ell}{i}\binom{\ell}{j}\frac{p_{\rho_1}\cdots p_{\rho_i}g_{\rho_{i+1}}{\!}^{\mu_{i+1}}\cdots g_{\rho_{\ell}}{\!}^{\mu_{\ell}}\qhat^{\symL\rho_1}\cdots\qhat^{\rho_j}g_{\nu_{j+1}}{\!}^{\rho_{j+1}}\cdots g_{\nu_{\ell}}{\!}^{\rho_{\ell}\symR}}{(\Delta)_i(\Delta-j-1)_{j+1}}
    \nn
    \\
    &\peq\xx
    \epsilon_{\mp\ell}^{\nunu{1}{\ell}}\frac{\partial}{\partial p_{\mu_1}}\cdots\frac{\partial}{\partial p_{\mu_i}}\frac{\partial}{\partial p^{\nu_1}}\cdots\frac{\partial}{\partial p^{\nu_j}}\frac{(-p^2)^{1-\Delta+j}}{(-\qhat \co p)^{2-\Delta+j}}
    \, .
\end{align}
Since the action of $\partial_{p^{\nu}}$ on $(-\qhat \co p)^{-\Delta}$ yields a factor proportional to $\qhat^{\nu}$, and all $\nu$-indices are contracted with the polarization tensor, the derivative $\partial_{p^{\nu}}$ can only act on $(-p^2)^{1-\Delta+j}$, resulting in
\begin{align}
    \shadow[\Phi_{m,\Delta,\pm\ell}^{\io,\mumu{1}{\ell}}]&=\sum_{i,j=0}^{\ell}\frac{\pi}{2^{2-\Delta}}\binom{\ell}{i}\binom{\ell}{j}\frac{p_{\symL\rho_1}\cdots p_{\rho_i}g_{\rho_{i+1}}{\!}^{\mu_{i+1}}\cdots g_{\rho_{\ell}\symR}{\!}^{\mu_{\ell}}}{(\Delta-1)_{i+1}}
    \nn
    \\
    &\peq\xx
    \epsilon_{\mp\ell}^{\nunu{1}{\ell}}\frac{\partial}{\partial p_{\mu_1}}\cdots\frac{\partial}{\partial p_{\mu_i}}\frac{(-p^2)^{1-\Delta}p_{\nu_1}\cdots p_{\nu_j}\qhat^{\rho_1}\cdots\qhat^{\rho_j}g_{\nu_{j+1}}{\!}^{\rho_{j+1}}\cdots g_{\nu_{\ell}}{\!}^{\rho_{\ell}}}{(-\qhat \co p)^{2-\Delta+j}}
    \, .
\end{align}
Since the polarization tensor is symmetric and transverse to $\qhat$, the sum over $j$ can be evaluated, yielding
\begin{align}
    \shadow[\Phi_{m,\Delta,\pm\ell}^{\io,\mumu{1}{\ell}}]&=\sum_{i=0}^{\ell}\frac{\pi}{2^{2-\Delta}}\binom{\ell}{i}\frac{p_{\rho_1}\cdots p_{\rho_i}g_{\rho_{i+1}}{\!}^{\mu_{i+1}}\cdots g_{\rho_{\ell}}{\!}^{\mu_{\ell}}}{(\Delta-1)_{i+1}}
    \nn
    \\
    &\peq\xx
    \epsilon_{\mp\ell}^{\nunu{1}{\ell}}\frac{\partial}{\partial p_{\mu_1}}\cdots\frac{\partial}{\partial p_{\mu_i}}\frac{(-p^2)^{1-\Delta}\cP_{\nu_1}{\!}^{\rho_1}(\qhat,\phat)\cdots\cP_{\nu_{\ell}}{\!}^{\rho_{\ell}}(\qhat,\phat)}{(-\qhat \co p)^{2-\Delta}}
    \, .
\end{align}
Finally, commuting the factors $p_{\rho_1}\cdots p_{\rho_i}$ with the differential operators and invoking the transversality of the projector, we yield
\begin{align}
    \shadow[\Phi_{m,\Delta,\pm\ell}^{\io,\mumu{1}{\ell}}]=\frac{(-1)^{\ell}2^{2\Delta-2}\pi m^{2-2\Delta}}{\Delta+\ell-1}\Phi_{m,2-\Delta,\mp\ell}^{\io,\mumu{1}{\ell}}
    \, .
\end{align}

\textbf{Tachyonic case.} The shadow transform of tachyonic bases can be computed following a similar procedure to the massive case. We first consider the case $\tsign{s}=\pm i$, where the tachyonic bases in \eqref{eq: CPWF tachyonic} with $J=\pm \ell$ simplify to
\begin{equation}\label{eq: CPWF tachyonic j=l}
    \begin{aligned}
        \Phi_{\Delta,\pm\ell}^{\tsign{\pm i},im,\mumu{1}{\ell}}(\hat{q},k)
        &=
        2^{\Delta-2}(-1)^{\ell}
        (-\qhat \co k\pm i\varepsilon)^{-\Delta}
        \cP_{\nu_{1}}{\!}^{\mu_{1}}(\qhat,\khat)\cdots\cP_{\nu_{\ell}}{\!}^{\mu_{\ell}}(\qhat,\khat)\,
        \epsilon_{\pm\ell}^{\nunu{1}{\ell}}
        \, .
    \end{aligned}
\end{equation}
The corresponding shadow basis takes the form
\begin{align}
    \shadow[\Phi_{\Delta,\pm\ell}^{\tsign{\pm i},im,\mumu{1}{\ell}}]=2^{\Delta-2}\epsilon_{\mp\ell}^{\nunu{1}{\ell}}(\qhat)\int d^2z'\frac{\cP_{\nu_1}{\!}^{\rho_1}(\qhat,\qhat')\cdots\cP_{\nu_\ell}{\!}^{\rho_\ell}(\qhat,\qhat')}{(-\frac{1}{2}\qhat\co \qhat')^{2-\Delta}(-\qhat' \co k\pm i\varepsilon)^{\Delta}}\cP_{\rho_{1}}{\!}^{\mu_{1}}(\qhat,\khat)\cdots\cP_{\rho_{\ell}}{\!}^{\mu_{\ell}}(\qhat,\khat)
    \, .
\end{align}
Using equation \eqref{eq: projector qk} together with the transversality properties of the projectors enables us to express the shadow bases as
\begin{align}
    \label{eq: tachyonic shadow bases with z integral}
    \shadow[\Phi_{\Delta,\pm\ell}^{\tsign{\pm i},im,\mumu{1}{\ell}}]&=\epsilon_{\mp\ell}^{\nunu{1}{\ell}}\sum_{i,j=0}^{\ell}\binom{\ell}{i}\binom{\ell}{j}\frac{k_{\rho_1}\cdots k_{\rho_i}g_{\rho_{i+1}}{\!}^{\mu_{i+1}}\cdots g_{\rho_{\ell}}{\!}^{\mu_{\ell}}\qhat^{\symL\rho_1}\cdots\qhat^{\rho_j}g_{\nu_{j+1}}{\!}^{\rho_{j+1}}\cdots g_{\nu_{\ell}}{\!}^{\rho_{\ell}\symR}}{(\Delta)_i(\Delta-j)_j}
    \nn
    \\
    &\peq\xx
    \frac{\partial}{\partial k_{\mu_1}}\cdots\frac{\partial}{\partial k_{\mu_i}}\frac{\partial}{\partial k^{\nu_1}}\cdots\frac{\partial}{\partial k^{\nu_j}}
    \intt{d^2z'}
    (-\qhat\co \qhat')^{\Delta-2-j}
    (-\qhat' \co k\pm i\varepsilon)^{-\Delta+j}
    \, ,
\end{align}
where it is be understood that all $\mu$-indices are symmetrized.

Then the $z'$-integral is the conformal two-point integral \eqref{eq: cft two-point integral qqk}, giving
\begin{align}
    \shadow[\Phi_{\Delta,\pm\ell}^{\tsign{\pm i},im,\mumu{1}{\ell}}]&=-\epsilon_{\mp\ell}^{\nunu{1}{\ell}}\sum_{i,j=0}^{\ell}\binom{\ell}{i}\binom{\ell}{j}\frac{k_{\rho_1}\cdots k_{\rho_i}g_{\rho_{i+1}}{\!}^{\mu_{i+1}}\cdots g_{\rho_{\ell}}{\!}^{\mu_{\ell}}\qhat^{\symL\rho_1}\cdots\qhat^{\rho_j}g_{\nu_{j+1}}{\!}^{\rho_{j+1}}\cdots g_{\nu_{\ell}}{\!}^{\rho_{\ell}\symR}}{(\Delta)_i(\Delta-j)_j}
    \nn
    \\
    &\peq\xx
    \frac{\partial}{\partial k_{\mu_1}}\cdots\frac{\partial}{\partial k_{\mu_i}}\frac{\partial}{\partial k^{\nu_1}}\cdots\frac{\partial}{\partial k^{\nu_j}}\frac{e^{\mp i\pi(\Delta-j)}2^{\Delta-j-2}\pi}{(\Delta-j-1)}\frac{(k^2)^{1-\Delta+j}}{(-\qhat \co k\pm i\varepsilon)^{2-\Delta+j}}
    \, .
\end{align}
We observe that the derivative $\partial_{k^{\nu}}$ can only act on $(k^2)^{1-\Delta+j}$, as all $\nu$-indices are contracted with the polarization tensor. This leads to the expression:
\begin{align}
    \shadow[\Phi_{\Delta,\pm\ell}^{\tsign{\pm i},im,\mumu{1}{\ell}}]&=-\epsilon_{\mp\ell}^{\nunu{1}{\ell}}\sum_{i,j=0}^{\ell}\binom{\ell}{i}\binom{\ell}{j}\frac{k_{\rho_1}\cdots k_{\rho_i}g_{\rho_{i+1}}{\!}^{\mu_{i+1}}\cdots g_{\rho_{\ell}}{\!}^{\mu_{\ell}}\qhat^{\symL\rho_1}\cdots\qhat^{\rho_j}g_{\nu_{j+1}}{\!}^{\rho_{j+1}}\cdots g_{\nu_{\ell}}{\!}^{\rho_{\ell}\symR}}{(\Delta)_i}
    \nn
    \\
    &\peq\xx
    \frac{\partial}{\partial k_{\mu_1}}\cdots\frac{\partial}{\partial k_{\mu_i}}\frac{e^{\mp i\pi\Delta}2^{\Delta-2}\pi}{(\Delta-1)}\frac{(k^2)^{1-\Delta}k_{\nu_1}\cdots k_{\nu_j}}{(-\qhat \co k\pm i\varepsilon)^{2-\Delta+j}}
    \, .
\end{align}
Given that the polarization tensor is symmetric and transverse to $\qhat$, the sum over $j$ can be evaluated, yielding
\begin{align}
    \shadow[\Phi_{\Delta,\pm\ell}^{\tsign{\pm i},im,\mumu{1}{\ell}}]&=-\frac{e^{\mp i\pi\Delta}\pi}{2^{2-\Delta}}\epsilon_{\mp\ell}^{\nunu{1}{\ell}}\sum_{i=0}^{\ell}\binom{\ell}{i}\frac{k_{\rho_1}\cdots k_{\rho_i}g_{\rho_{i+1}}{\!}^{\mu_{i+1}}\cdots g_{\rho_{\ell}}{\!}^{\mu_{\ell}}}{(\Delta-1)_{i+1}}
    \nn
    \\
    &\peq\xx
    \frac{\partial}{\partial k_{\mu_1}}\cdots\frac{\partial}{\partial k_{\mu_i}}\frac{(k^2)^{1-\Delta}\cP_{\nu_1}{\!}^{\rho_1}(\qhat,\khat)\cdots\cP_{\nu_{\ell}}{\!}^{\rho_{\ell}}(\qhat,\khat)}{(-\qhat \co k\pm i\varepsilon)^{2-\Delta}}
    \, .
\end{align}
Finally, commuting the factors $k_{\rho_1}\cdots k_{\rho_i}$ with the differential operators and invoking the transversality of the projector, we yield
\begin{align}
    \shadow[\Phi_{\Delta,\pm\ell}^{\tsign{\pm i},im,\mumu{1}{\ell}}]
    =
    -\frac{(-1)^{\ell}2^{2\Delta-2}\pi m^{2-2\Delta}e^{\mp i\pi\Delta}}{\Delta+\ell-1}
    \Phi_{2-\Delta,\mp\ell}^{\tsign{\pm i},im,\mumu{1}{\ell}}
    \, .
\end{align}
Then by the change of basis \eqref{eq: tachyon basis change}, we have
\begin{equation}
    \shadow[\Phi^{\tsign{\pm},im,\mumu{1}{\ell}}_{\Delta,\pm\ell}]
    =
    -\frac{(-1)^{\ell}2^{2\Delta-2}\pi m^{2-2\Delta}}{\Delta+\ell-1}\Phi^{\tsign{\mp},im,\mumu{1}{\ell}}_{2-\Delta,\mp\ell}
    \, .
\end{equation}


\newpage
\printbibliography


\end{document}